\documentclass[manuscript]{aastex}

\usepackage[dvips]{color}
\usepackage{amsmath}
\shorttitle{Multimodal DEM}
\shortauthors{Nuevo et al.}
\usepackage{natbib}
\def\Rsun{R_\odot}
\def\Eq#1{Equation (\ref{#1})}
\def\Eqs#1#2{Equations (\ref{#1})-(\ref{#2})}

\def\Tmin{T_{\rm min}}
\def\Tmax{T_{\rm max}}
\def\intmm{\int_{\Tmin}^{\Tmax}}
\def\bl{{\boldsymbol{\alpha}}}
\def\abund{{\bf a}_0}

\begin{document}

\title{Multimodal Differential Emission Measure in the Solar Corona}

\author{Federico A. Nuevo\altaffilmark{1}, Alberto M. V\'asquez\altaffilmark{1}, Enrico Landi\altaffilmark{2}, Richard Frazin\altaffilmark{2}}

\altaffiltext{1}{Instituto de Astronom\'{\i}a y F\'{\i}sica del Espacio (CONICET-UBA)
and FCEN (UBA), CC 67 - Suc 28, Ciudad de Buenos Aires, Argentina}
\altaffiltext{2}{Dept. of Atmospheric, Oceanic and Space Sciences, University
of Michigan, Ann Arbor, MI 48109}

\begin{abstract}
The Atmospheric Imaging Assembly (AIA) telescope on board the Solar Dynamics Observatory (SDO) provides coronal EUV imaging over a broader temperature sensitivity range than the previous generations of instruments (EUVI, EIT, and TRACE). Differential emission measure tomography (DEMT) of the solar corona based on AIA data is presented here for the first time. The main product of DEMT is the three-dimensional (3D) distribution of the local differential emission measure (LDEM). While in previous studies, based on EIT or EUVI data, there were 3 available EUV bands, with a sensitivity range $\sim 0.60 - 2.70$ MK, the present study is based on the 4 cooler AIA bands (aimed at studying the quiet sun), sensitive to the range $\sim 0.55 - 3.75$ MK. The AIA filters allow exploration of new parametric LDEM models. Since DEMT is better suited for lower activity periods, we use data from Carrington Rotation 2099, when the Sun was in its most quiescent state during the AIA mission. Also, we validate the parametric LDEM inversion technique by applying it to standard bi-dimensional (2D) differential emission measure (DEM) analysis on sets of simultaneous AIA images, and comparing the results with DEM curves obtained using other methods. Our study reveals a ubiquitous bimodal LDEM distribution in the quiet diffuse corona, which is stronger for denser regions. We argue that the nanoflare heating scenario is less likely to explain these results, and that alternative mechanisms, such as wave dissipation appear better supported by our results.
\end{abstract}


\section{Introduction}

Coronal heating is still a open research topic \citep{klimchuk, cranmer}. Different kinds of models have been proposed and these can be mainly classified in two broad classes: wave-driven models, in which wave dissipation transfers energy to heat the coronal plasma, and {nanoflare models, in which the magnetic energy is released on a small spatial scale in an impulsive way by different possible physical mechanisms (magnetic reconnection, turbulence, instabilities, etc).} To discriminate among the models, knowledge of the coronal \emph{differential emission measure} (DEM) is important, as different models predict it in a different way. Standard 2D determinations of DEM are affected by line-of sight (LOS) integration ambiguities. The tomographic 3D reconstruction of the DEM, known as local-DEM or LDEM, removes this problem, and allows studying {how the plasma is distributed in temperature within each tomographic grid voxel.}

\emph{Differential emission measure tomography} (DEMT) allows reconstruction of the plasma parameters (electron density and electron temperature) in 3D, using temporal series of extreme ultraviolet (EUV) images. In a first step, a temporal series of {narrow-band EUV images in a particular bandpass is used to determine the 3D distribution of the plasma emission in that channel, which we will refer to as \emph{filter band emissivity} (FBE)}. {The 3D FBE is found for every band of the EUV telescope independently.} In a second step, the FBEs in each tomographic grid voxel are used to determine the  LDEM, which describes the thermal distribution of the plasma within the voxel. Finally, moments of the LDEM are taken in each voxel to determine the mean electron density and the mean electron temperature. {Previously published DEMT works are based on images from the \emph{Extreme Ultraviolet Imager} (EUVI), on board the \emph{Solar TErrestrial RElations Observatory} (STEREO) mission, a telescope that has 3 coronal bands covering a temperature sensitivity range between 0.6 and 2.7 MK. \citet{prominences} produced the first observational 3D analysis of coronal prominence cavities, while \citet{CR2077, vasquez11} analyzed the global structure of the solar corona during the minimum of activity between solar cycles 23 and 24. \citet{huang12} and \citet{nuevo13} combined DEMT with \emph{potential field source surface} (PFSS) models of the global magnetic field to study the temperature structure of the quiet diffuse corona. Their studies revealed the ubiquitous presence of magnetic loops with downward gradients of temperature, and showed that they are present only during the solar minimum epoch. DEMT has been also succesfully used as a validation tool for MHD modeling of the the solar corona \citep{jin_12, evans} and the solar wind \citep{rona}.
 
The \emph{Atmospheric Imaging Assembly} (AIA) telescope on board the \emph{Solar Dynamics Observatory} (SDO) provides 6 coronal bands covering a temperature sensitivity range between 0.5 and 15 MK. The {increased number of filters and their broader temperature range covered compared to EUVI}, allows exploration of new models for the LDEM (or standard 2D DEM), providing more accurate estimates of the plasma parameters. As DEMT focuses on the slowly evolving quiet Sun, the AIA filters more appropriate for DEMT studies are those of 171, 193, 211, and 335 \AA, with maximum sensitivity temperatures in the range 0.9 to 2.5 MK. {The 94 and 131 \AA\ bands have been primarily designed to record very hot flaring plasma, and thus exhibit a much smaller signal-to-noise (S/N) ratio in the quiet diffuse corona, which leads to more noisy FBE reconstructions. In the case of the 94 \AA\ filter, its \emph{temperature response function} (TRF) includes a peak at about 7 MK and another one at about 1.25 MK \citep{boerner12, lemen12} but its sensitivity at the lower temperatures is poorly known \citep{aschwanden13}. These bands could be used as well, but this is deferred to a future publication. Finally, the 304 \AA\ emission, that corresponds to He II {and Si XI}, is optically thick on the disk and hence not useful for quantitative tomography (see Frazin et al. 2009)}.}

{DEM inversion problems are difficult because of their ill-posed nature. When the DEM is inverted using spectra as constraint, such as those provided for example by the \emph{Extreme-ultraviolet Imaging Spectrograph} (EIS) on board \emph{Hinode}, \emph{Monte Carlo Markov Chain} (MCMC) approaches (Kashyap \& Drake, 1998), or regularized inversion techniques (Hannah \& Kontar, 2012) may be applied, with no need to prescribe a parametric functional form for the DEM. {These methods have too many free parameters to apply to data produced by narrow band filters. The capabilities of MCMC inversion techniques were recently investigated by \citet{testa} using synthetic EIS and AIA images computed from a 3D radiative MHD simulation. They found that the inversion method is able to reproduce some global properties of the thermal distribution when using EIS images. The inversion is less accurate when using AIA images due to the more limited number of constraints and the broad nature of the TRF of its filters. Similar conclusions were found by \citet{delzanna2013}, who explored the limitations of applying MCMC methods to real AIA images. Also, using synthetic spectra computed from assumed DEM distributions, Landi et al. (2012) studied the ability of MCMC methods to reconstruct nearly isothermal plasmas, and found them unable to resolve them from multithermal plasmas with width smaller than $\Delta{\rm log}\,T=0.05$.}

{In the case of narrow band images, parametrization of the DEM is a useful approach, as it provides simple, computationally efficient solutions. Parametric DEM studies, related to active regions, have been recently carried out by Aschwanden \& Boerner (2012), Plowman et al. (2013), and Del Zanna et al. (2013). Parametric inversion has been implemented in DEMT to find the LDEM using the FBE values as constraints (V\'asquez et al. 2009; Frazin et al. 2009). In this case, the values of the free parameters of the model are found as to minimize a functional that measures the mean square difference between the tomographic and synthesized FBEs. This is performed independently for each voxel of the tomographic grid. In previous DEMT works the LDEM was modeled by a single normal (Gaussian) function constrained by the FBEs provided by the 3 bands of the EUVI telescopes. In this work, a variety of parametric models for the LDEM is explored to best reproduce the tomographic FBEs of the 4 AIA bands selected for DEMT.} 

The article is organized as follows: In section 2, we describe the DEMT technique and the LDEM parametrizations that are tested. Sections 3.1 and 3.2 include a study of the LDEM inversion from simulated data and a comparison of DEMT results based on data from EUVI and AIA, respectively. Section 3.3 analyses the DEMT results using different parametric models. Section 3.4 shows an analysis of the bimodal properties of the LDEM in different coronal structures. Section 3.5 is 2D DEM validation study of the parametric inversion and the bimodal model. In Section 4 a discussion and summary of the results of this work is presented.

\section{{Methods}}

\subsection{The DEMT technique}
\label{DEMT}

{DEMT consists of two steps. In a first step, a time series of EUV images is used to tomographically determine the 3D distribution the plasma emission in a given band (the FBE), formally defined in Equation (1) below.} To that end the corona is discretized in a spherical computational grid. This first task is independently performed for each EUV band. {In a second step, the FBE values of all bands are used as constraints to the inversion of the local temperature distribution (the LDEM) of the coronal plasma. This second task is independently performed for each voxel of the tomographic grid.} From the resulting LDEM distributions, 3D maps of the coronal electron density and temperature can be derived. This section provides a brief description of the key concepts of DEMT that are needed in this work. For a detailed explanation the reader should consult Frazin et al. (2009).}

\begin{figure}[ht]
  \begin{center}
  \includegraphics[height=0.28\linewidth]{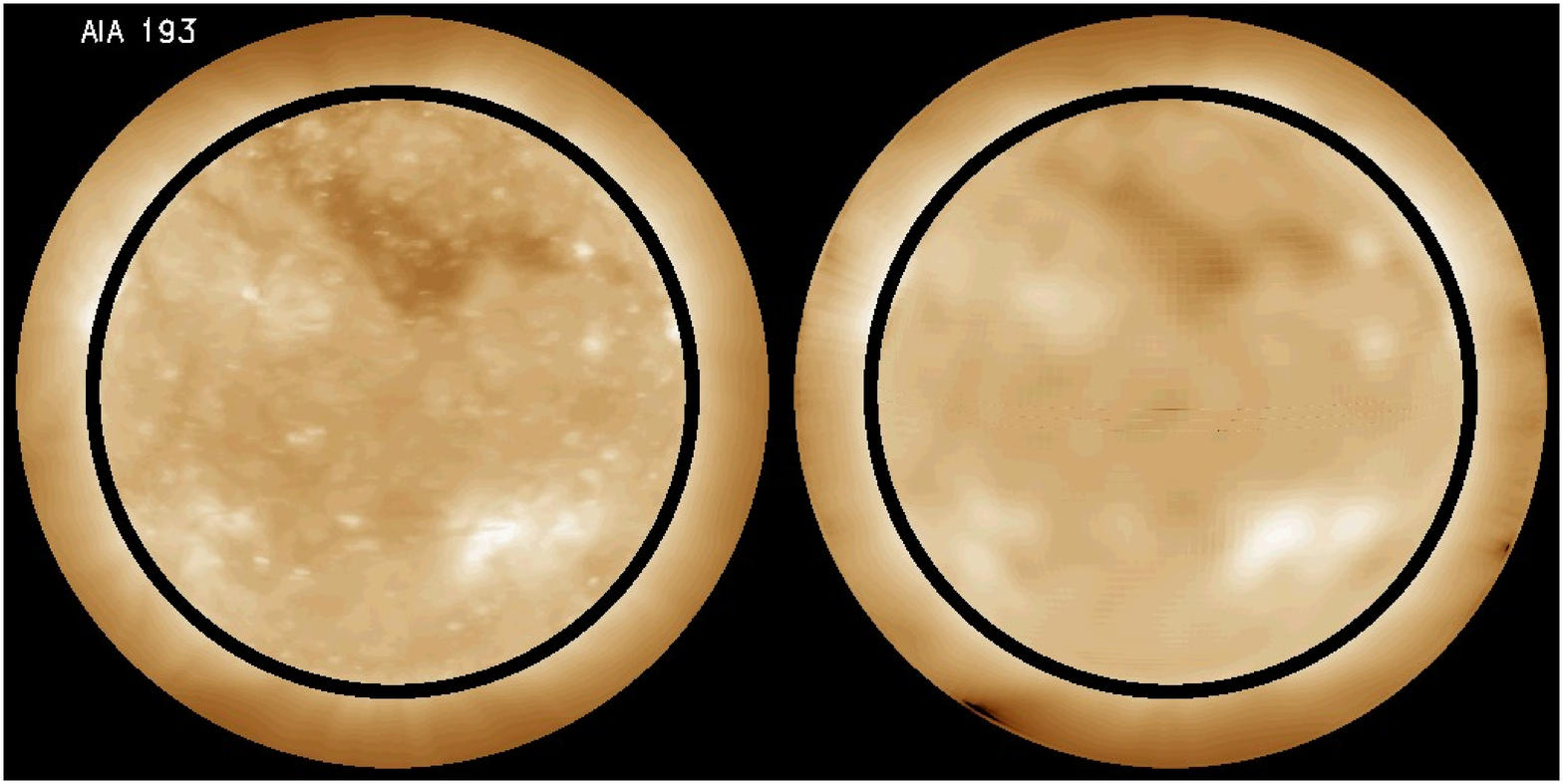}	
  \includegraphics[height=0.28\linewidth]{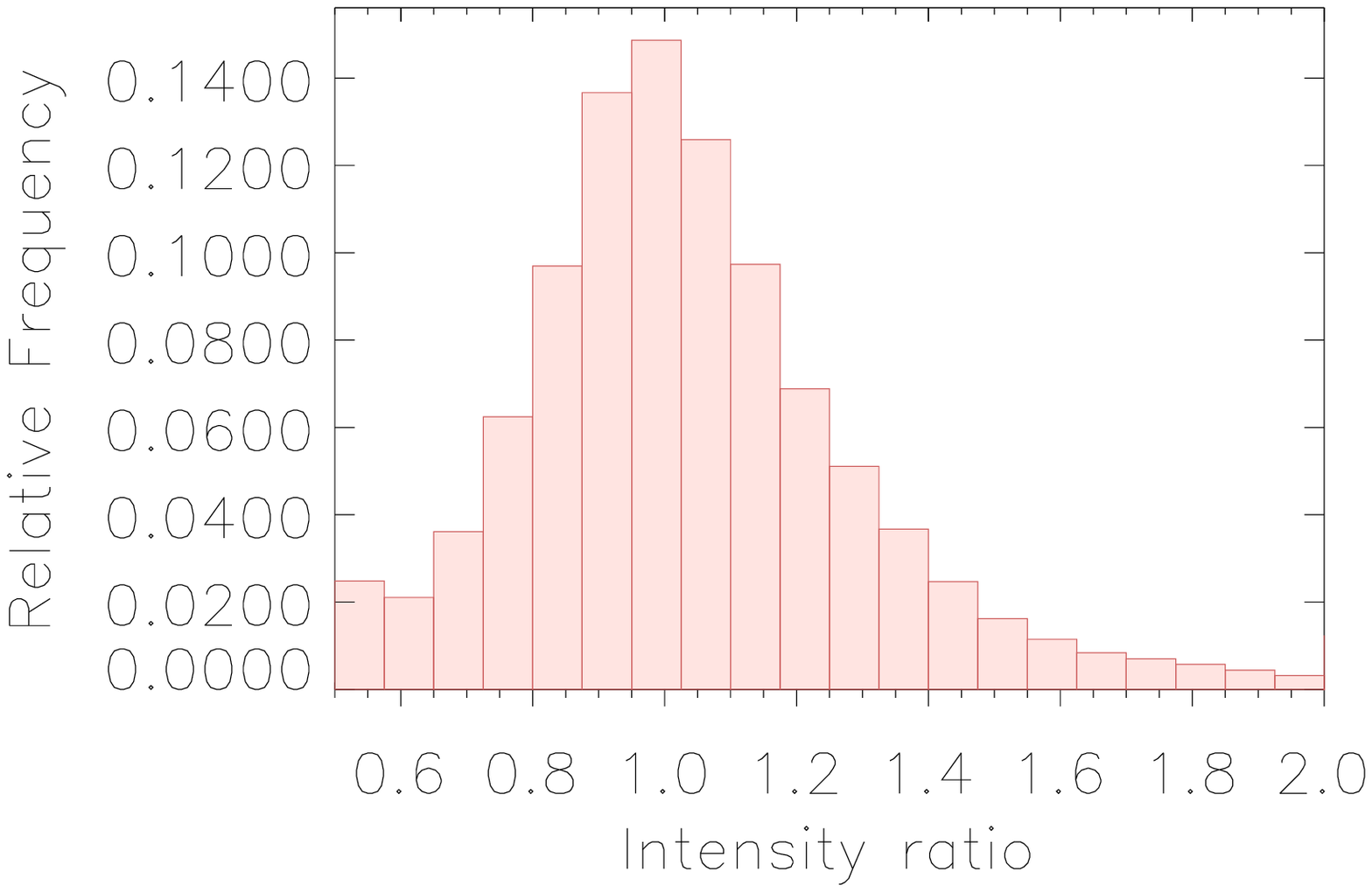}\\
   \end{center}
  \caption{{Left panel: image taken by the 193 \AA\ band of the AIA instrument (left), and corresponding synthetic image computed by LOS-integration of the tomographic FBE (right). Right panel: frequency histogram of the synthetic to observed intensity ratio for every corresponding pair of pixels in the two images.}}
  \label{tomo}
\end{figure}

{The AIA instrument images the off-limb corona up to about 1.25 $\Rsun$. To perform EUV tomography, the inner corona volume in the {height range 1.00-1.25 $\Rsun$ is discretized on a 25$\times$90$\times$180 (radial $\times$ latitudinal $\times$ longitudinal) spherical grid. Due to optical depth issues and EUV S/N levels, the DEMT results are reliable in the height range from 1.03 to 1.20 $\Rsun$}. For each band $k$, a series of images covering a full solar rotation is tomographically inverted to find, for each tomographic cell $i$, the value of the FBE} {$\zeta_i^{(k)}$}, defined by \citep{frazin09}

\begin{equation}
{\zeta^{(k)}_i=\int d\lambda \, \phi_k(\lambda) \, \eta_i(\lambda) \,,}
\end{equation}

\noindent
where $\eta_i(\lambda)$ is the {plasma emissivity volumetrically averaged over the $i$-th tomographic cell} and $\phi_k(\lambda)$ is the {normalized bandpass function of the band $k$. Due to temporal changes in the corona, tomographic reconstructions exhibit artifacts such as smearings and negative values of the reconstructed FBEs}, or zero when the solution is constrained to positive values. These are called zero-density artifacts (ZDAs, see Frazin et al. 2009). {Synthetic images based on the tomographic reconstruction can be produced by numerically computing line of sight (LOS) integrals of the 3D FBE}. {As an example, the left panel in Figure \ref{tomo} shows an image taken with the 193 \AA\ AIA filter (left) and the corresponding synthetic image computed from the tomographic FBE (right). The right panel shows the frequency histogram of the ratio of synthetic to observed intensity ratio for every corresponding pair of pixels in the two images. The relative difference between the synthetic and observed values in each pixel is below 10, 20 and 30\%, for 37, 63 and 79\% of the pixels, respectively, and virtually the same statistics hold when off-limb or on-disk pixels are considered separately. A similar level of agreement is found for all AIA bands. The tomographic model provides then a quite detailed reliable description of the average global corona during the reconstructed period.}

{Within any given tomographic voxel, the LDEM $\xi_i(T)$ is a measure of the thermal distribution within the $i^{\rm th}$ voxel. Following Frazin et al. (2009), the FBE and the LDEM are related by}

\begin{equation}
 \zeta_{k,i} = \intmm dT\ Q_k(T) \ \xi_i(T)\,,
\label{FBE_LDEM}
\end{equation}

\noindent
where the \emph{temperature response function} (TRF) $Q_k(T)$ is defined as

\begin{equation}
Q_k(T) \equiv \int d\lambda \  \phi_k (\lambda)\,     \eta(\lambda;T,N_{e0},\abund)\,/\,N_{e0}^2 \,,
\label{Qkl_def}
\end{equation}
in which $N_{e0}$ is a reference electron density and $\abund$ is a reference set of abundance used to computed the plasma emissivity from an optically thin plasma emission model such as CHIANTI (Dere et al. 1997).

Due to the limited number of available bands, $K$, the inversion of LDEM function is under determined. The solution is implemented by modeling the LDEM with a family of functions $\xi_i (T) = F \,(T,\bl_i)$, depending on a vector of parameters $\bl_i$, with $L\sim K $ elements. In each tomographic cell $i$ the problem consists of finding the values of the parameters $\bl_i$ that allow to best reproduce the $K$ tomographically reconstructed values of FBE in that cell.  To do so an objective function is defined, that measures the quadratic differences between tomographically determined FBEs and the FBEs synthesized from the modeled LDEM, i.e.

\begin{equation}
{ \Phi\left(\bl_i\right) \equiv \sum_{k=1}^{K} C_k^{-2} \left[\zeta_{k,i}\ - 
\intmm dT\ Q_k(T)\ F\left(T,\bl_i\right)\right]^2,}
\label{phi}
\end{equation}

\noindent
{where $K$ normalization coefficients $C_k$, each given by the median value of $\zeta_{k,i}$ in the whole reconstructed coronal volume, are included to make the objective function dimensionless. 
The minimization of the objective function is numerically implemented using the conjugate gradient method \citep{press2007}.}

At each cell $i$ the following score is then computed, as a measure of the degree of success of the LDEM in reproducing the tomographically reconstructed FBEs

\begin{equation}\label{chisqr}
R_i \, \equiv \ (1/K) \, \sum_{k=1}^K \left| \, 1 - \zeta^{(k)}_{i,\rm syn} / \,\zeta^{(k)}_{i,\rm tom}\, \right| \,,
\end{equation}

\noindent
where $\zeta^{(k)}_{i,\rm tom}$ y $\zeta^{(k)}_{i,\rm syn}$ are the tomographic and synthetic FBEs. A perfect fit implies $R=0$, and the higher the score the poorer the fit.

{Once the LDEM is found at each tomographic cell $i$, we can derive 3D maps of the plasma parameters, by taking moments of the LDEM, specifically
\begin{eqnarray}
 N_{e,i}^2 &=& \intmm dT\ \xi_i(T)\,,
\label{Ne2} \\
 T_{m,i} &=& N_{e,i}^{-2}\ \intmm dT\  \xi_i(T)\ T\,,
\label{Tm} 
\end{eqnarray}
that are the square electron density and the mean electron temperature, respectively.}
 
{The same formalism can also be used to perform a parametric 2D DEM analysis of EUV image sets. The relationship between the intensity of band $k$ in pixel $i$, $I_{k,i}$, and the DEM along the LOS associated with that pixel, $\psi_i(T)$ is, similarly to  Equation (\ref{FBE_LDEM}),
\begin{equation}
I_{k,i} = \intmm dT\ Q_k(T) \ \psi_i(T)\,,
\label{I_DEM}
\end{equation}
In the analysis below the DEM is described by the same parametric models used for the LDEM, and the optimal value of the parameter vector for each pixel is found in a similar fashion. In this case, the zeroth and first moments of the DEM distribution $\psi_i(T)$ provide the emission measure (EM) and the mean electron temperature along the LOS associated with pixel $i$, respectively. In Section \ref{DEM} a parametric 2D DEM analysis of an \emph{active region} (AR) and the diffuse region around it is developed, and the results are compared to those obtained by other authors.}

\subsection{Temperature Response Functions}
\label{TRF}

Figure \ref{fqkl} shows the temperature response function (TRF) of the {4 coronal bands of the AIA instrument used in this work, as well as those of the 3 coronal bands of the EUVI. We used the latest effective areas released by the AIA team in the Solar Soft package, which includes a broad sensitivity plateau, roughly between 400 and 900 \AA, that affects the low temperature sensitivity of all bands. Calculations were performed with version 7.1 of the CHIANTI atomic data base (Landi et al. 2013), using the abundance set {\tt sun\_coronal\_feldman\_1992\_ext.abund} \citep{feldman1992,landi1992} and the ionization equilibrium calculations set {\tt chianti.ioneq}. An important change in the spectrum introduced in CHIANTI v 7.1 is found in the wavelength range below 170 \AA, implying a more accurate calculation of the TRF of the 171 \AA\ band.  With these settings we revisited the calculation of the response function of all instruments. When compared to previous calculations,\citep{lemen12, nuevo2012b} the most important change is found in the very low temperature sensitivity of the AIA 335 \AA~ band, below the coronal temperatures considered in this work. The most noticeable change at coronal temperatures is found in the TRF of the 171 band of both instruments, becoming narrower {and thus less sensitive to temperatures below $\sim 0.5$ MK.} All the TRFs are mainly characterized by a maximum sensitivity temperature ($T_0$). Table \ref{tabQklsensit} summarizes the values of $T_0$, as well as the full width half maximum  (${\rm FWHM}_0$). {This information serves as a simple characterization of the sensitivity temperature range of each telescope band, and will be useful for setting up different aspects of the parametric} LDEM inversion, as we describe in the next Section. {Note that, in the case of the 335 \AA\ band, there is a secondary high sensitivity range that overlaps that of the 171 \AA.}

\begin{figure}[ht] 
  \begin{center} 
    \includegraphics[width=\linewidth]{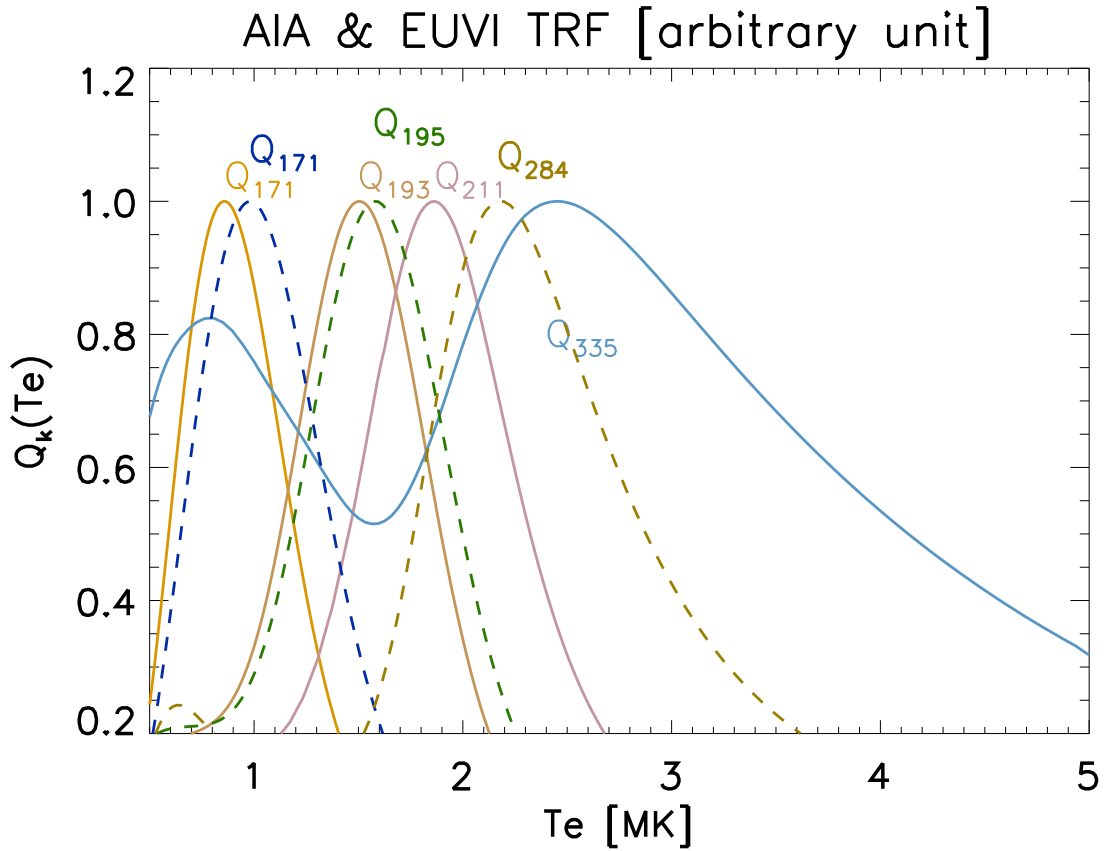}
  \end{center}
  \caption{Temperature response functions of AIA (solid) and EUVI (dashed).} 
  \label{fqkl}
\end{figure}

\begin{table}
\begin{center}
\begin{tabular}{rccccc}
\hline
Instrument & $\lambda_0$ [\AA] & 
$T_0$ [MK] & ${\rm FWHM}_0$ [MK] \\
\hline
AIA   & 171 & 0.86   &0.60 \\
      & 193 & 1.50   &0.78 \\
      & 211 & 1.87   &0.87 \\
      & 335 & 2.47   &2.58 \\
      
\hline
EUVI & 171 & 0.98    & 0.74\\
     & 195 & 1.57    & 0.82\\
     & 284 & 2.17    & 1.10\\
\hline
\end{tabular}
\end{center}
\caption{{Maximum sensitivity temperature ($T_0$) and full width half maximum  (${\rm FWHM}_0$) for SDO/AIA and STEREO/EUVI.}}
\label{tabQklsensit}
\end{table}

\subsection{Filter Sets and LDEM Parametric Models}
\label{parametrization}

In this section we describe and discuss the parametric models used to determine the LDEM.

\begin{itemize}

\item 
{Single normal model (named ``N1'' hereafter):
\begin{equation}
\xi(T) =  A\  \mathcal{N}\left(T;[T_0, \sigma_T]\right)\equiv \frac{A}{\sqrt{2\pi}\sigma_T}\, {\rm exp}\left[-\frac{1}{2}\left(\frac{T-T_0}{\sigma_T}\right)^2\right]\,.
\label{singlegaussian}
\end{equation}
\noindent
In this model, the free parameters are the centroid $T_0$, the standard deviation ${\sigma_T}$ and the amplitude $A$.} This model will be applied when using the 3 AIA bands: 171, 193 and 211 \AA\, {a set of filters we will refer to as ``AIA-3'' hereafter. This is the same parametric model used for all previous DEMT work based on data provided by the three bands of the EUVI instrument (e.g., V\'asquez et al., 2011). The temperature range considered for the inversion, indicated as the limits in the integral of Equation (\ref{FBE_LDEM}), is determined with the following criteria.} The minimum temperature is taken to be roughly  $\Tmin \approx T_0 - {\rm FWHM}_0 / 2$, where $T_0$ and ${\rm FWHM}_0$ {are the values of Table \ref{tabQklsensit} for the coolest band that is considered,  or 171 \AA\ for AIA-3}. Similarly, we take $\Tmax \approx T_0 + {\rm FWHM}_0 /2$, where $T_0$ and ${\rm FWHM}_0$ are the values of Table \ref{tabQklsensit} of the hottest band that is considered, or 211 \AA\ for AIA-3. In the case of both the AIA-3 and the EUVI sets this criteria leads to consider the temperature range {$[0.5,2.3]$} MK.

\item
Double normal model (named ``N2'' hereafter): 
\begin{equation}
\xi(T) = A_1 \ \mathcal{N}\left(T;[T_{0,1}, \sigma_{T1}]\right) 
\ + \ A_2 \  \mathcal{N}\left(T;[T_{0,2}, \sigma_{T2}]\right)\,, 
\label{doublegaussian}
\end{equation}
being a superposition of two normal functions. The first one has 3 free parameters ({as for the N1 model}). The second one has 2 free parameters (amplitude and centroid), with its standard deviation set at $\sigma_{T2} \approx 0.25~ {\rm MK}$ (which is half the temperature step between the peak response of the consecutive AIA bands 211 and 335 \AA). This model has then a total of 5 free parameters, and will be applied when using the 4 AIA bands: 171, 193, 211 and 335 \AA, a set of filters named "AIA-4" hereafter. In this case, even if the number of free parameters is larger than the number of constraints, the approach is justified as it is posed as a global optimization problem. To select the temperature range to be considered for the LDEM inversion we apply the same criteria {applied to N1}, which leads to consider the range {$[0.5,3.8]$} MK. {The strategy of this N2 model is to allow the first fully-free normal distribution to mainly describe the plasma below about 2 MK (detected by the 171, 193 and 211 \AA~bands, as well as by the low temperature end of the TRF of the 335 \AA~band), while the second normal aims at mainly describing the plasma with temperatures around 2.5 MK (detected by the main sensitivity peak of the 335 \AA~band). Accordingly, the standard deviation value set for the second normal function is of the order of the temperature difference that the ``neighbouring" 211 and 335 \AA\, bands can resolve.}

\item 
{Quadruple normal model (named ``N4'' hereafter):}
\begin{equation}
\xi(T) =  \sum_{j=1}^4 A_j\  \mathcal{N}\left(T;[T_{0,j}, \sigma_{T,j}]\right) \,,
\label{fourgaussian}
\end{equation}
{being a superposition of four normal functions with their centroids and widths set at prescribed values, with 4 free parameters that are the amplitudes of each normal function. The fixed centroids are set to the $T_0$ values of the four TRFs (see table \ref{tabQklsensit}) and the fixed widths are set equal to half the temperature difference between the $T_0$ values of consecutive bands. The strategy of this parametrization, applied when using AIA-4, is for each normal function to predict the emissivity of the respective band.  
\item
{
Other unimodal models: For AIA-4 we experimented several alternate unimodal models, specifically: 
\begin{enumerate}
 \item[a)] The N1 model.
 \item[b)] The ``top hat'' (TH) model,
 \begin{equation}
 \xi(T) = {\rm TH}\left(T;[A,T_0, \sigma_T,a]\right) \equiv 
 \frac{A}{\sqrt{2\pi}\sigma_T}\, 
 {\rm exp}\left[-\frac{1}{2}\left(\frac{T-T_0}{\sigma_T}\right)^{2a}\right]\, ,
 \label{TH}
 \end{equation}
 characterized by 4 free parameters: an amplitude $A$, a centroid $T_0$, a width $\sigma_T$, and a parameter $a$ controling the gradient of the sensitivity drop at both the cooler and the hotter temperature ends of the model. Note that for $a=1$ the model becomes normal, while for larger values of $a$ the shape of the model resembles that of a top hat function.
 \item[c)] The asymmetric ``top-hat'' (ATH) model,
 \begin{equation}
 \xi(T) = {\rm ATH\left(T;[A,T_0, \sigma_T,a,p]\right)} \equiv {\rm TH}\left(T;[A,T_0, \sigma_T,a]\right) \ T^p\, ,
 \label{ATH}
 \end{equation}
 which multiplies the TH model by a power law, adding a 5$^{\rm th}$ free (signed) parameter $p$ that allows to achieve asymmetrical distributions. 
\end{enumerate}
}
}
\end{itemize}

{Figure \ref{ejemplos_LDEM}  shows examples for all the models described above. As mentioned in the introduction, similar models have been used by recent parametric DEM studies. \citet{aschwanden11} performed DEM analysis along bright loops in ARs based on AIA images, modeling the DEM with a combination of different number of normal distributions in order to reproduce the emission observed in the six bands of AIA. Del Zanna (2013) also used AIA data to study the DEM in the core of an AR, modeling the DEM in each pixel with $\sim 10$ normal distributions spaced across the temperature range from 0.5 to 4.0 MK.}

\section{Results}

\subsection{Inversion of Simulated Data}\label{simul}

{With the aim of performing a controlled study of the LDEM inversion, and understanding the capabilities and limitations of the parametric models, we implemented an ensemble of assumed LDEM curves and derived synthetic FBE data. To generate the synthetic data {we adopt for the LDEM a ``top-hat'' model with a very steep gradient at both ends of the distribtion. The free parameters of the model are set so that the resulting distribution is} roughly constant between a fixed minimum temperature $\sim 0.8$ MK (a coronal temperature a bit lower than the maximum temperature sensitivity of the 171 \AA\ band), and a maximum temperature that {we change for each model}. The ensemble is generated by changing the maximum temperature of the simulation, that was set at a minimum value of $\sim 1.2$ MK, and incremented in steps of 0.01 MK up to maximum value of $\sim 10$ MK. The simulation with the lowest maximum temperature has all the plasma in the sensitivity range of the 171 \AA\ band, while the successive simulations add plasma at progressively larger temperatures, in the range of sensitivity of all the bands of the  AIA-4 filter set, and beyond.}

{For every assumed LDEM of the ensemble we calculated its moments using \Eqs{Ne2}{Tm}. Using the TRFs showed in section \ref{TRF} we calculated synthetic values of the FBE for all bands using \Eq{FBE_LDEM}. We then used this synthetic data to perform the  LDEM inversion, as described in section \ref{DEMT}, using the parametrizations N1 and N2 explained in section \ref{parametrization}. Finally, we calculated the moments of the inverted LDEM. We are now in a position to compare, for all simulations in the ensemble, the moments of the assumed LDEM and those of the {inverted} LDEM. Note that the way the simulations are designed, the mean temperature $T_m$ of the simulations monotonically increases with the density $N_e$, which is important to bear in mind when analyzing the results below.}

{Using the AIA-3 data set and the N1 model for the LDEM  (AIA-3/N1, hereafter), the top panels of Figure \ref{scat_sim_aia3vsaia4} show scatter plots of the moments of the assumed LDEM versus those of the inverted LDEM. The bottom panels show the same comparisons when using the AIA-4 data set and the N2 model for the LDEM (AIA-4/N2, hereafter).}

\begin{figure}[ht] 
\includegraphics[width=0.49\linewidth]{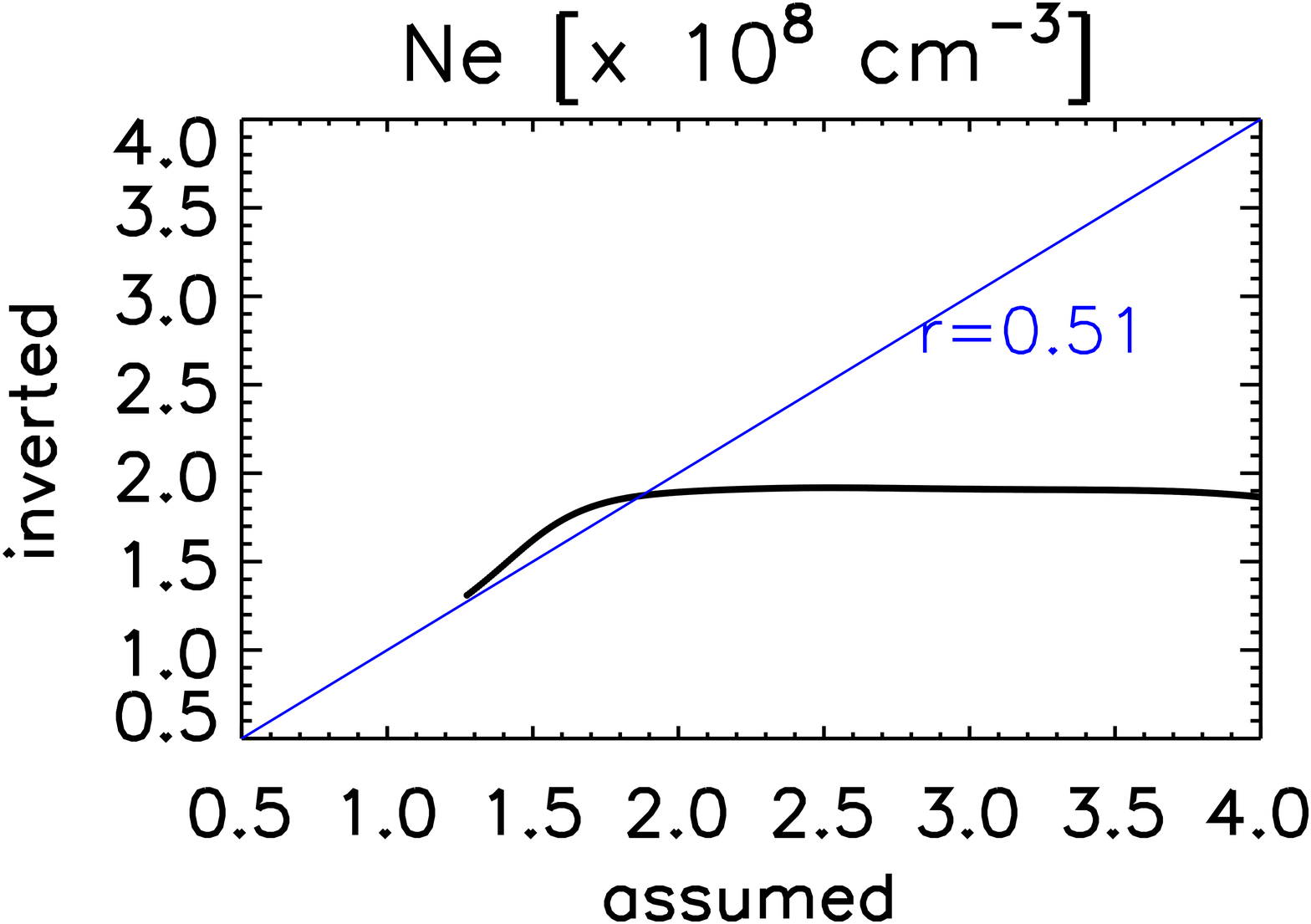}
\includegraphics[width=0.49\linewidth]{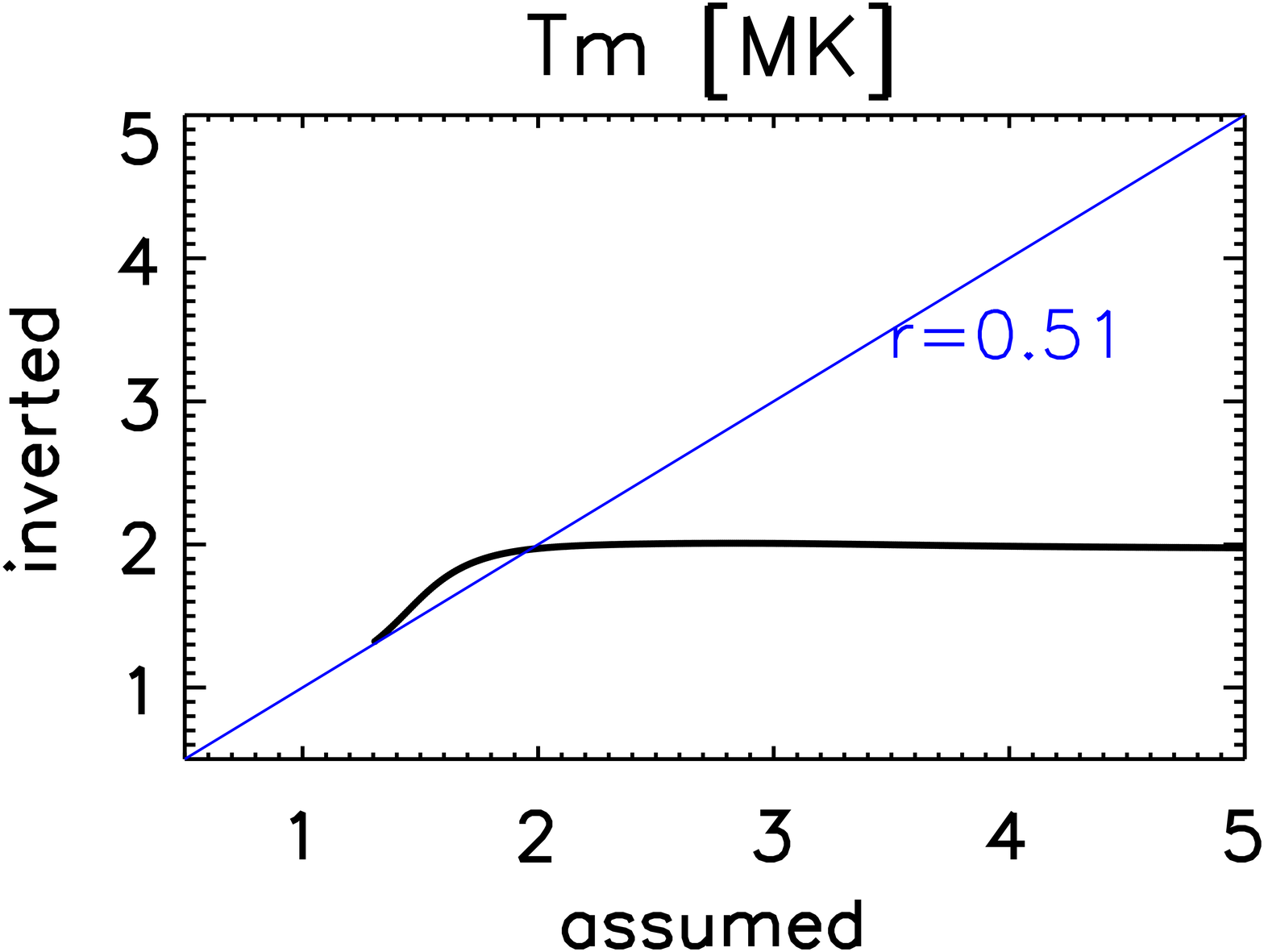}
\includegraphics[width=0.49\linewidth]{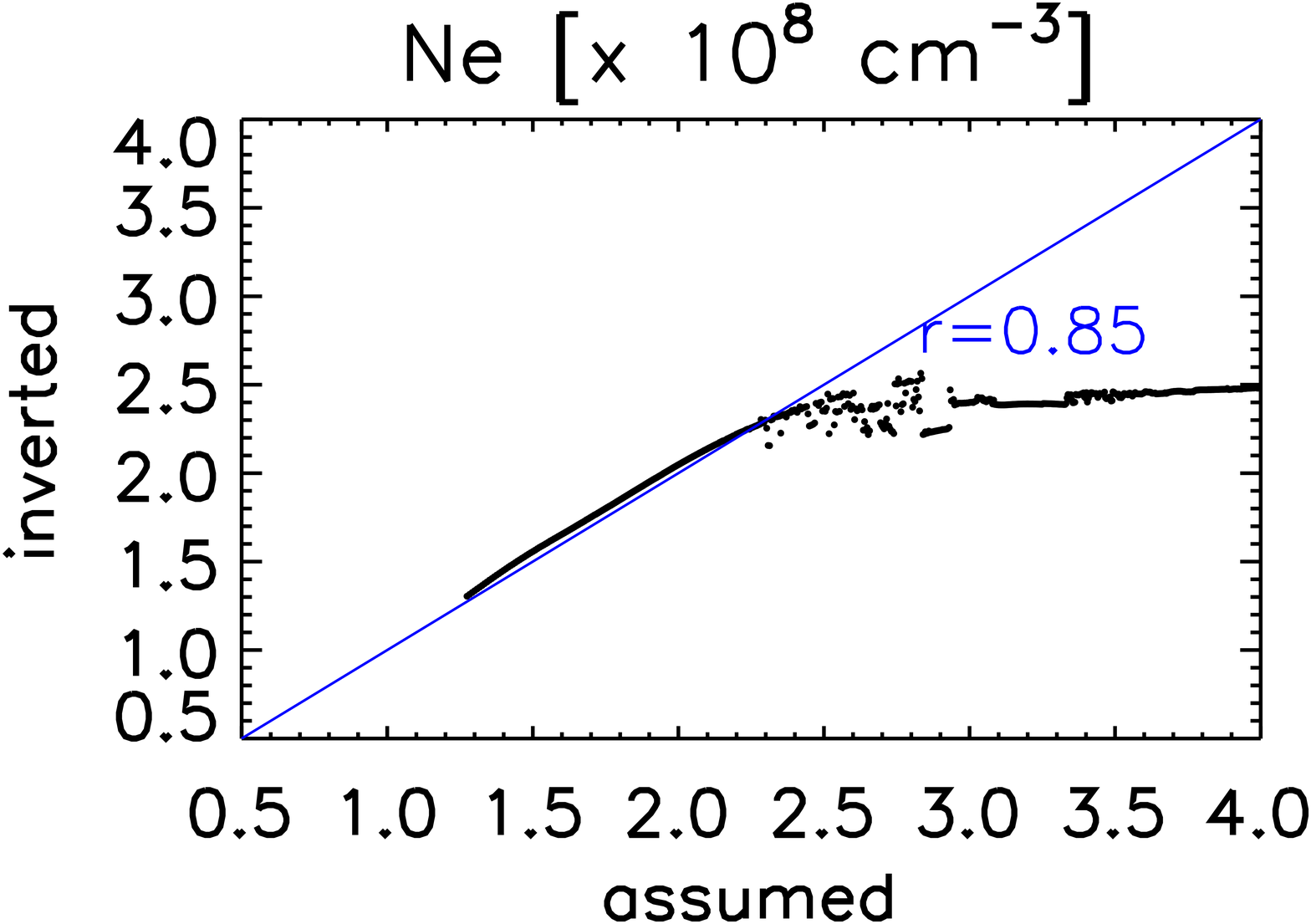}
\includegraphics[width=0.49\linewidth]{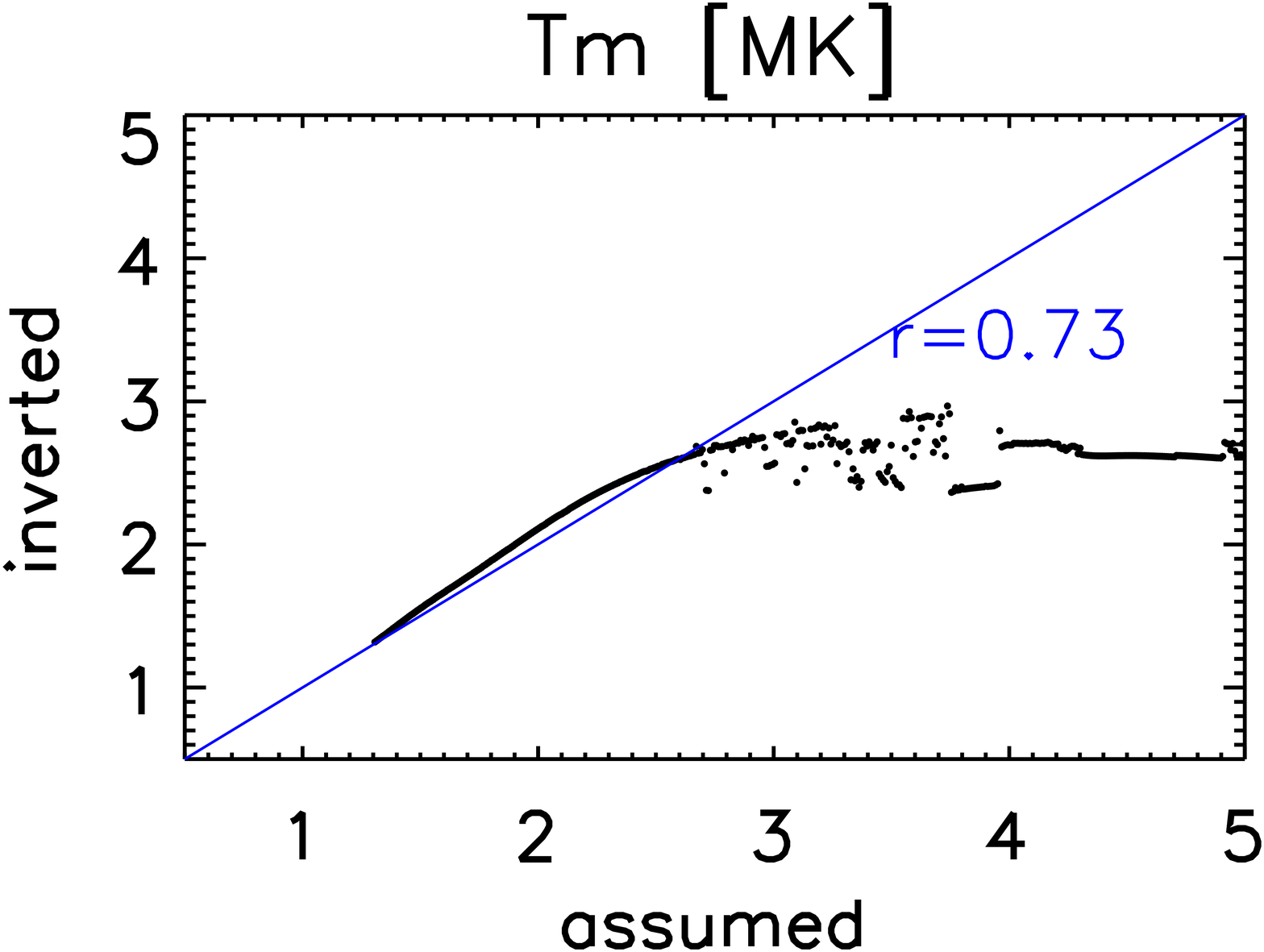}
\caption{Scatter plots of the assumed LDEM moments versus the inverted LDEM moments for AIA-3/N1 (first row) and AIA-4/N2 (second row).} 
\label{scat_sim_aia3vsaia4}
\end{figure}

{For AIA-3/N1, the simulations with $T_m < 1.7$  MK show a very high correlation between {the moments of the simulated and the inverted LDEM}, while for simulations with $T_m \gtrsim 1.8$ MK  the {moments of the} inverted LDEM are approximately constant. The reason for this is that for top-hat LDEM simulations with $T_m \gtrsim 1.8$ MK the maximum temperature becomes $T_{\rm MAX}\gtrsim 2.8$MK, so that the plasma that is added by models with increasing $T_m$ is well beyond the upper limit of the high sensitivity range of the 211 \AA~ band, which, from Table 1 is roughly $T_0+{\rm FWHM_0}/2 \sim 2.3\,{\rm MK}$.}

{Similarly, for AIA4/N2 (second row of panels) there is a good correlation between the {moments of the simulated and the inverted LDEM} for simulations with $T_m \lesssim 2.7$ MK, while for simulations with $T_m \gtrsim 2.8$ MK  the {moments of the inverted LDEM} are approximately constant. The reason for this is that for top-hat LDEM simulations with  $T_m \gtrsim 2.8$ MK the maximum temperature becomes $T_{\rm MAX}\gtrsim 4.8$MK,  so that the plasma that is added by models with increasing $T_m$ is well beyond the upper limit of the high sensitivity range of the 335 \AA~ band, which, from Table 1 is roughly $T_0+{\rm FWHM_0}/2 \sim 3.75\,{\rm MK}$. }

As previous DEMT works have been based on the EUVI instrument, we are interested in testing how the results of inverting the LDEM based on data from that instrument compares with those of inverting the LDEM based on data from AIA-3. Towards that end we have also performed a similar analysis of inversion of simulated data, as described above, but using the TRFs of the EUVI-B instrument and the N1 model for the LDEM (EUVI-B/N1, hereafter). {For every simulation of the ensemble, we compared the moments of the LDEM inverted from EUVI-B data against those of the LDEM inverted from AIA-3 data, using the N1 model in both cases. The inferred electron density differ in less than $3\%$ in all cases, while the mean temperature differ in less than $5\%$.} 

\subsection{DEMT Results with Three Bands}
\label{euvivsaia3}

We performed the DEMT analysis of the Carrington Rotation (CR) 2099 (2010, 13 July through 9 august), during the early rising phase of solar cycle 24. This period is one of the {earliest ones for which AIA data is available for the full rotation, and the most quiet period ever} observed by that instrument, which is preferred for global DEMT analysis. {As all previously published work on EUV tomography has been based on EUVI data,} we also performed reconstructions based on EUVI data for the same period, for comparison purposes. During CR-2099 the angular separation between SDO and STEREO was about $\sim 72^\circ$. As a consequence, differences observed between the results using AIA and EUVI are partly caused by solar dynamics. As the DEMT analysis is not well suited to study ARs, these were excluded from it. In the quantitative analyses shown below, tomographic voxels belonging to the open corona, the closed quiet sun (QS) corona, and the ARs where carefully separated. To this end, ARs were identified in the catalogue provided by the National Oceanic and Atmospheric Administration (NOAA) Space Weather Prediction Center\footnote{\tt www.swpc.noaa.gov/ftpmenu/warehouse/2010.html}, and verified that they correspond to the highest density values in the tomographic reconstruction. Indeed, we were able to indentify all ARs by finding the regions where the tomographic density value was above a suitable threshold value. {A threshold was established at each height of the tomographic grid independently, since the density decreases rapidly with height}. For example, at 1.075 $\Rsun$ a good way to characterize ARs in the tomographic reconstruction is to look for tomographic 
density values above $\sim 1.4 \times 10^8\,{\rm cm^{-3}}$.}

\begin{figure}[ht] 
\includegraphics[width=0.49\linewidth]{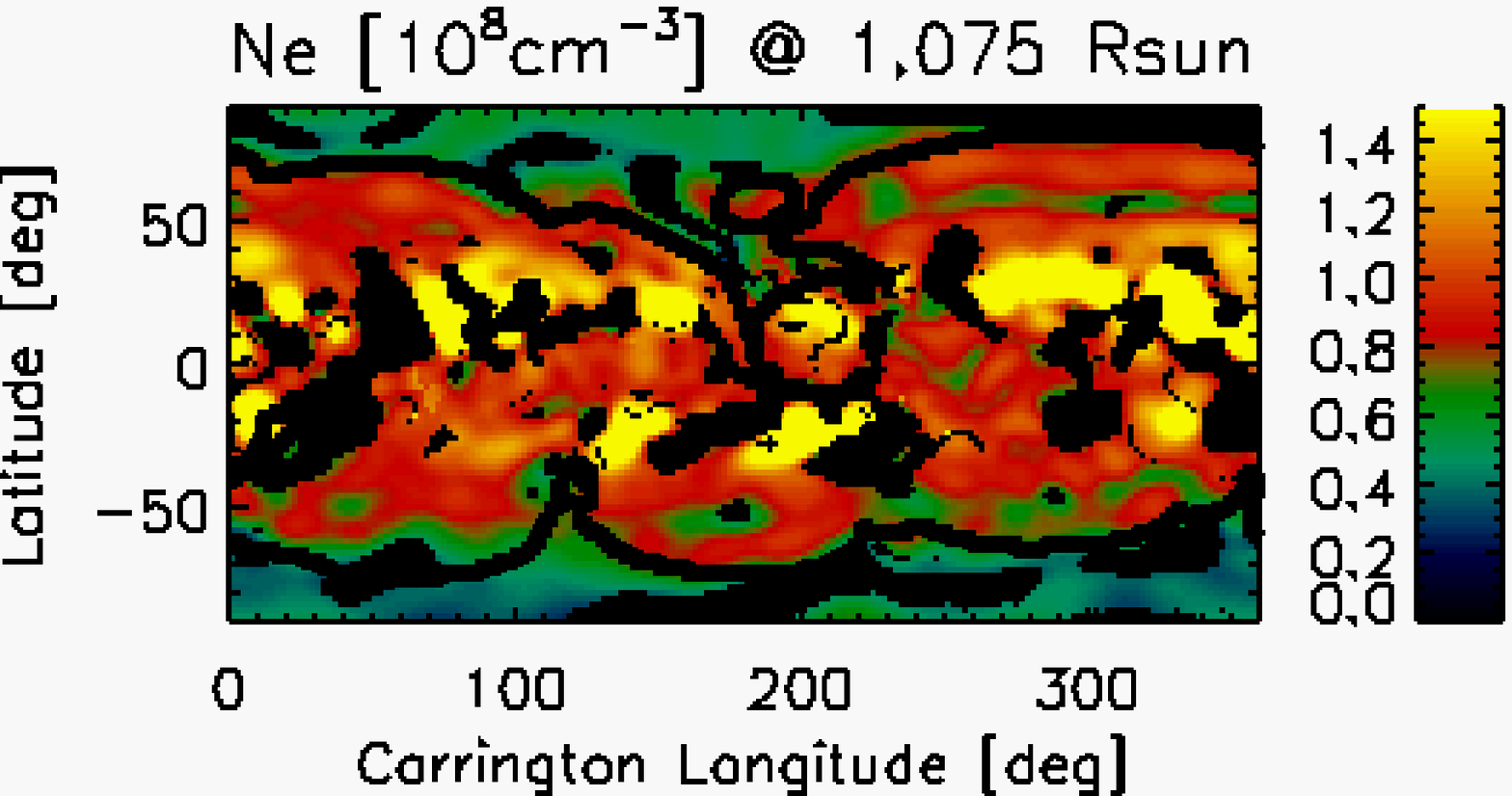}
\includegraphics[width=0.49\linewidth]{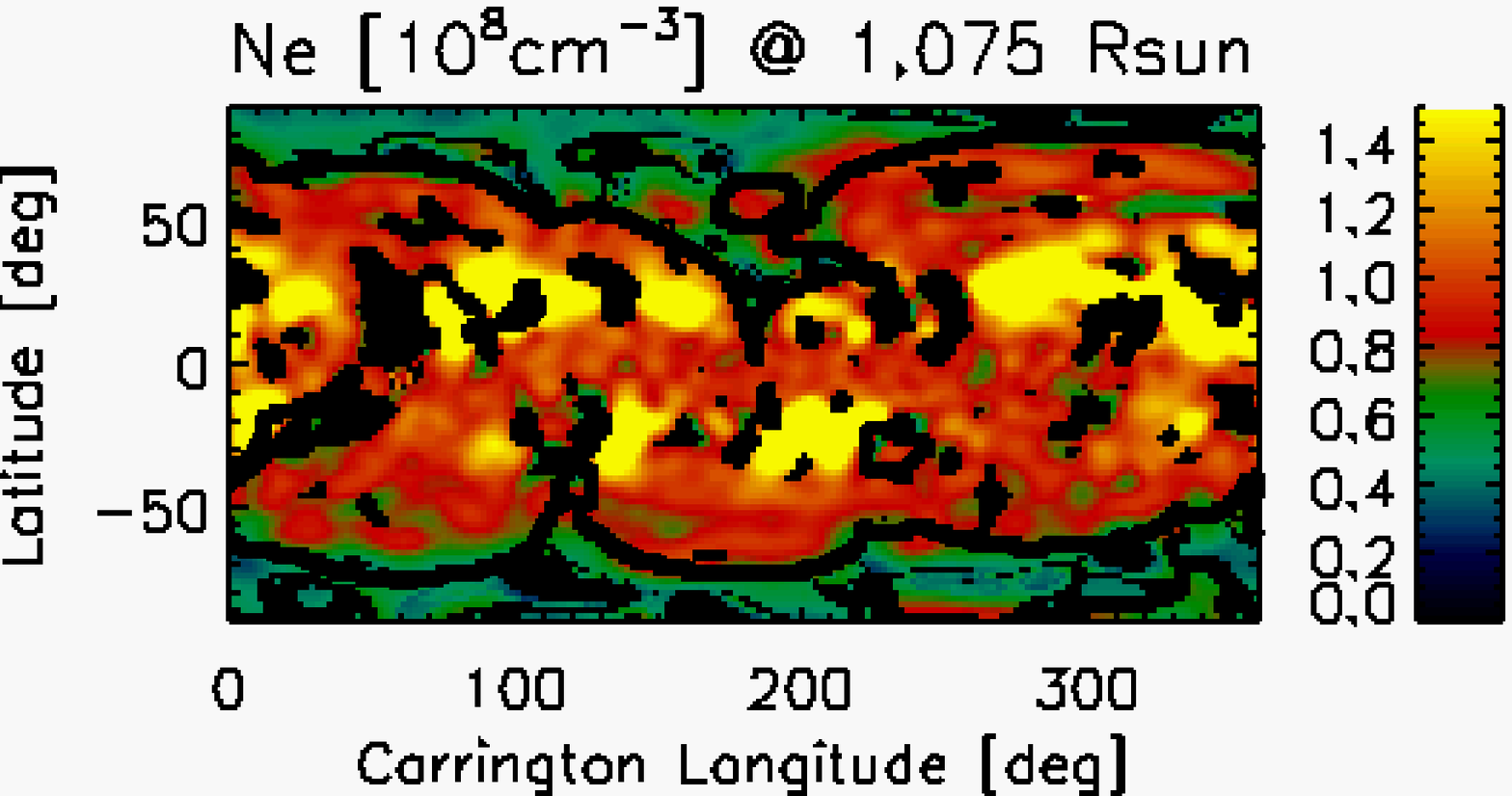}
\includegraphics[width=0.49\linewidth]{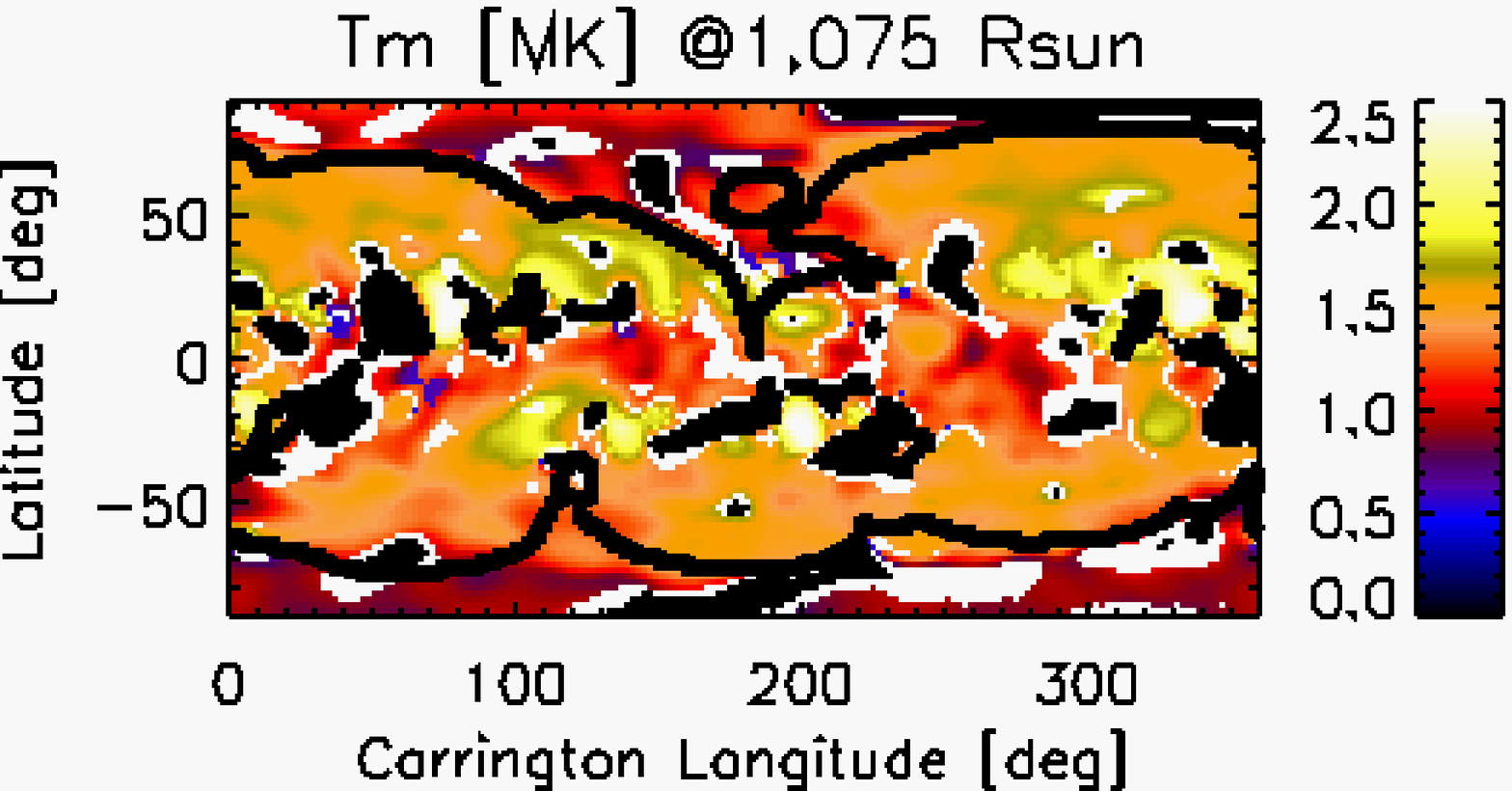}
\includegraphics[width=0.49\linewidth]{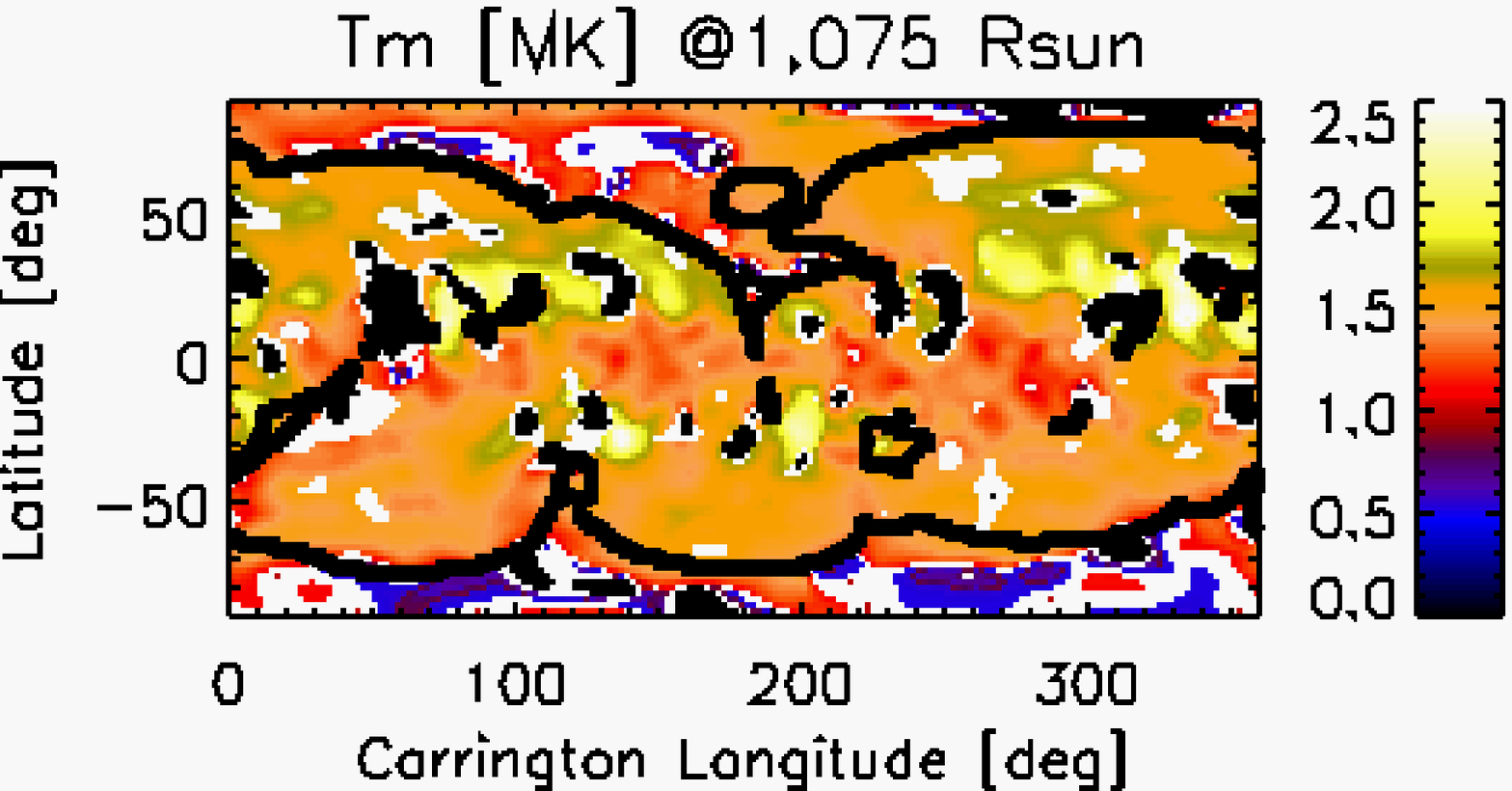}
\includegraphics[width=0.49\linewidth]{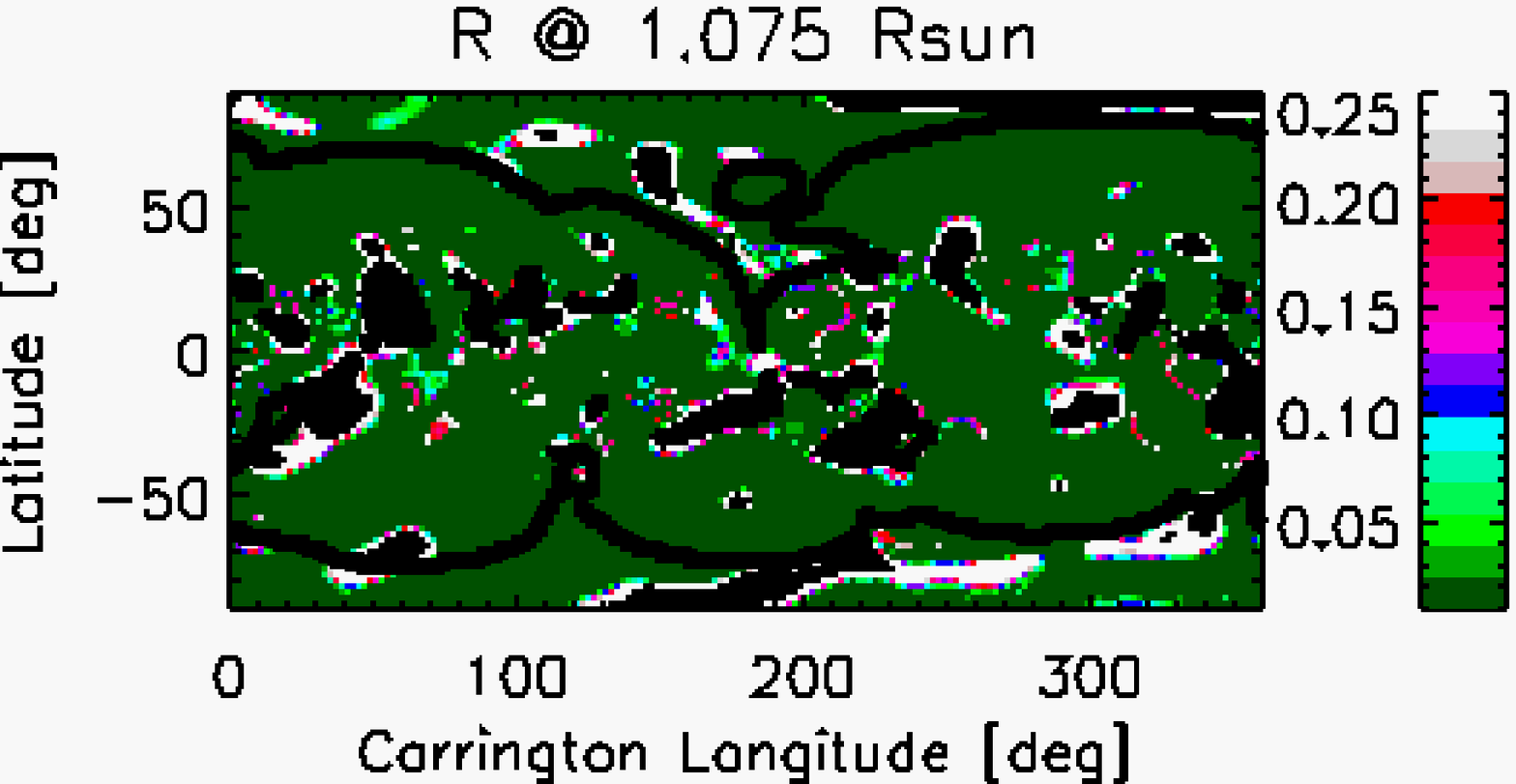}
\includegraphics[width=0.49\linewidth]{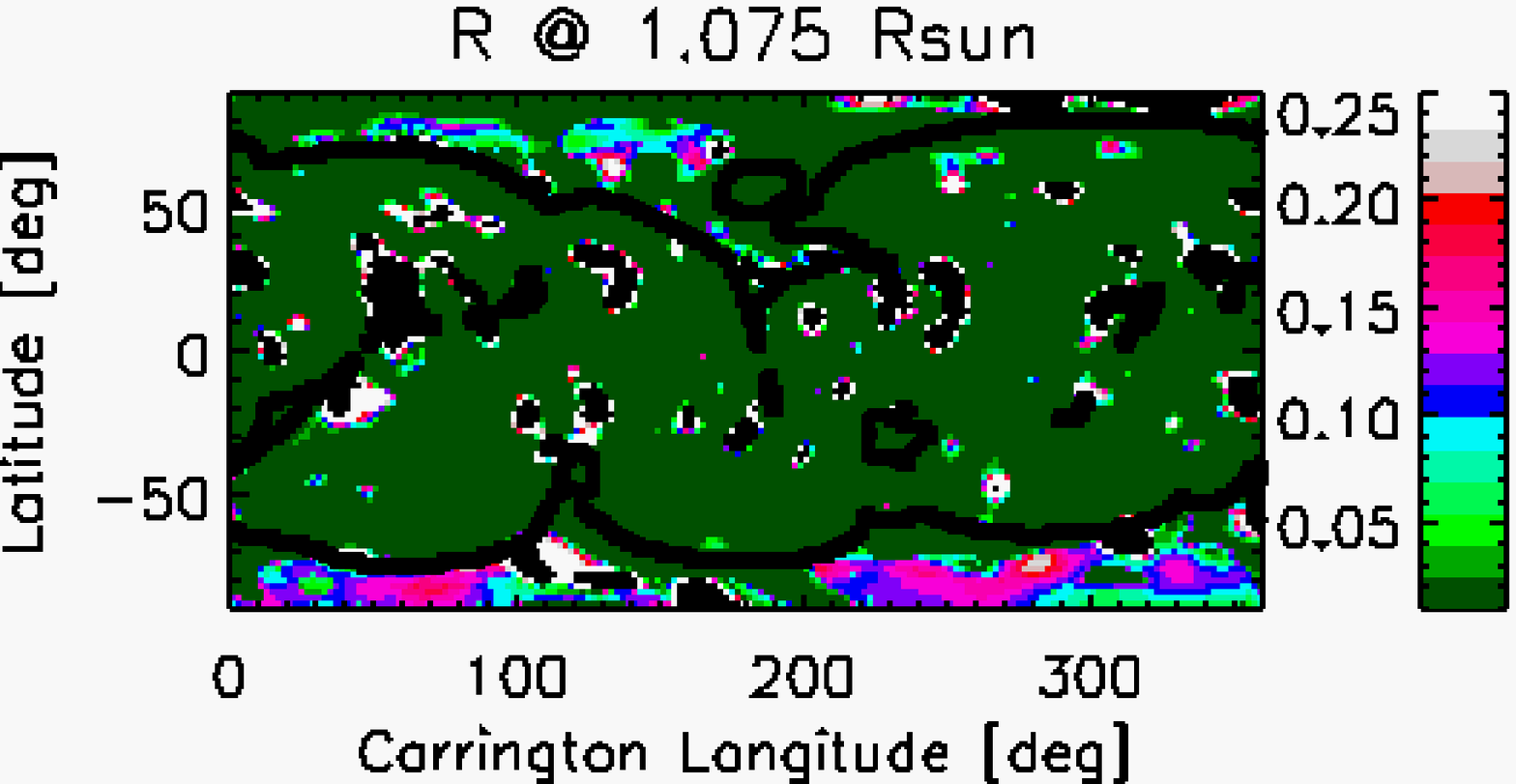}
\caption{CR-2099. Carrington maps of $N_e$, $T_m$ and the score $R$ at 1.075 $\Rsun$ using data of EUVI-B/N1 (left) and AIA-3/N1 (right). In all maps we overplot the boundary between magnetically open and closed regions as a thick black curve. ZDAs and AEVs (see text for a definition) are indicated as black and white cells, respectively, in the $T_m$ maps, and as black cells in the $N_e$ maps. In the $R$ maps the ZDAs are indicated as black cells.}
\label{maps3band}
\end{figure}

Figure \ref{maps3band} shows the DEMT results at a sample height of 1.075 $\Rsun$ using both instrumental sets (with 3 bands) and the N1 LDEM parametrization: EUVI-B/N1 and AIA-3/N1. Similar maps are also obtained at all 25 height bins of the tomographic grid. Using a PFSS extrapolation we overplot the boundaries between magnetically open and closed regions as a thick black curve. The PFSS model is computed using the finite-difference iterative solver {FDIPS} by Toth et al. (2011) on a $150\times180\times360$ spherical grid, covering 1.0 to 2.5  $\Rsun$. While the morphology of the corona is clearly more complex compared to that of solar minimum, {there is an overall good agreement between the open/closed boundary of the PFSS model and the density and temperature structures of the DEMT analysis.} {There is an overall high consistency between the results of the PFSS and the tomographic models. Figure \ref{maps3band} shows that the location of the open/closed boundary of the PFSS model is characterized by a very high transverse gradient in both the electron density and the mean temperature maps derived from the DEMT analysis. This has been previously found and analyzed in detail in DEMT studies of solar minimum rotations \citep{CR2077, vasquez11}.}

{In the mean temperature maps, black voxels indicate ZDAs, the regions of zero emissivity described in Section \ref{DEMT}. In the same maps, white voxels indicate cells for which the parametric LDEM possesses  a score $R > 0.2$, in these cases, we consider that the parametric LDEM performs poorly in reproducing the tomographic FBEs. This happens in locations where the set of FBE values is unusual, and we dub these \emph{anomalous emissivity voxels} (AEVs). In the electron density maps, both ZDAs and AEVs are indicated as black. In the score $R$ maps, ZDAs are indicated as black, while AEVs are not highlighted.}

\begin{figure}[ht]
\includegraphics[width=0.49\linewidth]{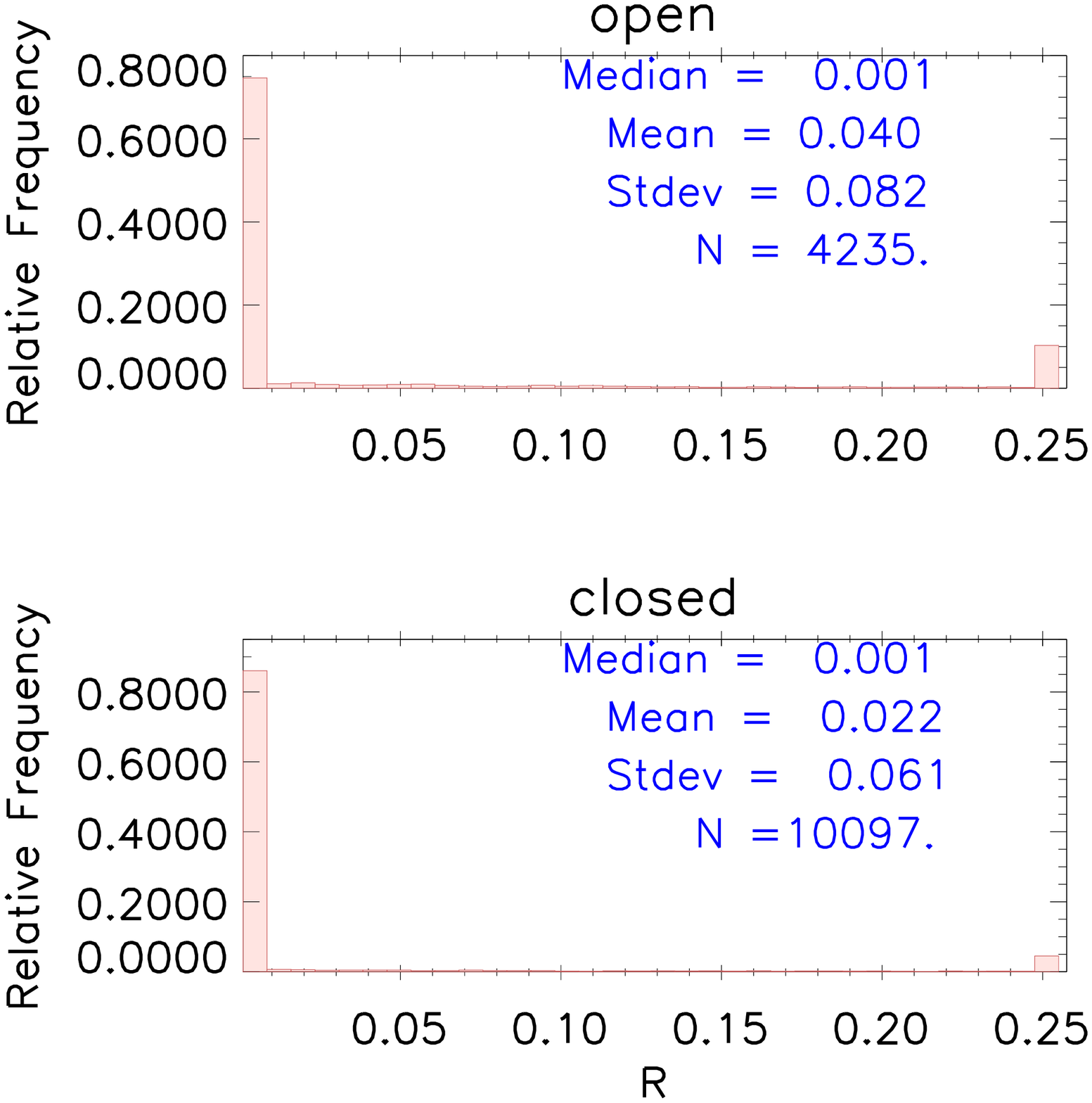}
\includegraphics[width=0.49\linewidth]{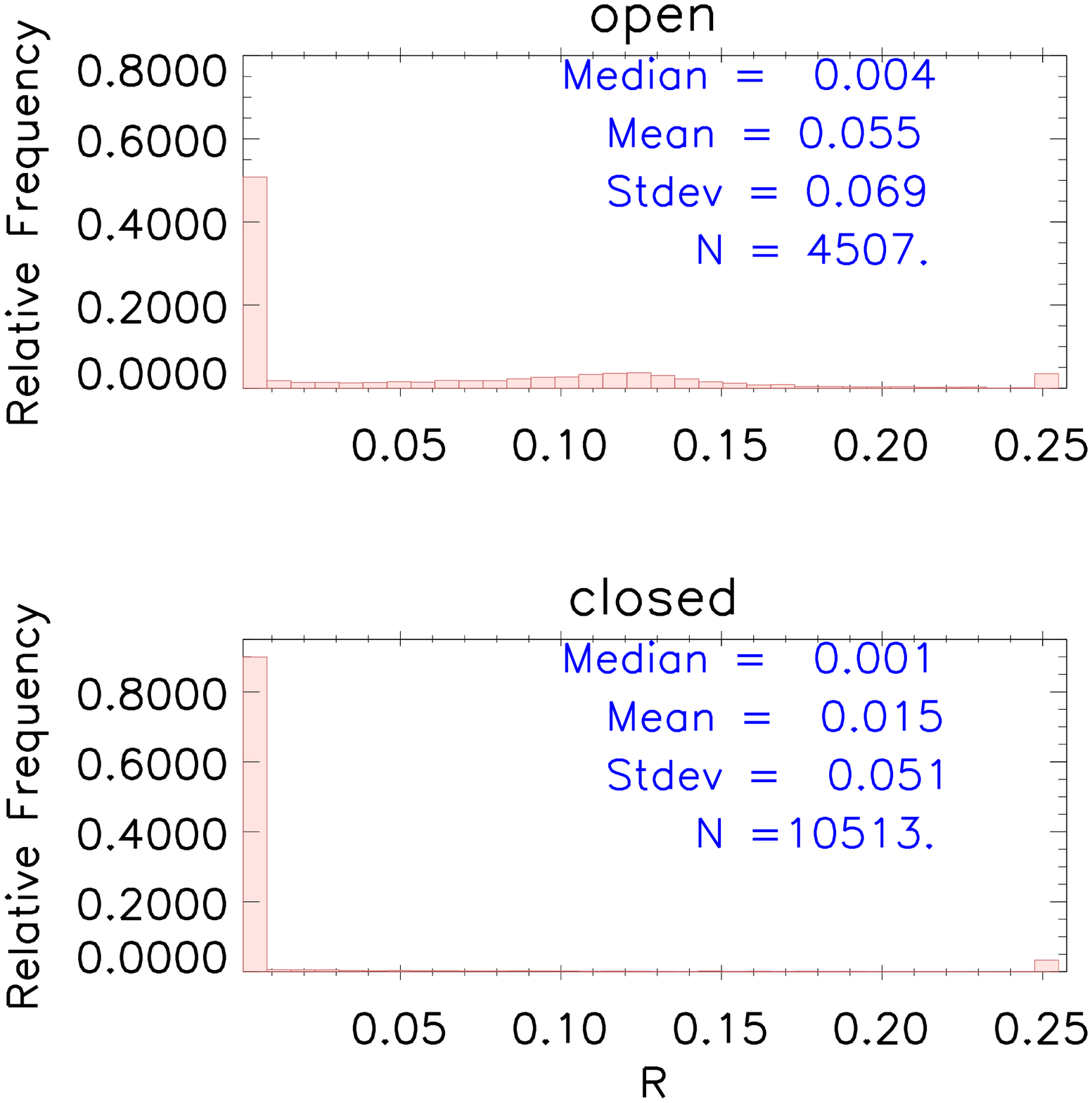}
\caption{Histograms of $R$ at 1.075 $\Rsun$ in the open region (top panels) and in the closed region (bottom panels) using  EUVI-B/N1 (left) and AIA-3/N1 (right).}
\label{histo3band}
\end{figure}

Before comparing the similarities and differences between the DEMT results obtained with the different instrumental sets, we will comment on the quality of the results as measured by the score $R$ defined in Equation (5). Figure \ref{histo3band} shows histograms for the values of $R$ shown in the bottom panels of Figure \ref{maps3band}, for the open and closed region separately. {The vast majority of the tomographic voxels possess a score $R < 10^{-2}$, meaning that the tomographic and synthetic band emissivities differ less than 1\%, in average.}

\begin{figure}[ht]
\includegraphics[width=0.49\linewidth]{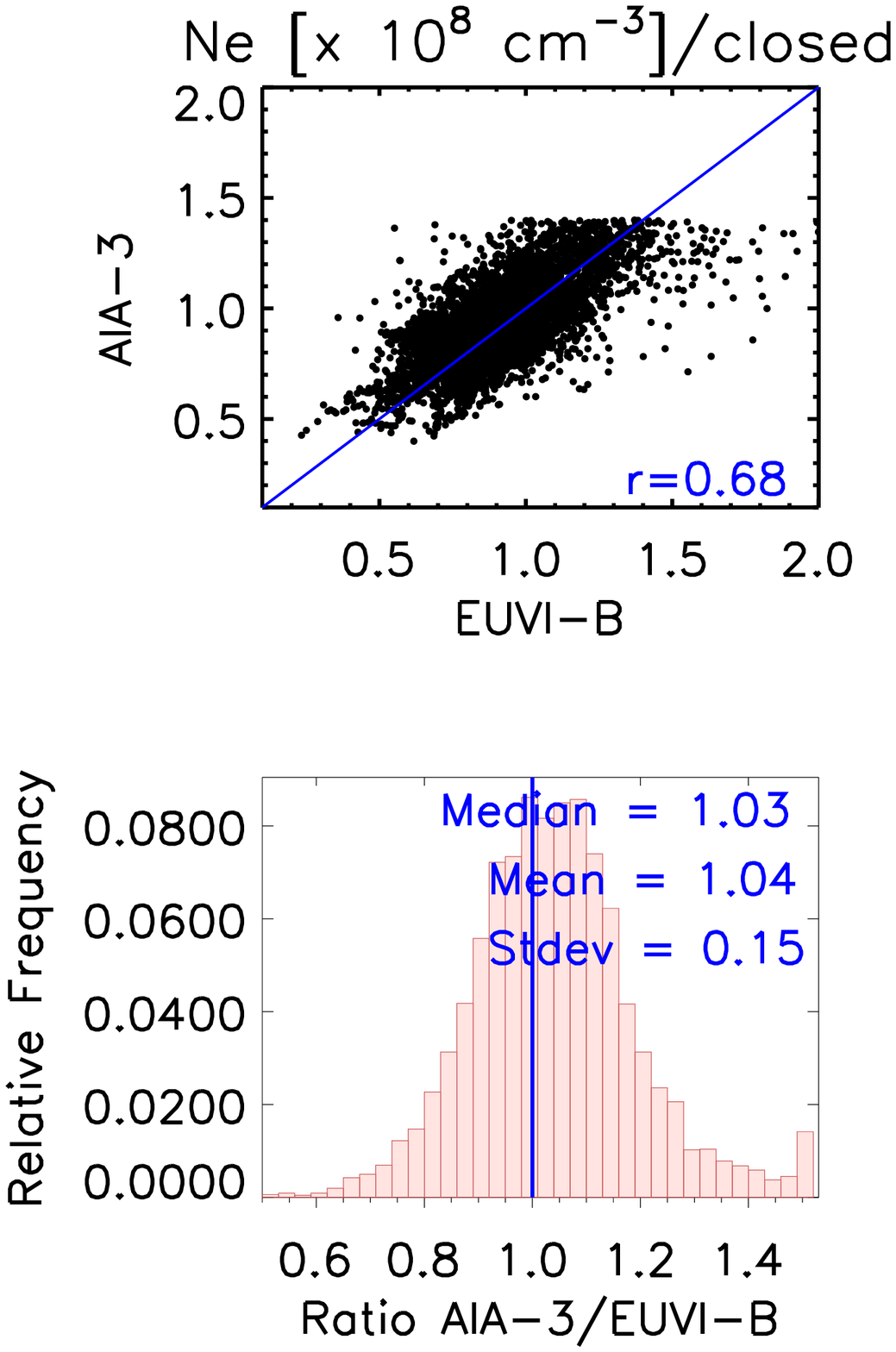}
\includegraphics[width=0.49\linewidth]{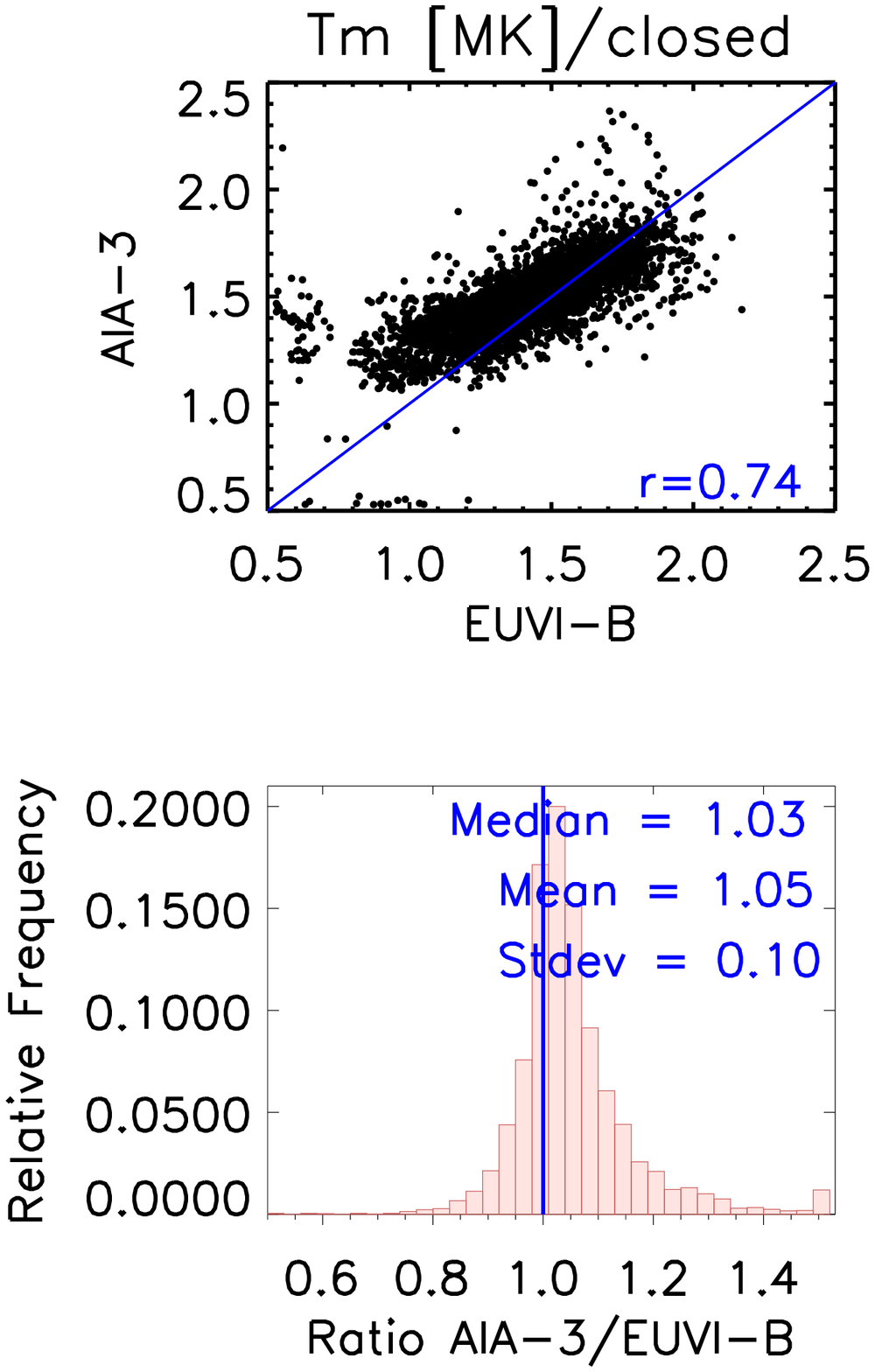}
\caption{Comparison of DEMT results in closed magnetic region between EUVI-B/N1 (x-axis) and AIA-3/N1 (y-axis) at 1.075 $\Rsun$. Top panels: scatter plots for $N_e$ (left) and $T_m$ (right), comparing in each panel the results obtained with both instruments. Bottom panels: histograms of the ratio of the results shown in the scatter plots.}
\label{scat3band}
\end{figure}

As seen in Figure \ref{maps3band}, the results obtained with EUVI/N1 and AIA-3/N1 are very consistent in the closed magnetic region. To evaluate differences quantitatively, Figure \ref{scat3band} shows scatter plots and histograms comparing the DEMT results obtained with EUVI/N1 and AIA-3/N1 in  the closed region. There are small systematic differences, with a mean temperature and density increase of order $\sim 5\%$ for the AIA results respect to those of EUVI. The differences are due to differences in the TRFs of both instrumental sets, mainly between the AIA 211 \AA\ band  and EUVI 284 \AA\ band TRFs. There are also non-systematic differences, reflected by the standard deviation of the histograms, which are to be expected due to solar dynamics effects as the time series of data used for each instrument are not synchronous.

\subsection{{DEMT Results with Four Bands}}
\label{aia4}

{In this section we analyze the results of DEMT based on the AIA-4 data set, which adds the 335 \AA\ band to the AIA-3 set. The additional band has a maximum sensitivity at temperature 2.47 MK.}

First, the LDEM is modeled with the N1 parametrization. The results are displayed in the left column of Figure \ref{4band_unimodal} which shows, from top to bottom, the Carrington maps of $N_e$, $T_m$ and $R$ at 1.075 $\Rsun$, and the frequency histograms of the $R$ values at the same height, separated in the open and closed region. A comparison of the results for $R$ in this column against those in the right column of Figure \ref{maps3band}, and also those in Figure \ref{histo3band}, shows that the N1 model can not explain the emisivity in the 4 bands as succesfully as it does for 3 bands. {It is the shape of the N1 model and the fact that it depends on only 3 parameters which makes it unsuccessful in coping with the wider temperature sensitivity range of the AIA-4 data set.}

{Next, we test the TH parametric model, {which by adding a $4^{\rm th}$ parameter has the flexibility of widening the range of temperatures over which the LDEM is near its maximum value, as compared to the N1 model}, while still being a simple symmetric unimodal distribution. The middle column in Figure \ref{4band_unimodal} shows the results using this model. {Even if the number of free parameters has been increased by 1 compared to the N1 model, the histograms of $R$ do not show improvement over the previous model.}

{Next, we test the ATH model, which allows for both negative and positive temperature gradients, depending on the sign of the exponent of the multiplicative power law, allocating more plasma at the low or high temperature end, as needed.} The results of using this unimodal assymetric parametrization are shown in the right column of Figure \ref{4band_unimodal}. The histograms of $R$ show a {slight improvement {over the N1 and TH models, with the smallest median and mean values for the score $R$}, while still performing significantly worse than the N1 model for the AIA-3 set.}

\begin{figure}[ht] 
\begin{center}
\includegraphics[width=0.32\linewidth]{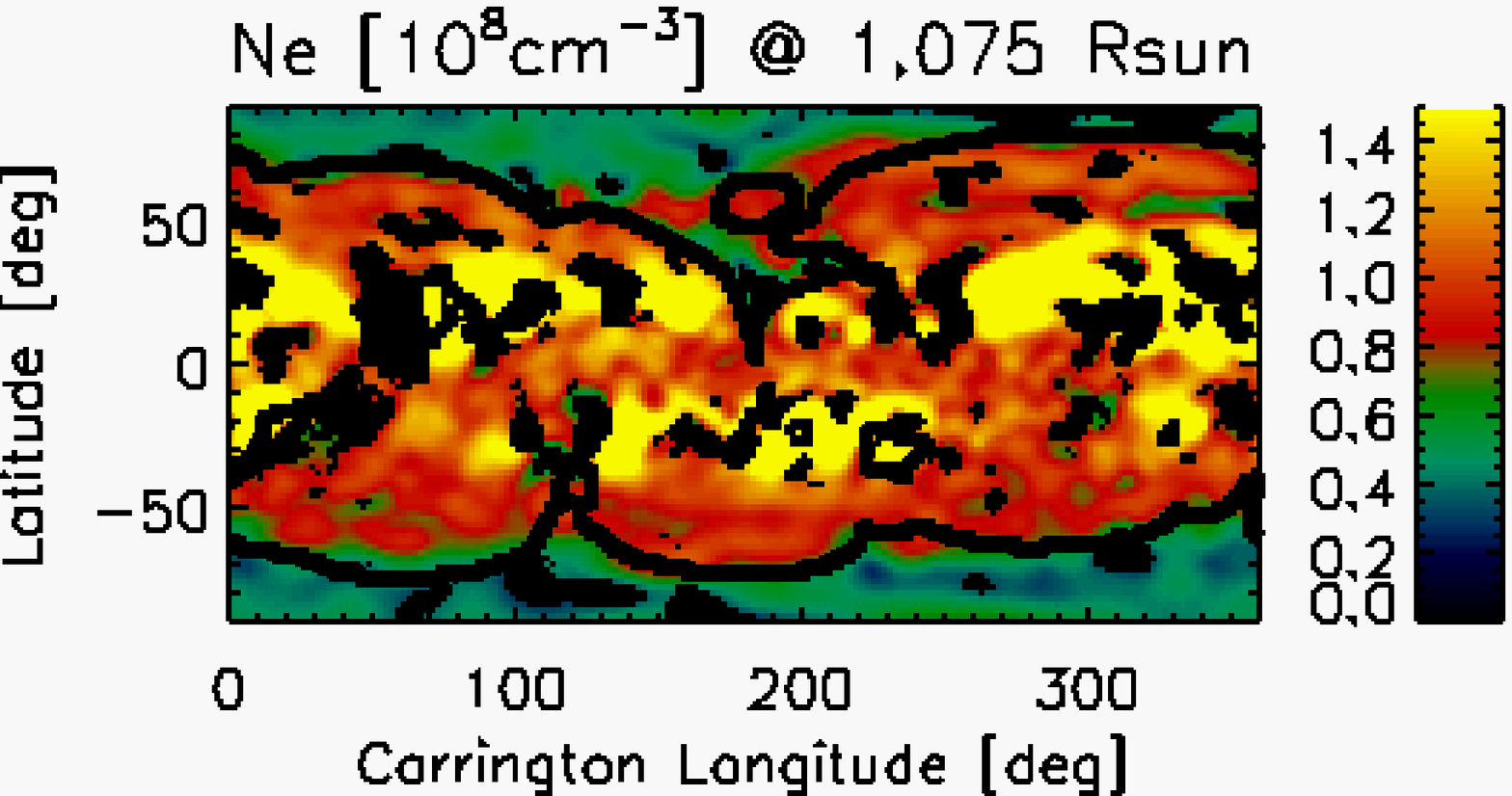}
\includegraphics[width=0.32\linewidth]{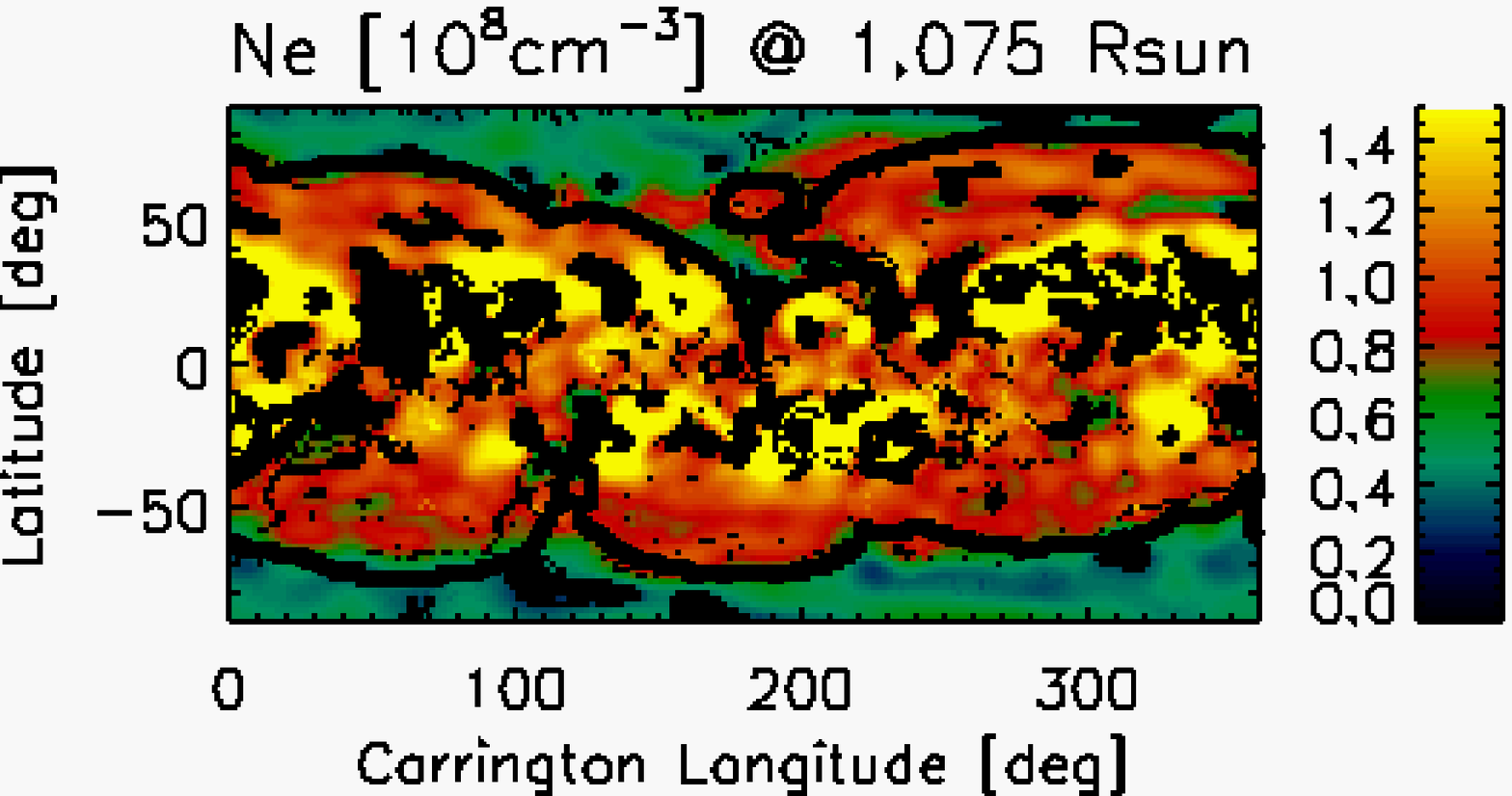}
\includegraphics[width=0.32\linewidth]{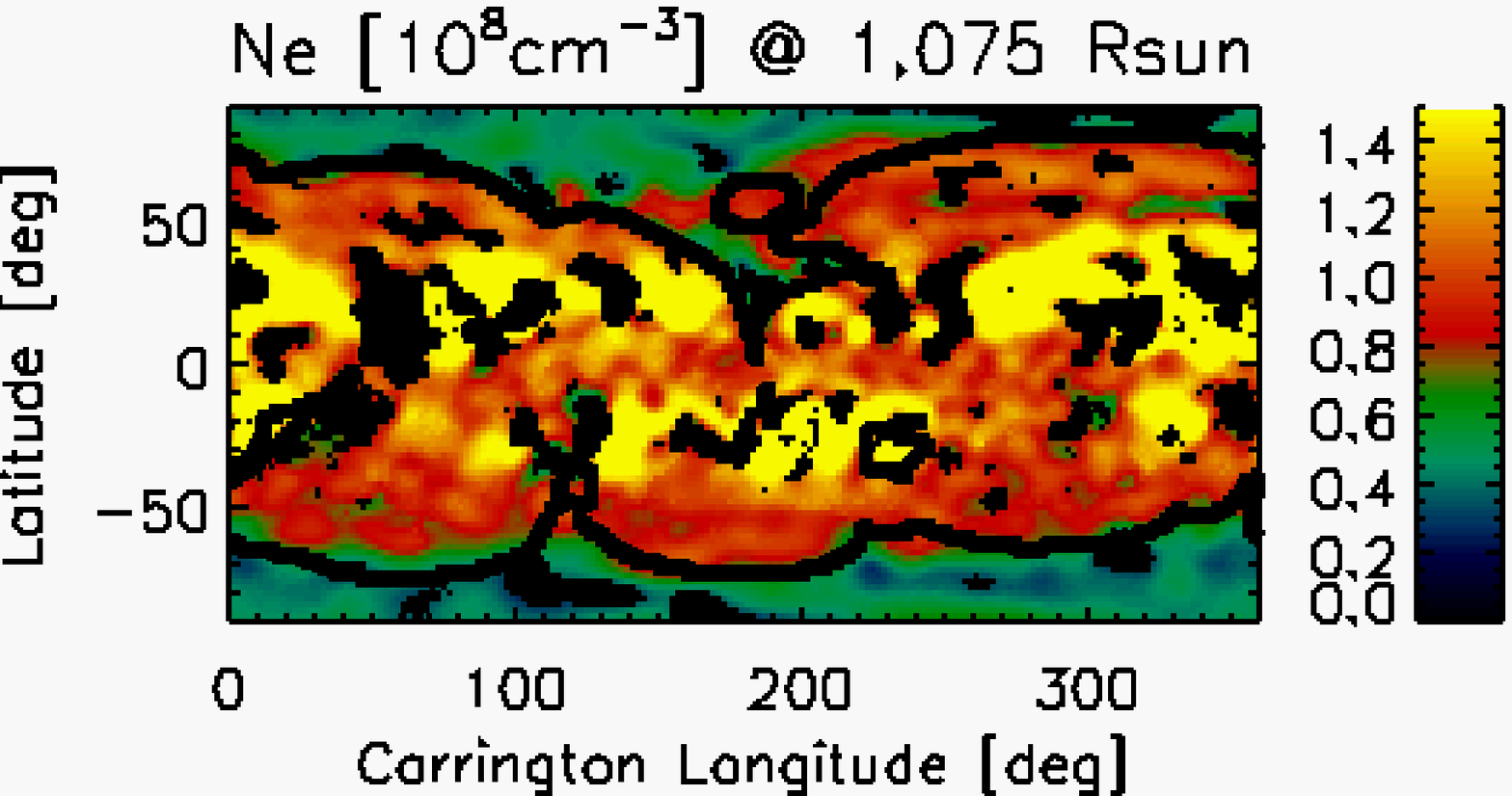}\\
\includegraphics[width=0.32\linewidth]{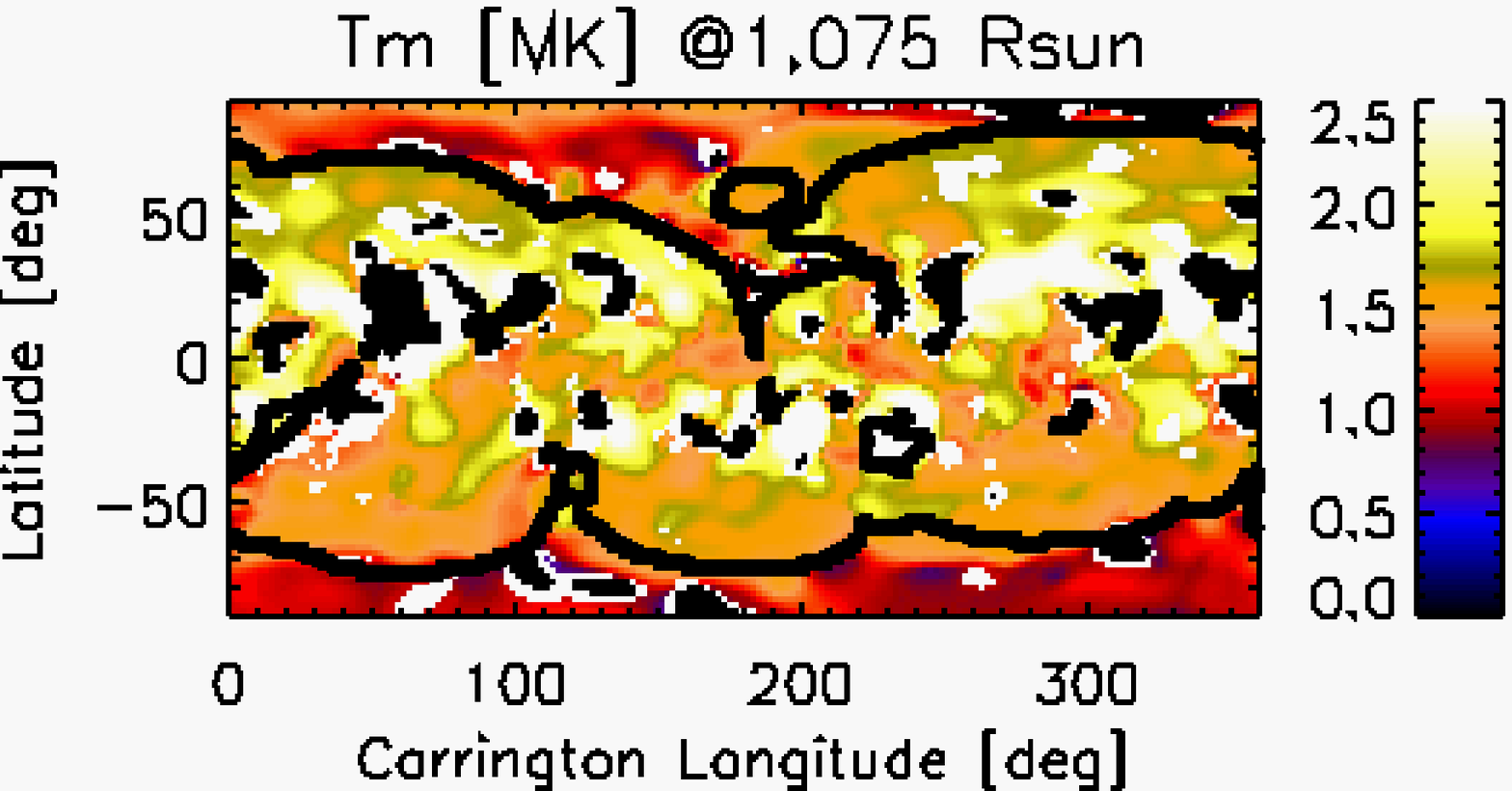}
\includegraphics[width=0.32\linewidth]{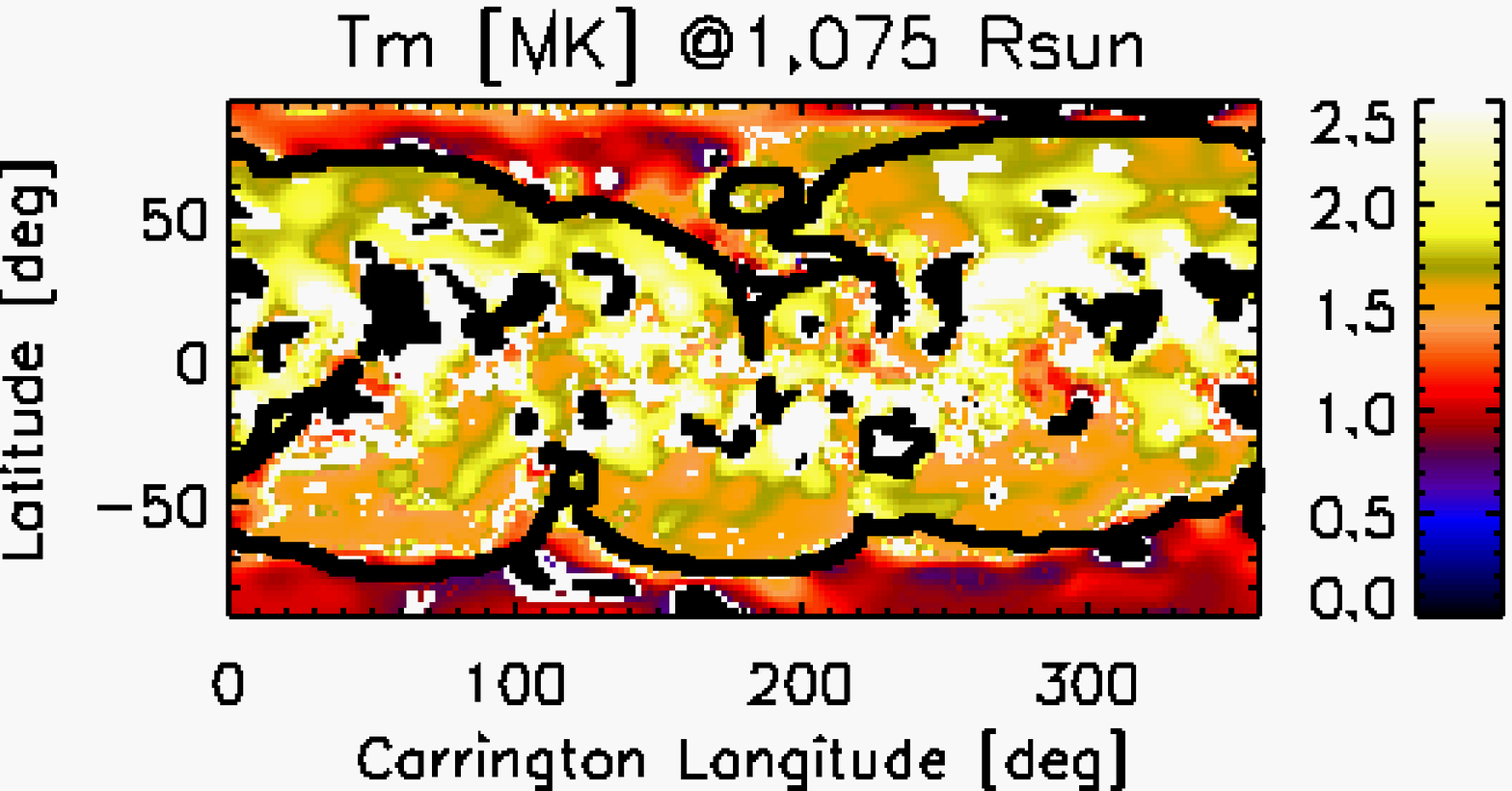}
\includegraphics[width=0.32\linewidth]{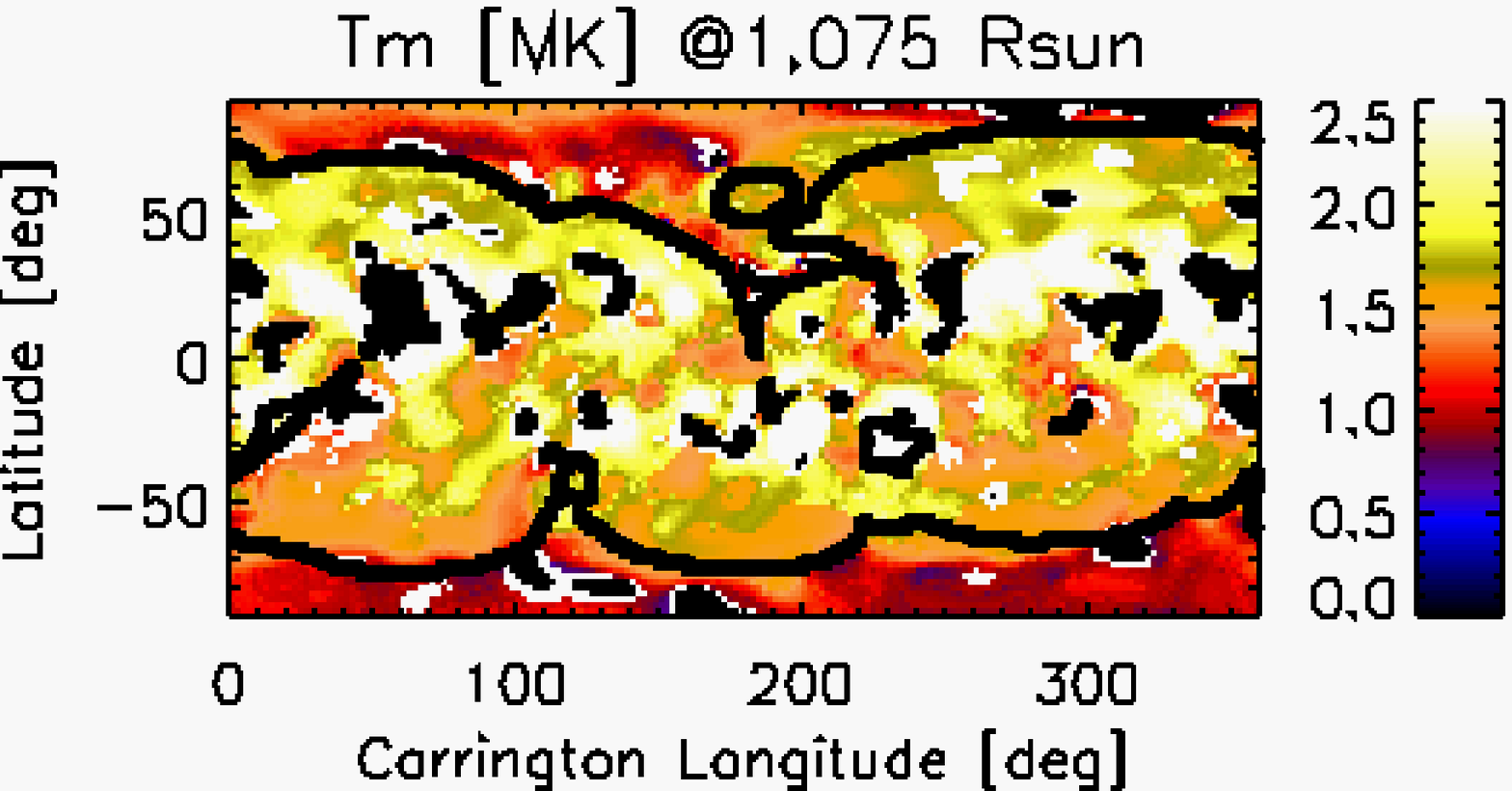}\\
\includegraphics[width=0.32\linewidth]{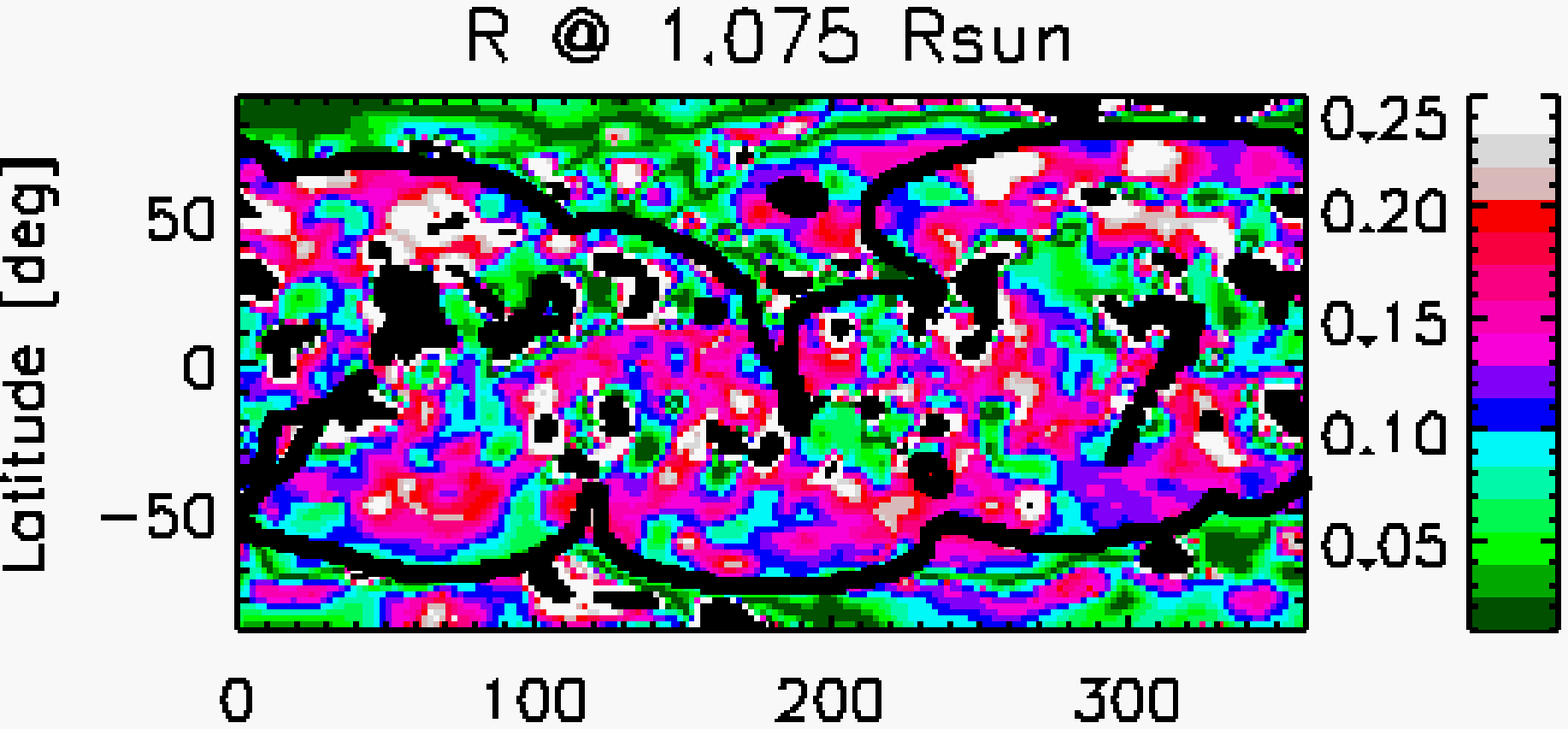}
\includegraphics[width=0.32\linewidth]{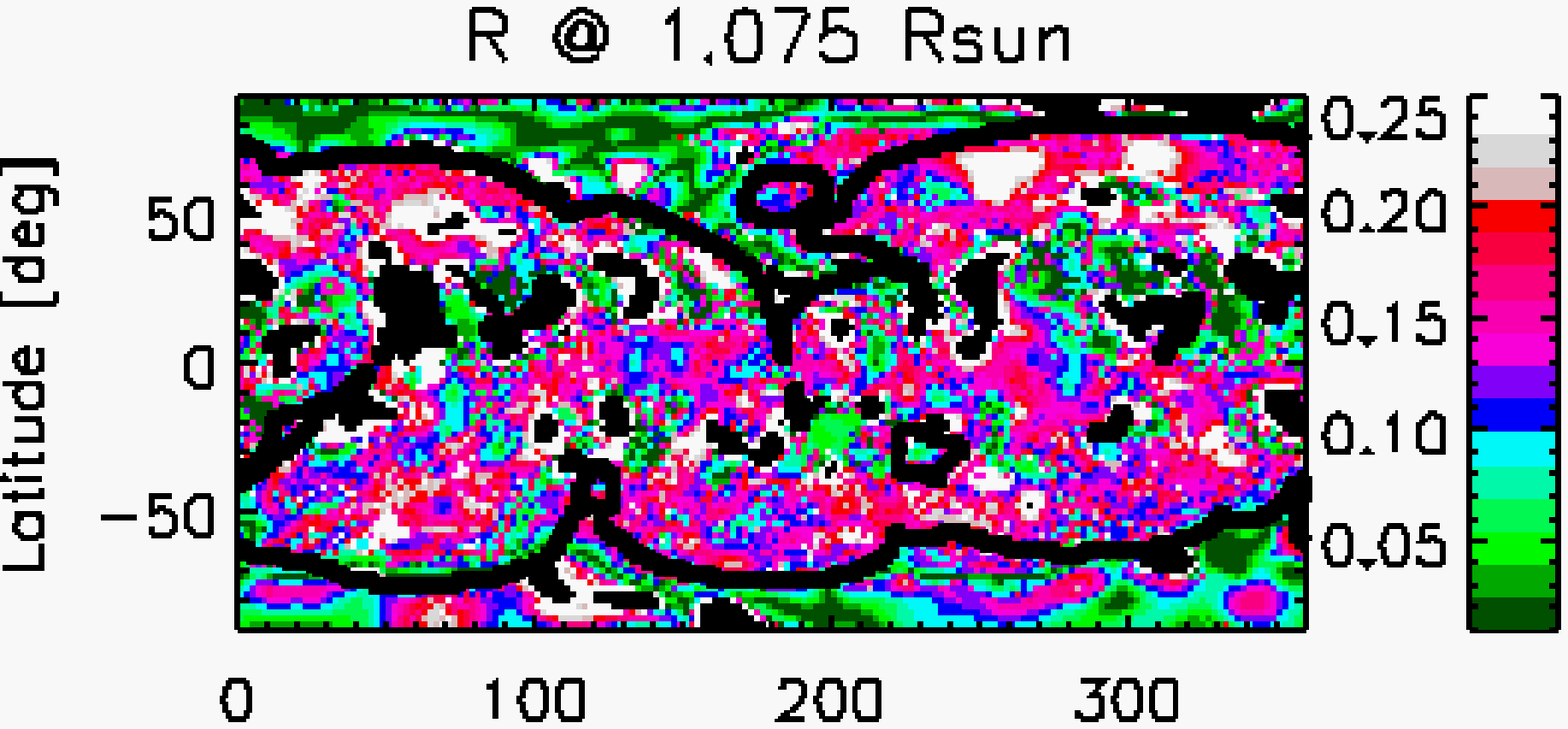}
\includegraphics[width=0.32\linewidth]{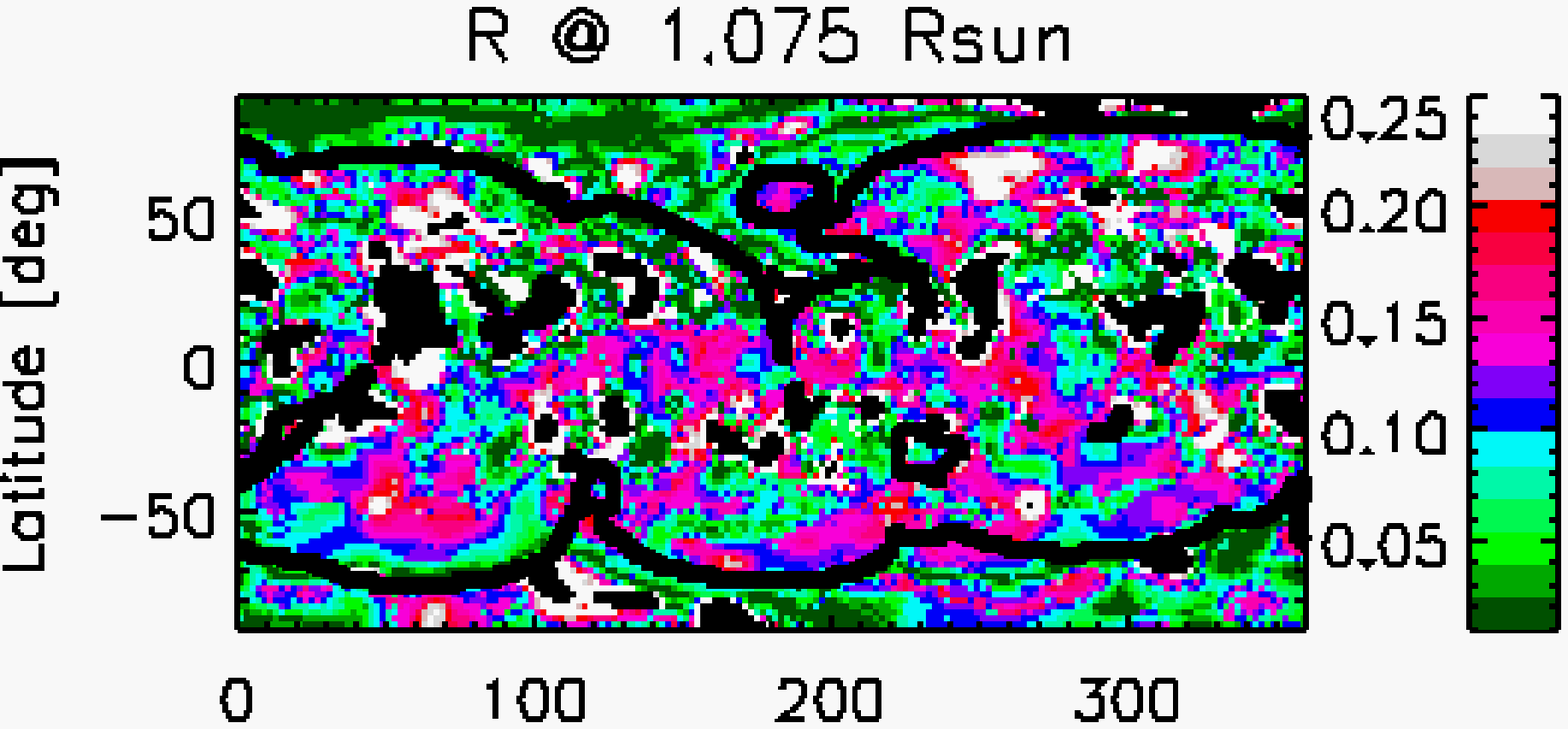}\\
\includegraphics[width=0.32\linewidth]{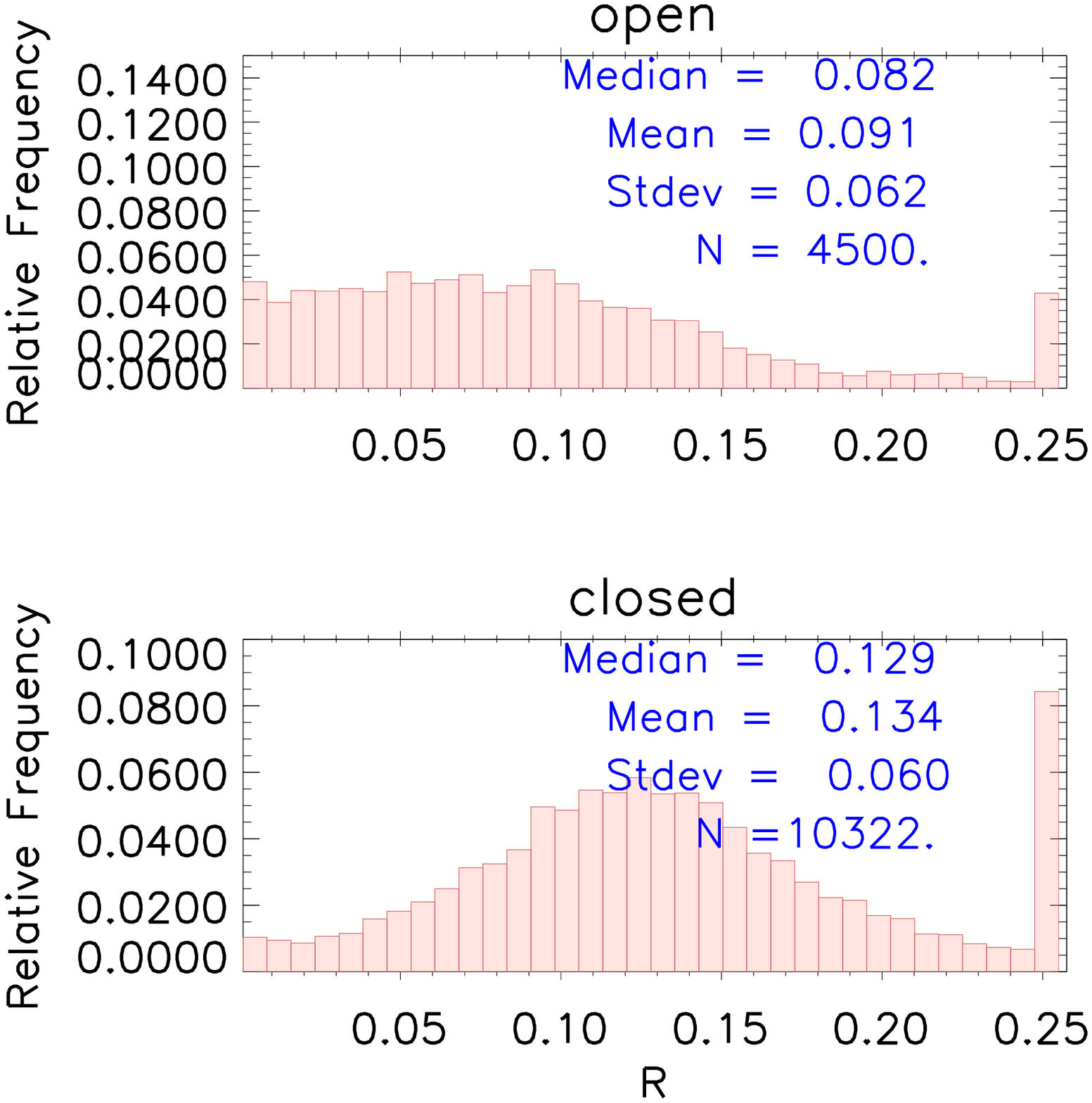}
\includegraphics[width=0.32\linewidth]{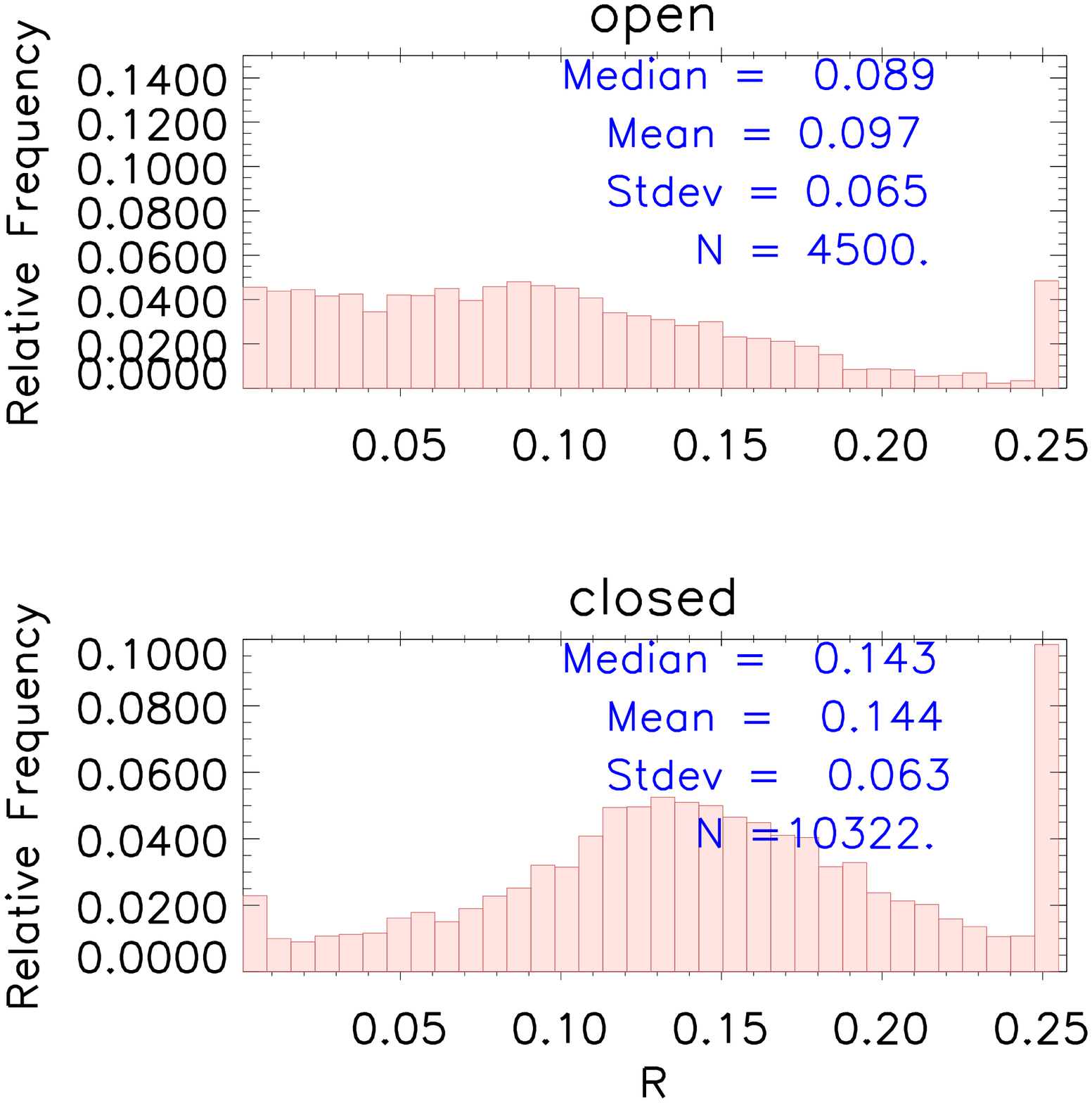}
\includegraphics[width=0.32\linewidth]{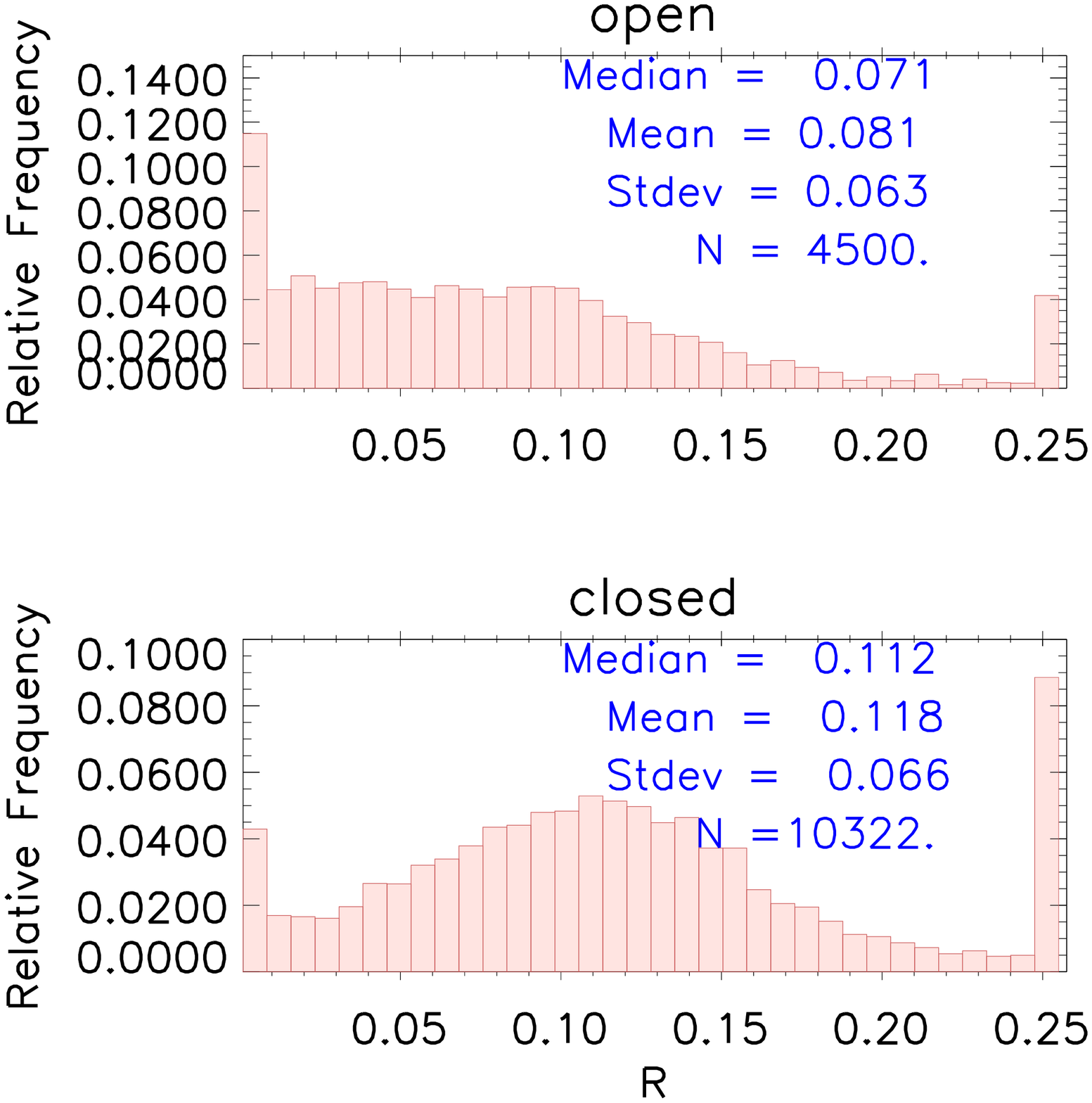}\\
 \end{center}
 \caption{{Results of inverting the AIA-4 data set using unimodal parametric models. Top three panels: Carrington maps of $N_e$, $T_m$ and $R$, at 1.075 $\Rsun$. ZDAs and AEVs are marked in the same way as in Figure \ref{maps3band}. Bottom two panels: histograms of the score $R$ separating the voxels in the open and closed magnetic region. The three columns correspond to the different models. {Left:} N1 model (3 free parameters), {Middle:} TH model (4 free parameters), {Right:} ATH model (5 free parameters). See the text for details on the models.}}
\label{4band_unimodal}
\end{figure}

Figure \ref{4band_unimodal}} shows that, in order to explain the emissivities of the AIA-4 set none of the unimodal models, symmetric or asymmetric, performs as well as the N1 model does for the AIA-3 set (Figure \ref{maps3band}). Looking for better agreement, we attempt the multimodal distributions N2 and N4 described in Section \ref{parametrization}. The maps of $N_e$, $T_m$ and $R$  are shown in Figure \ref{maps4band}, and the histograms of $R$ are shown in Figure \ref{histo4band}.

\begin{figure}[ht] 
\begin{center} 
\includegraphics[width=0.49\linewidth]{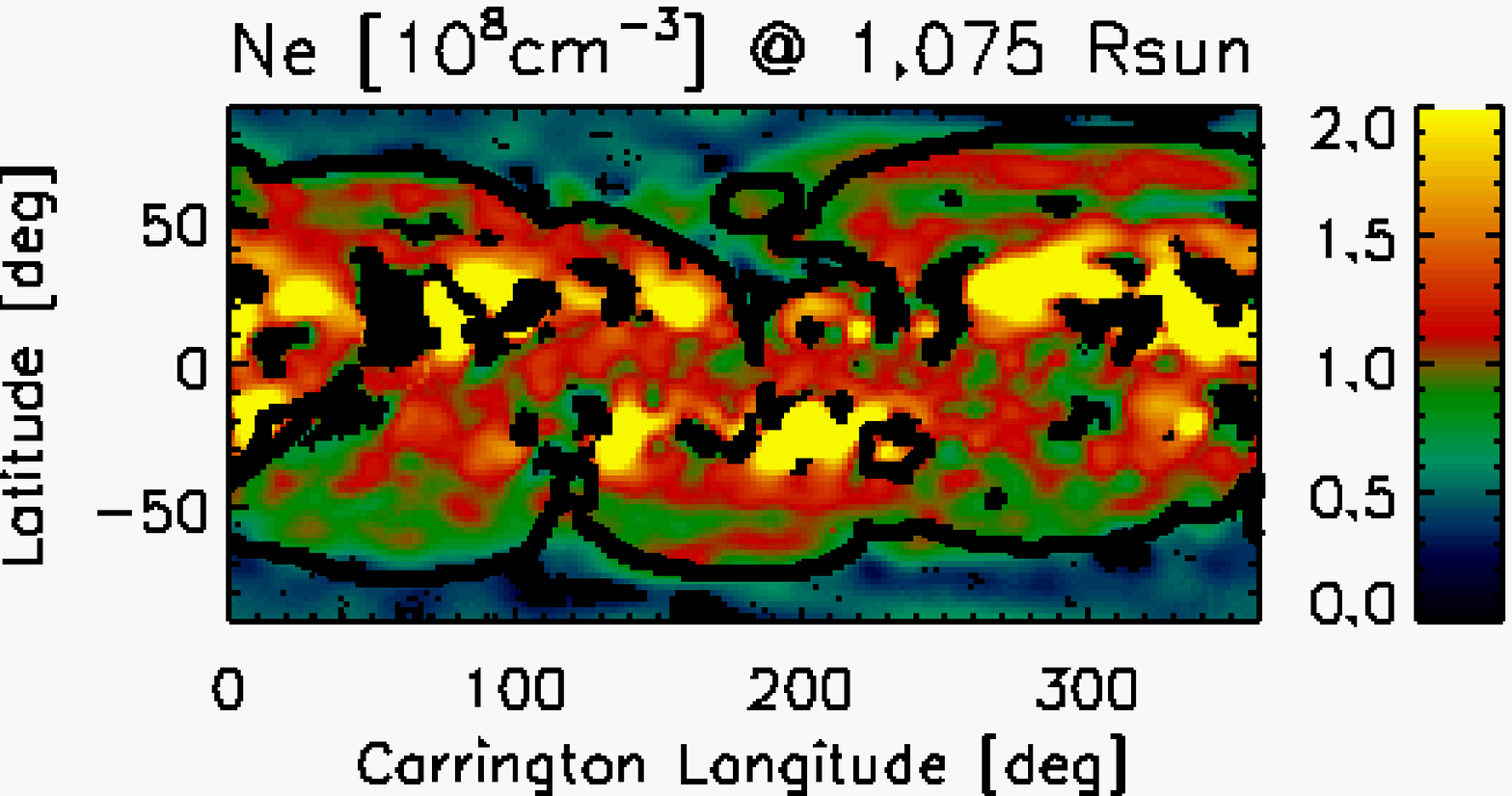}
\includegraphics[width=0.49\linewidth]{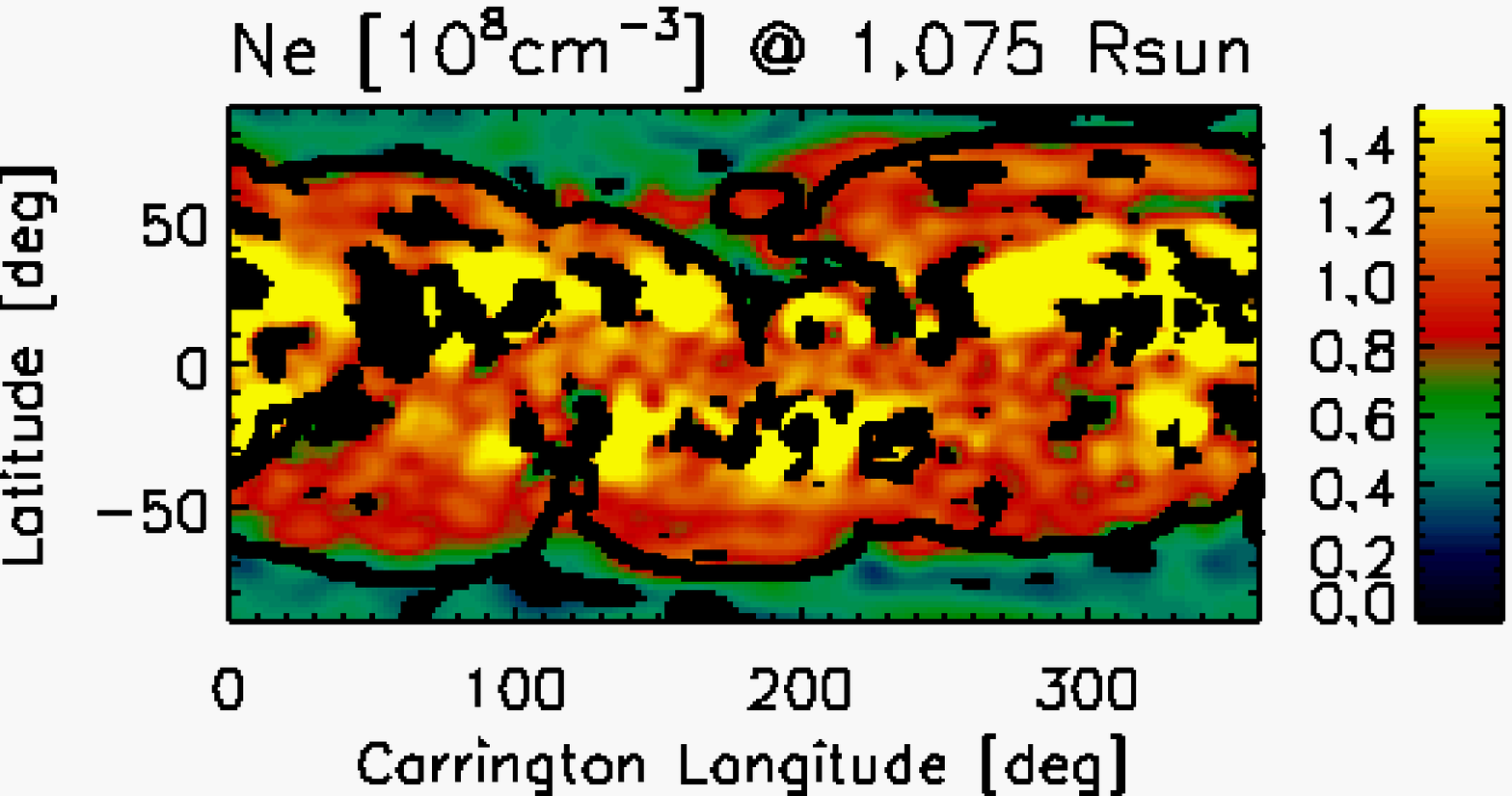}\\
\includegraphics[width=0.49\linewidth]{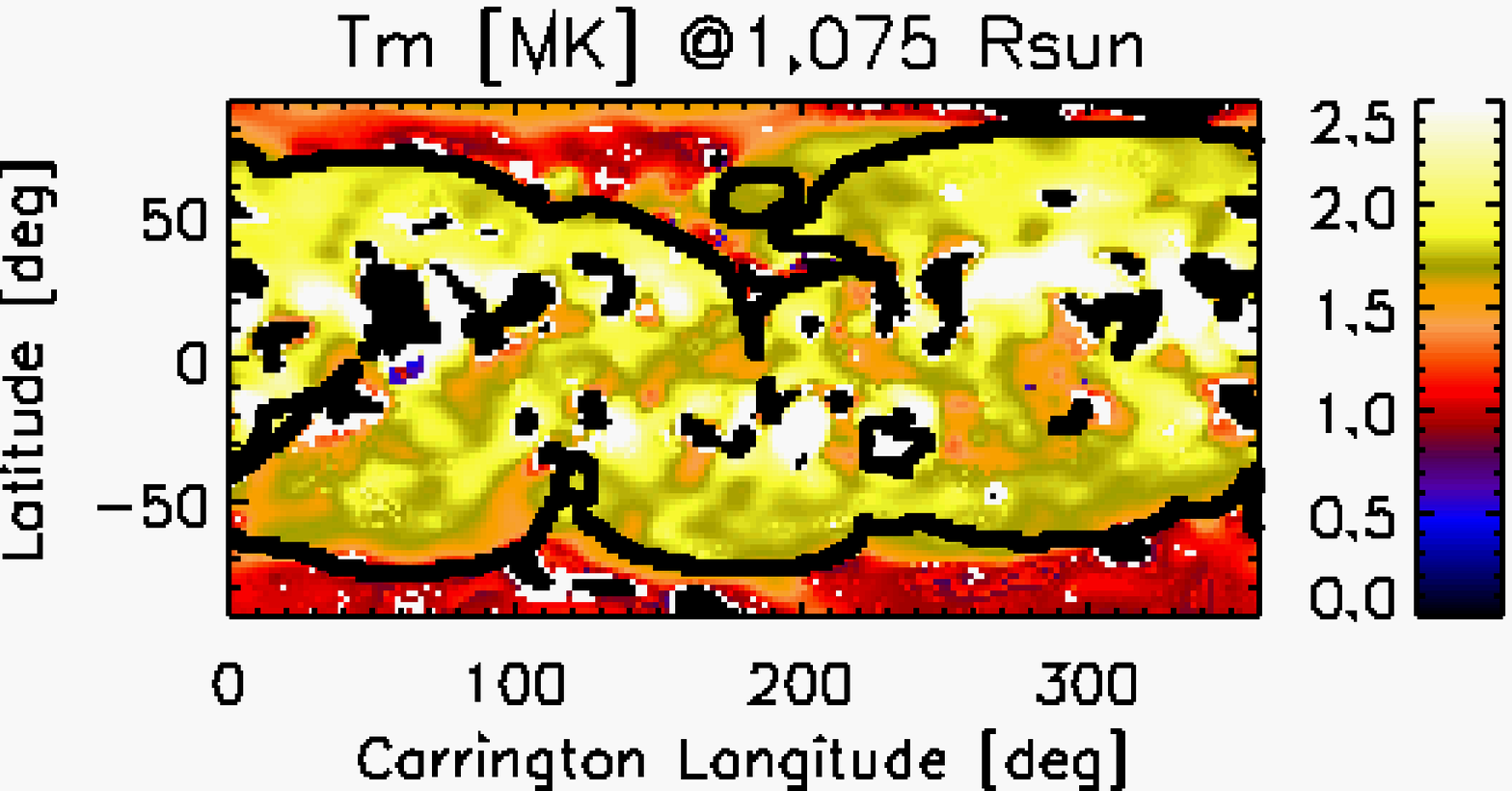}
\includegraphics[width=0.49\linewidth]{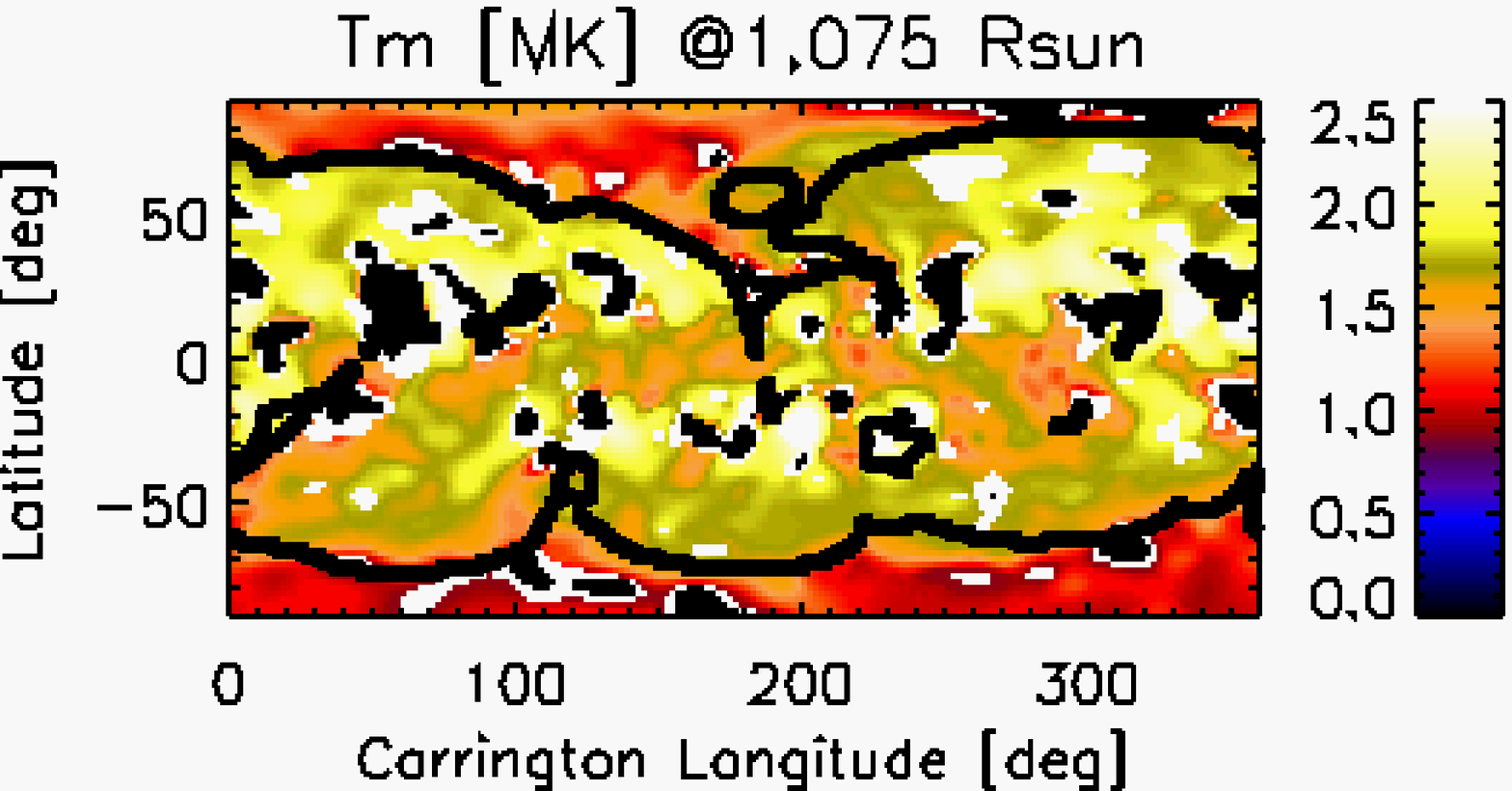}\\
\includegraphics[width=0.49\linewidth]{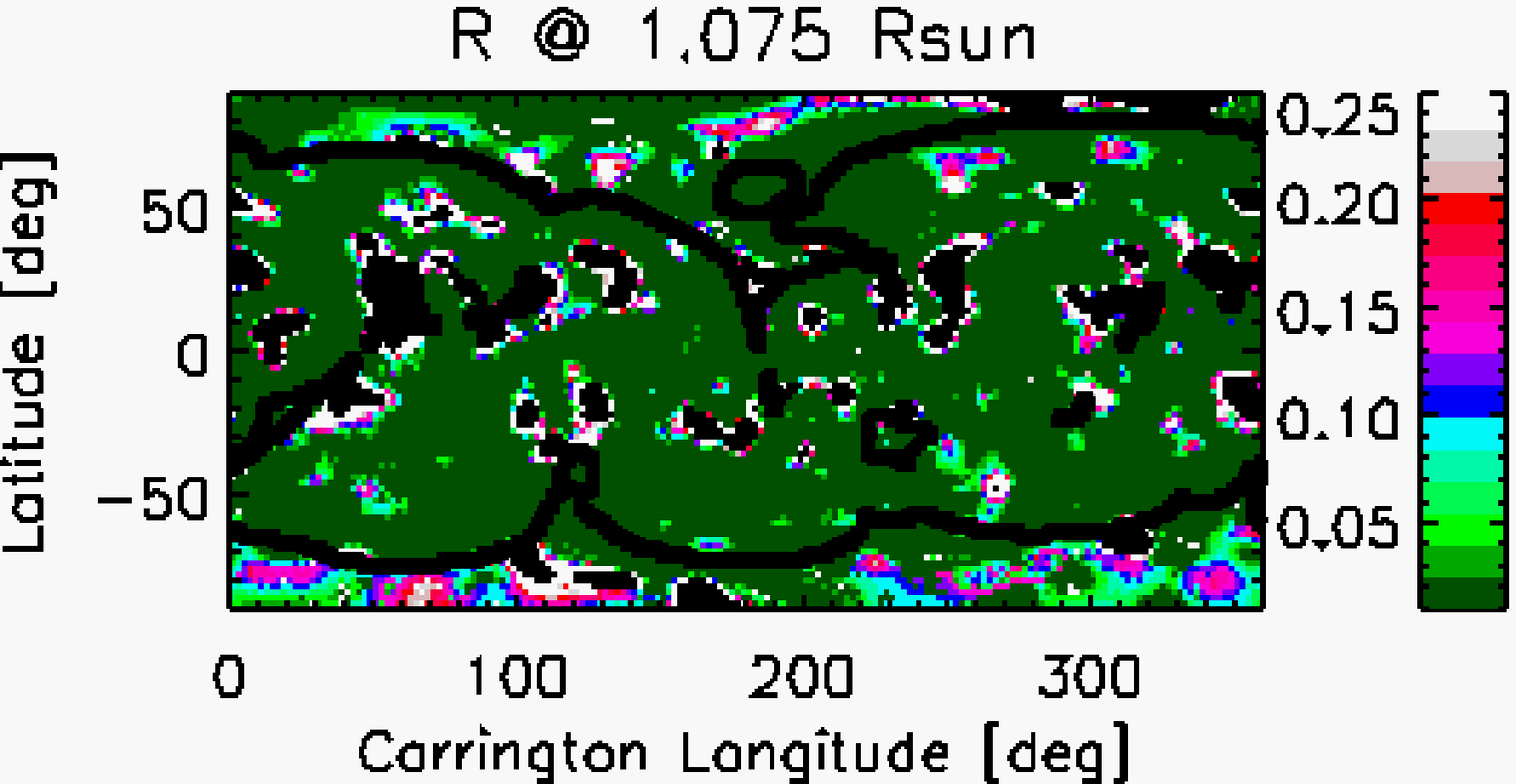}
\includegraphics[width=0.49\linewidth]{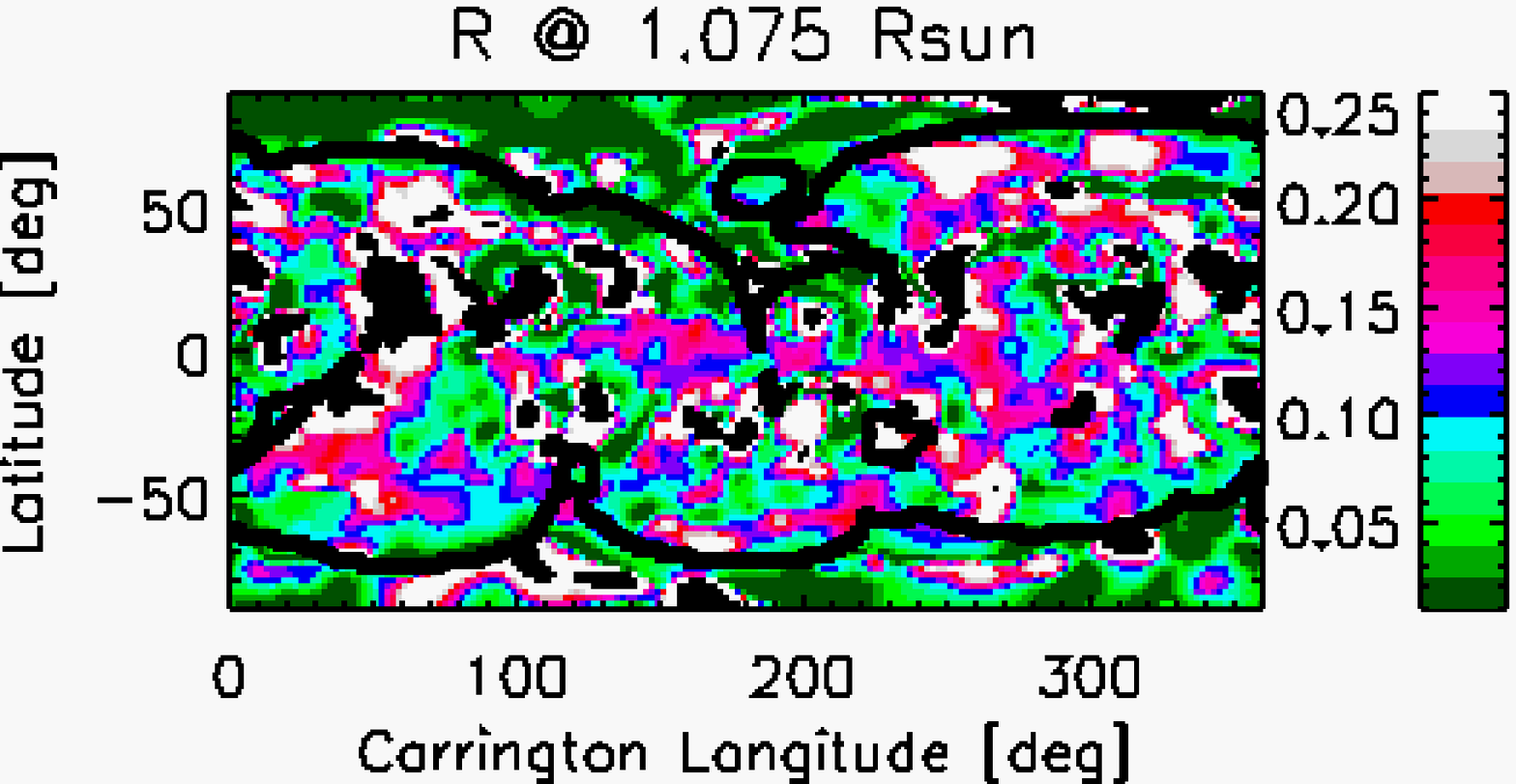}\\
\end{center}
\caption{CR-2099. Carrington maps of $N_e$, $T_m$ and
$R$ at 1.075 $\Rsun$ using data of AIA-4/N2(left) and AIA-4/N4 (right). The boundary between the magnetically open and closed region is overploted as a thick black curves in all maps. ZDAs and AEVs are marked in the same way as in Figure \ref{maps3band}.}
\label{maps4band}
\end{figure}

\begin{figure}[ht]
\includegraphics[width=0.49\linewidth]{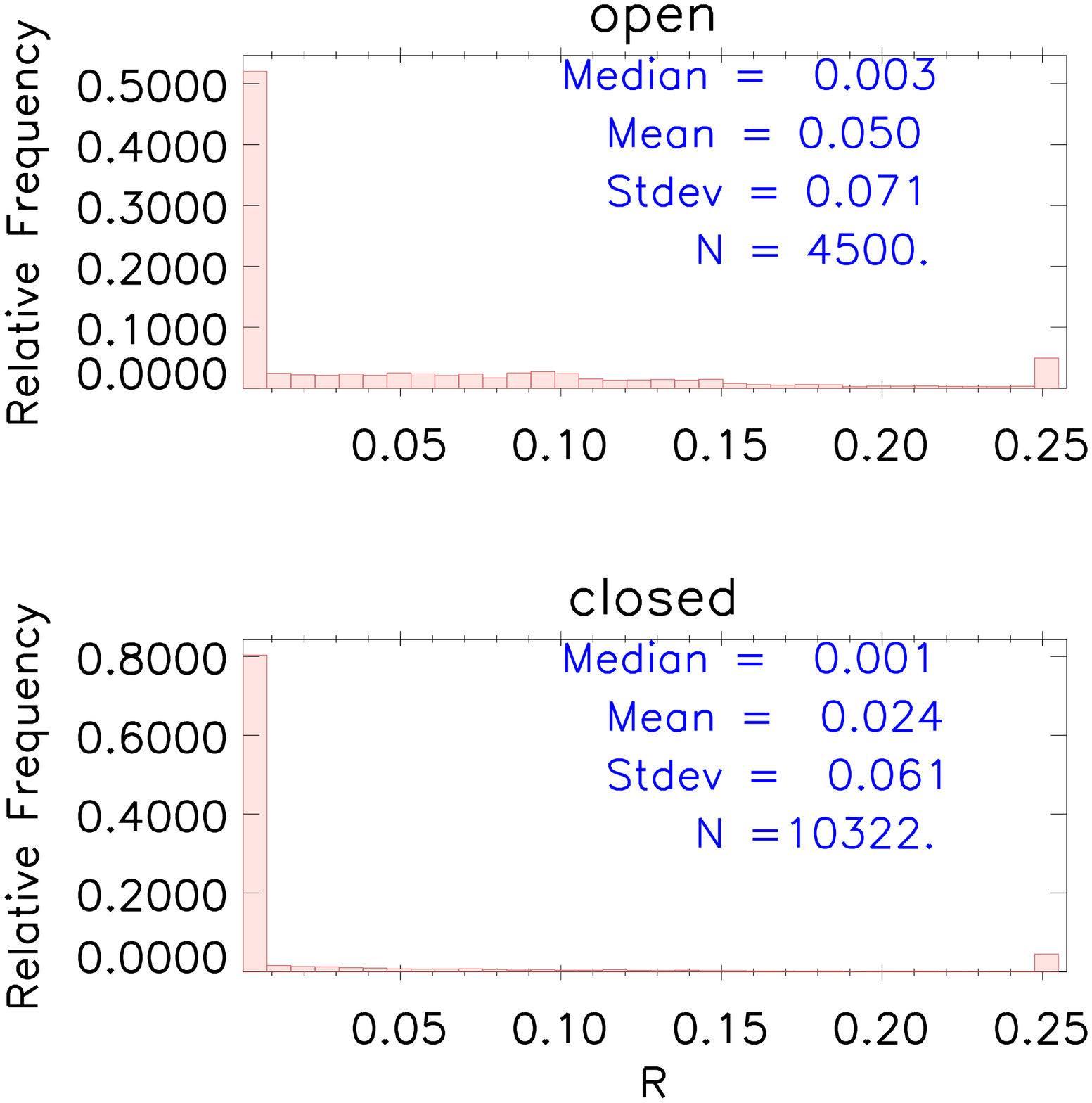}
\includegraphics[width=0.49\linewidth]{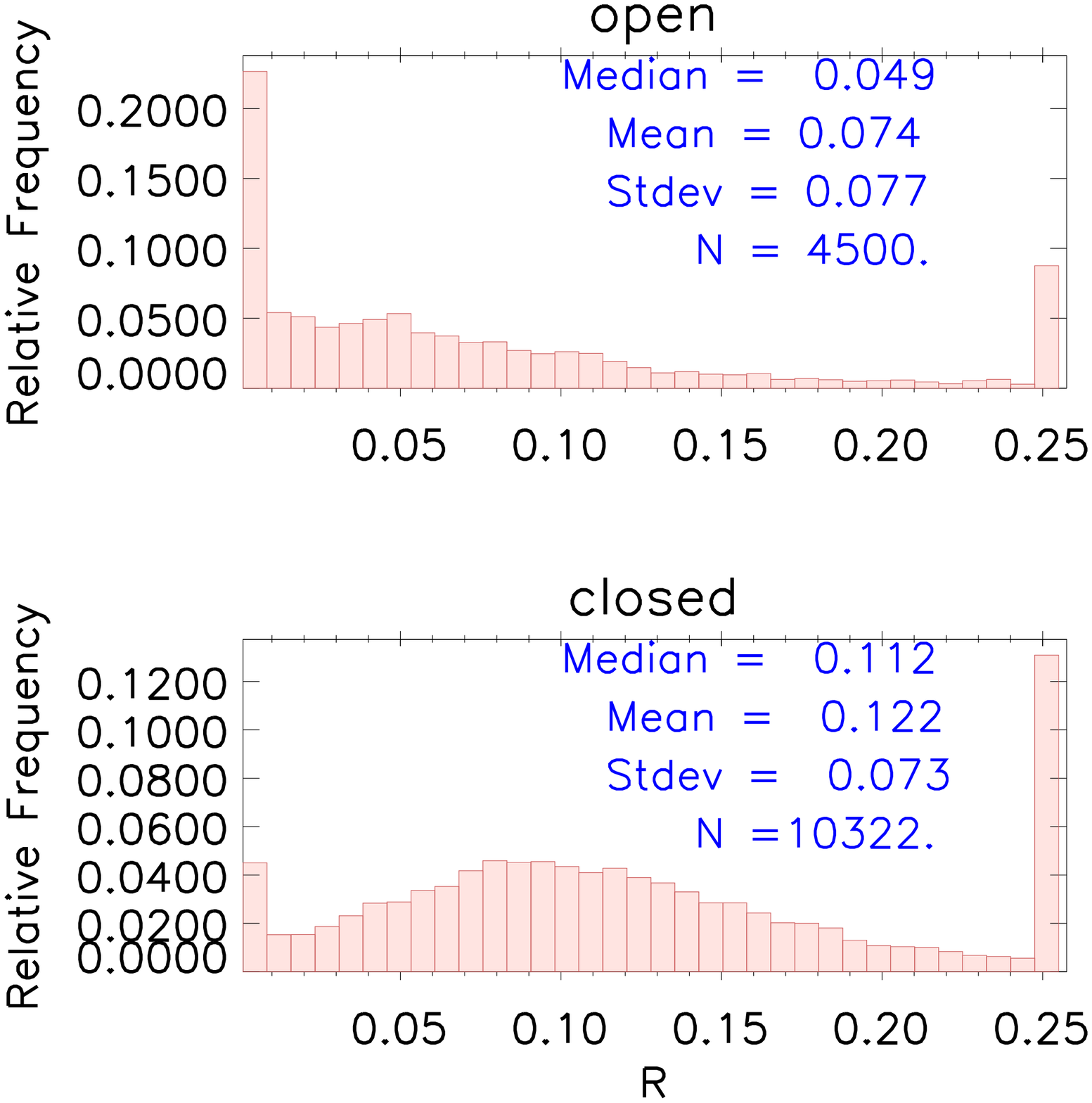}
\caption{Histograms of $R$ at 1.075 $\Rsun$ in the open region (top panels) and in the closed region (bottom panels) using AIA-4/N2 (left) and AIA-4/N4 (right).}
\label{histo4band}
\end{figure}

\begin{figure}[ht]
\includegraphics[width=\linewidth]{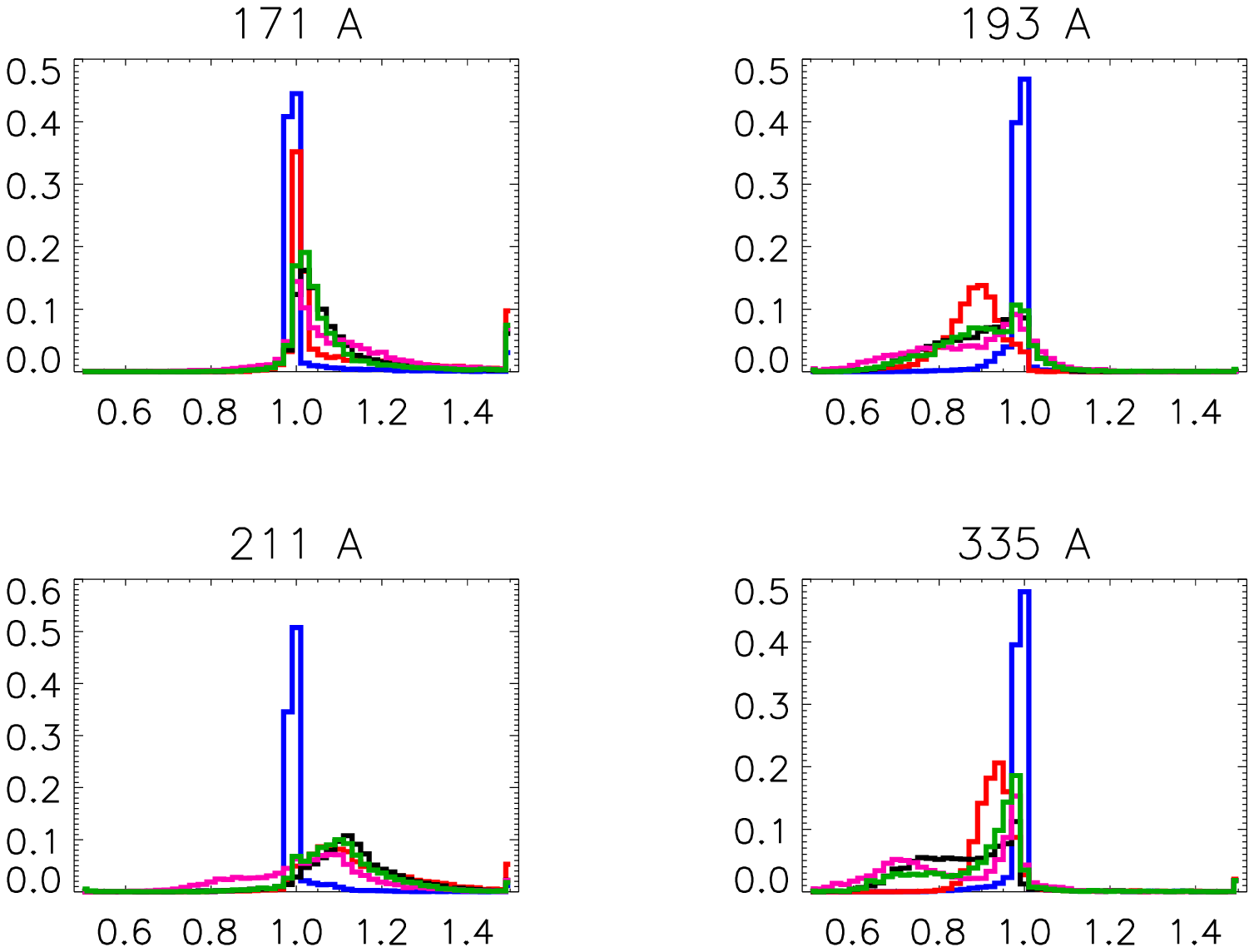}
\caption{{Histograms of the ratio $\zeta_{k}^{\rm(synth)}/\zeta_{k}^{\rm(tom)}$ for the four bands $k$ of AIA-4, for the different parametric models used for the LDEM: N2 (blue), N4 (red), N1 (black), TH (pink), and ATH (green).}}
\label{histoband}
\end{figure}

{A qualitative comparison between the results of the N2 and N4 models (left and right panels of Figure \ref{maps4band}) shows that both models have similar values of $N_e$ and $T_m$. {Indeed, a quantitative comparison (not shown to save space)} reveals that the median LDEM moments differ in less than 4\%. Even so, the N2 model performs better than the N4 model, as it achieves the best score $R$. Indeed, a comparison of the histogram of the left panel of Figure \ref{histo4band} with the histograms of Figure \ref{histo3band}, reveals that the score $R$ of the N2 model when using the AIA-4 data set is as good as the one achieved by the N1 model when using the AIA-3 data set.
Figure \ref{histoband} shows the histograms of the ratio $\zeta_{k}^{\rm(synth)}/\zeta_{k}^{\rm(tom)}$ for the 4 bands $k$ of AIA-4, for the different parametric models used for the LDEM, at 1.075 $\Rsun$. It is readily verified the much higher degree of suceess of the N2 model for each band independently.}

{The high degree of success of the N2 parametrization of the LDEM to reproduce the FBEs observed with all 4 bands of the AIA-4 instrumental set is due to the assumed bimodality. {Our results show then that the quiet solar corona local (3D) DEM is multimodal at the spatio-temporal resolution of the DEMT and the temperature resolution of the TRFs of the AIA filters.}

The differences seen in Figures \ref{maps3band} and  \ref{maps4band} between the results obtained with AIA-3/N1 and with AIA-4/N2 data sets are mainly due to the inclusion of AIA 335 \AA\ band, sensitive to temperatures higher than the other three bands. As can be qualitatively seen in the right panels of Figure \ref{maps3band} and the left panels of Figure \ref{maps4band}, the AIA-4/N2 results have systematically larger values of $N_e$ and $T_m$ respect to the results obtained with AIA-3/N1. The systematic increase in the LDEM moments is more significant in the closed region, where the hotter plasma is located. {To evaluate differences quantitatively, Figure \ref{scat4band} shows scatter plots comparing the plasma parameters obtained in the closed region with N2/AIA-4 and those obtained with N1/AIA-3, as well as histograms of the respective ratios. When the 335 \AA\ band is included the LDEM inversion leads to an increase in the electron density and mean temperature of about 14 and 18\%, respectively, in the closed region. A similar analysis, but considering the voxels in the open region (not included here to save space) indicates that the increase in the density and temperature are about 3 and 10\%, respectively.}

\begin{figure}[ht] 
\includegraphics[width=0.49\linewidth]{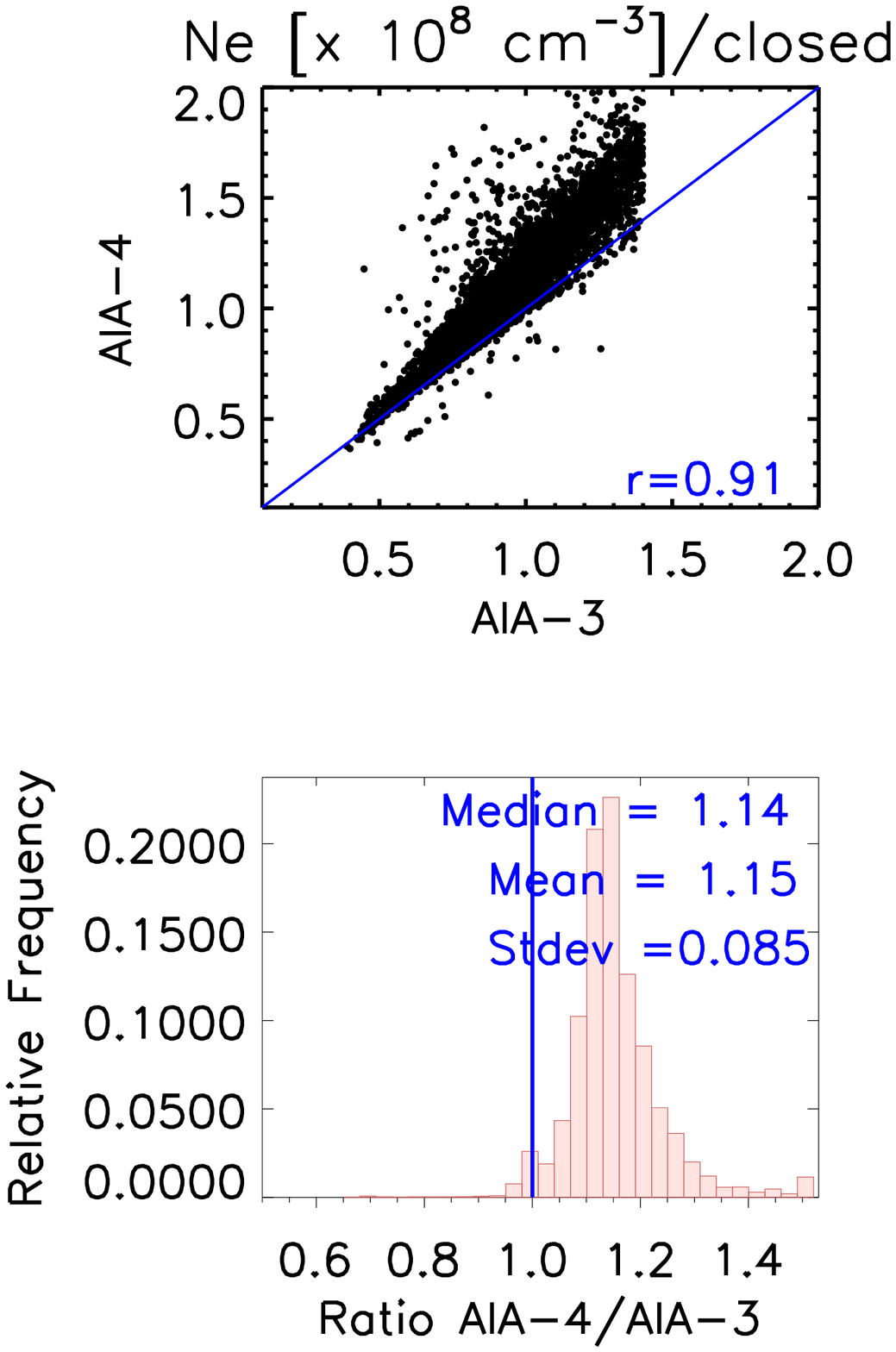}
\includegraphics[width=0.49\linewidth]{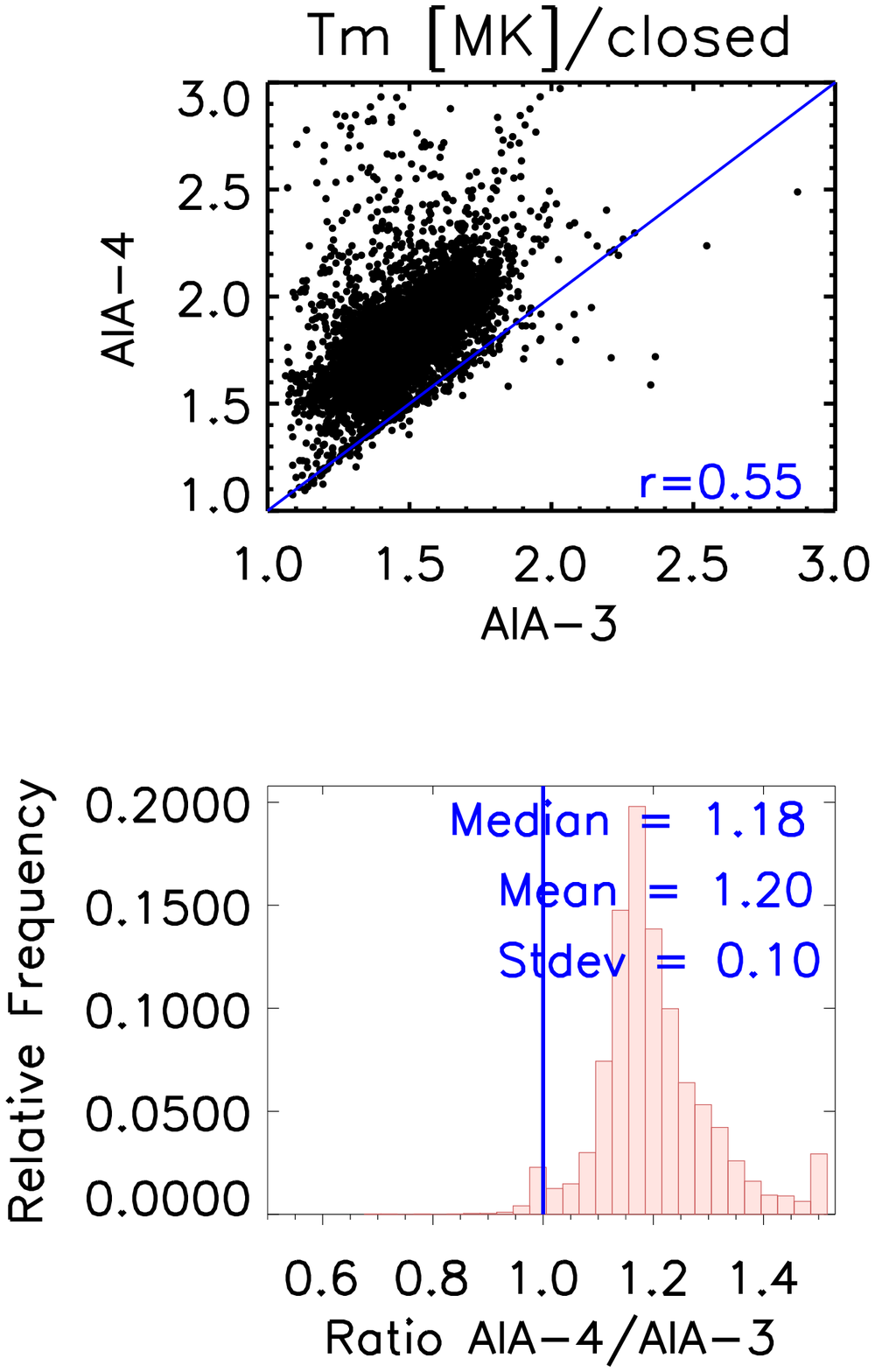}
\caption{Comparition of DEMT results in closed magnetic region between AIA-3 (x-axis) and AIA-4 (y-axis). Similar to Figure \ref{scat3band}.}
\label{scat4band}
\end{figure}

{So far we have compared global maps of the moments and score $R$ that are obtained with all the different parametrizations. It is interesting to also compare all the parametric distributions. For two sample tomographic voxels, one belonging to a hotter and denser coronal region than the other, Figure \ref{ejemplos_LDEM} shows the LDEM obtained using the different parametric models, and compares their performances.}

\begin{figure}[ht]
\begin{center}
\includegraphics[width=0.75\linewidth]{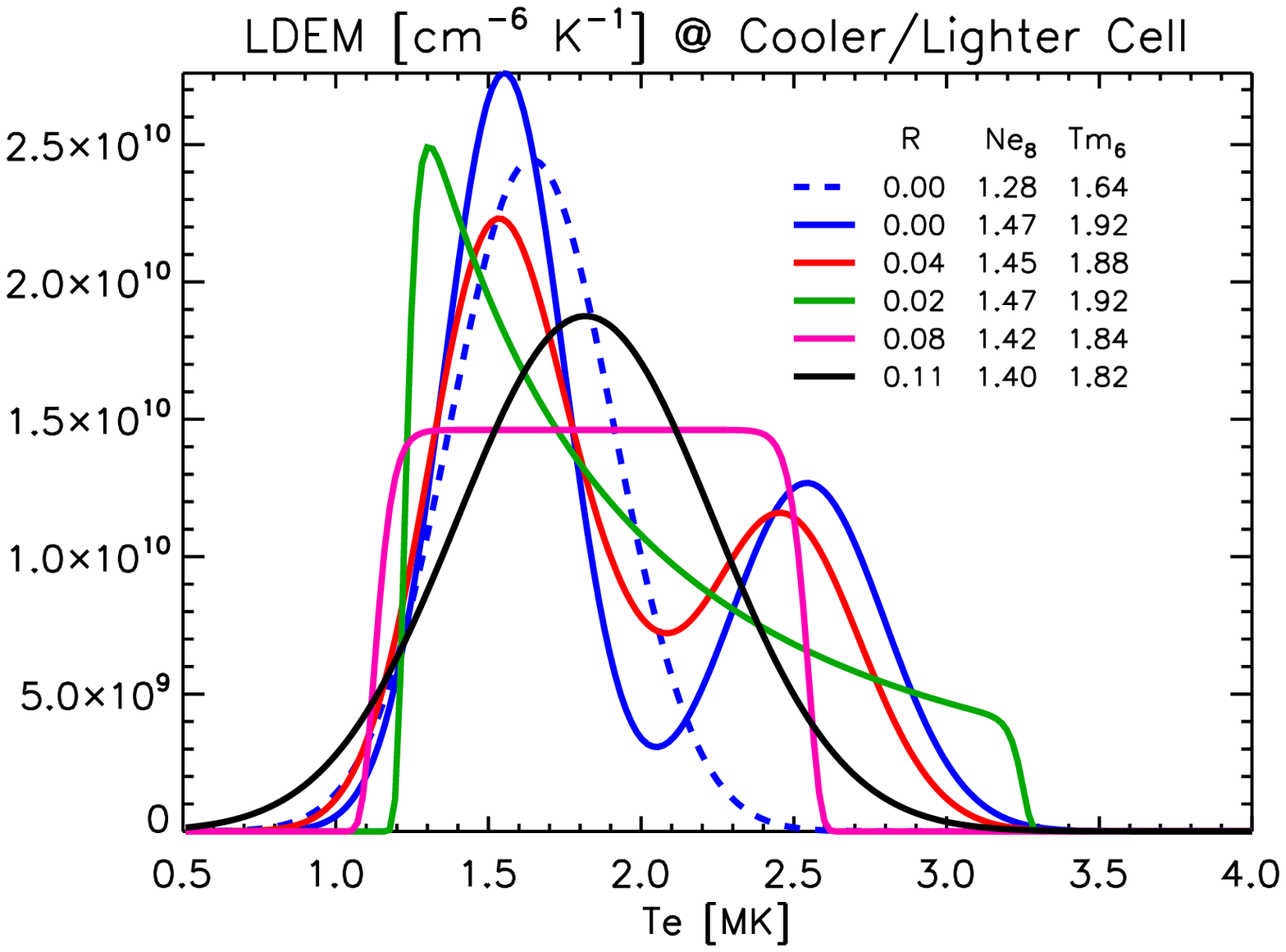}
\includegraphics[width=0.75\linewidth]{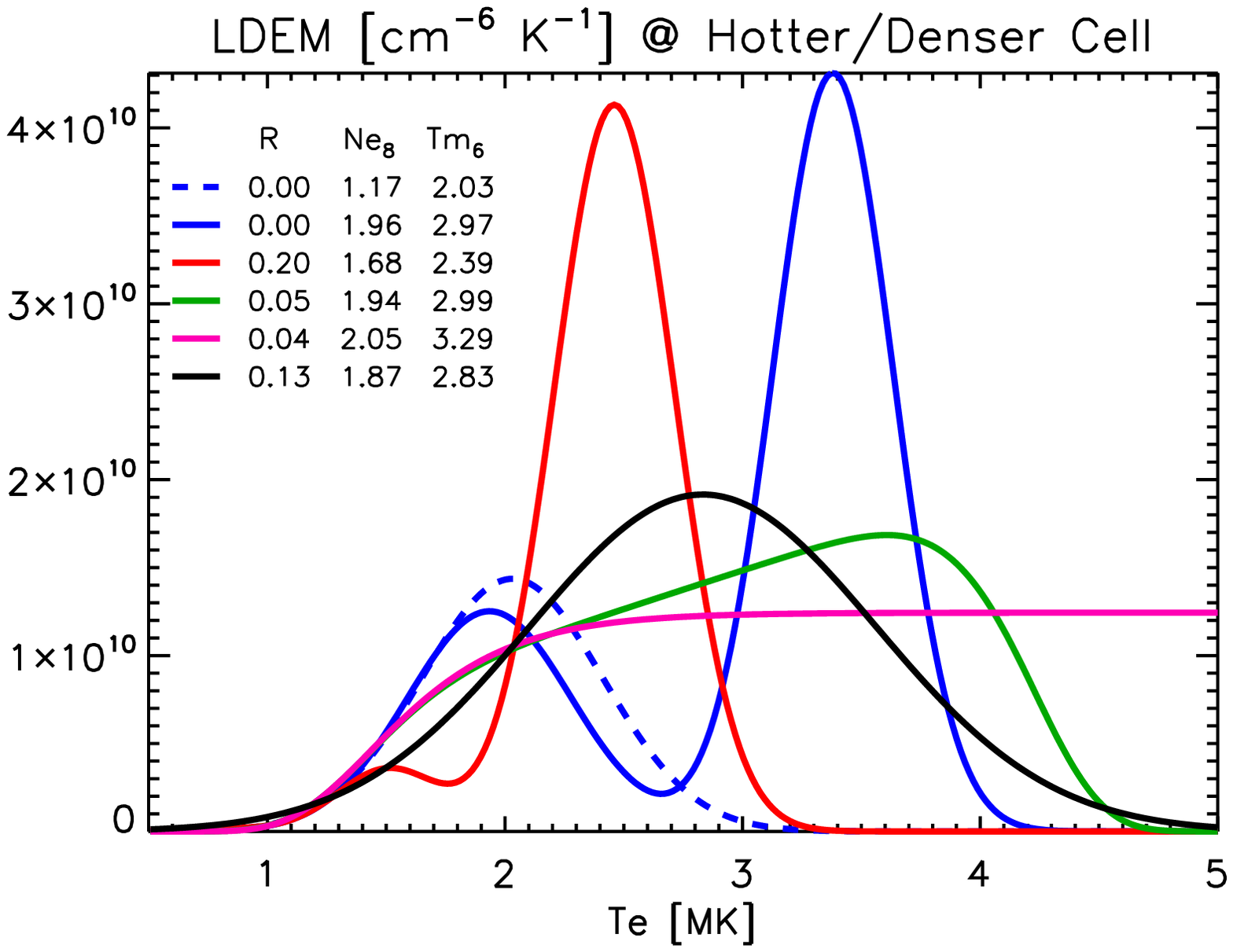}
\end{center}
\caption{All LDEM models for two sample cells of the magnetically closed corona, one belonging to a hotter and denser region than the other. {Inversion of AIA-3:} N1 (Dahed-Blue). {Inversion of 4 bands:} N2 (solid-blue), N4 (red), ATH (green), TH (pink) and N1 (black). The values of $R$, $N_e$ and $T_m$ for each model is shown in the top region of the Figure.}
\label{ejemplos_LDEM}
\end{figure}

We first highlight that the N1/AIA-3 (dashed-blue) and N2/AIA-4 (solid-blue) results achieve $R<10^{-2}$ in both coronal regions. Among the rest of the models, the one that best performs for AIA-4 in both voxels is the ATH model (solid-green). The reason is that by adjusting a positive or negative temperature gradient, this model is able to allocate more plasma in the large temperature end for the hotter cell, and in the low temperature end for the cooler one. In the case of the N1 model for AIA-4 data, the single normal adjusts its centroid to an intermediate temperature between the two modes of the N2 model, but it is of course unable to allocate plasma preferently in one temperature end or the other. Interestingly enough the N4 model does not perform so well. The reason for this is that this model can only adjust the 4 free amplitudes to allocate more or less plasma at each fixed centroid temperature, which affects neighboring bands as their respectives TRFs overlap. The reason the N2 model performs so well is that the three bands 171-193-211 \AA\ are well explained by the cooler component, while the sensitivity at the maximum sensitivity of the 335 \AA\ band is explained by the hotter component. The low-temperature sensitivity in the 335 \AA\ band (that overlaps that of the 171 \AA\ band) is also captured by the cooler component. 

{Finally, it is very interesting to note that the electron density and mean temperature predicted by the N2 model are almost the same as the values predicted by the ATH model, over the whole corona. This is shown in detail in Figure \ref{scateN2vsATH}, which compares the electron density and mean electron temperature obtained with both models. Even if the choice of a reasonably low $R$ parametrization is not unique, they achieve very similar 3D maps of the electron density and mean temperature. In this regard, the AIA-4 set is able to reveal extra plasma not previously detected by AIA-3, EUVI, or EIT, which accounts for an extra 14\% of coronal mass in the closed region, and a mean temperature 18\% higher.
} 

\begin{figure}[ht]
\includegraphics[width=0.49\linewidth]{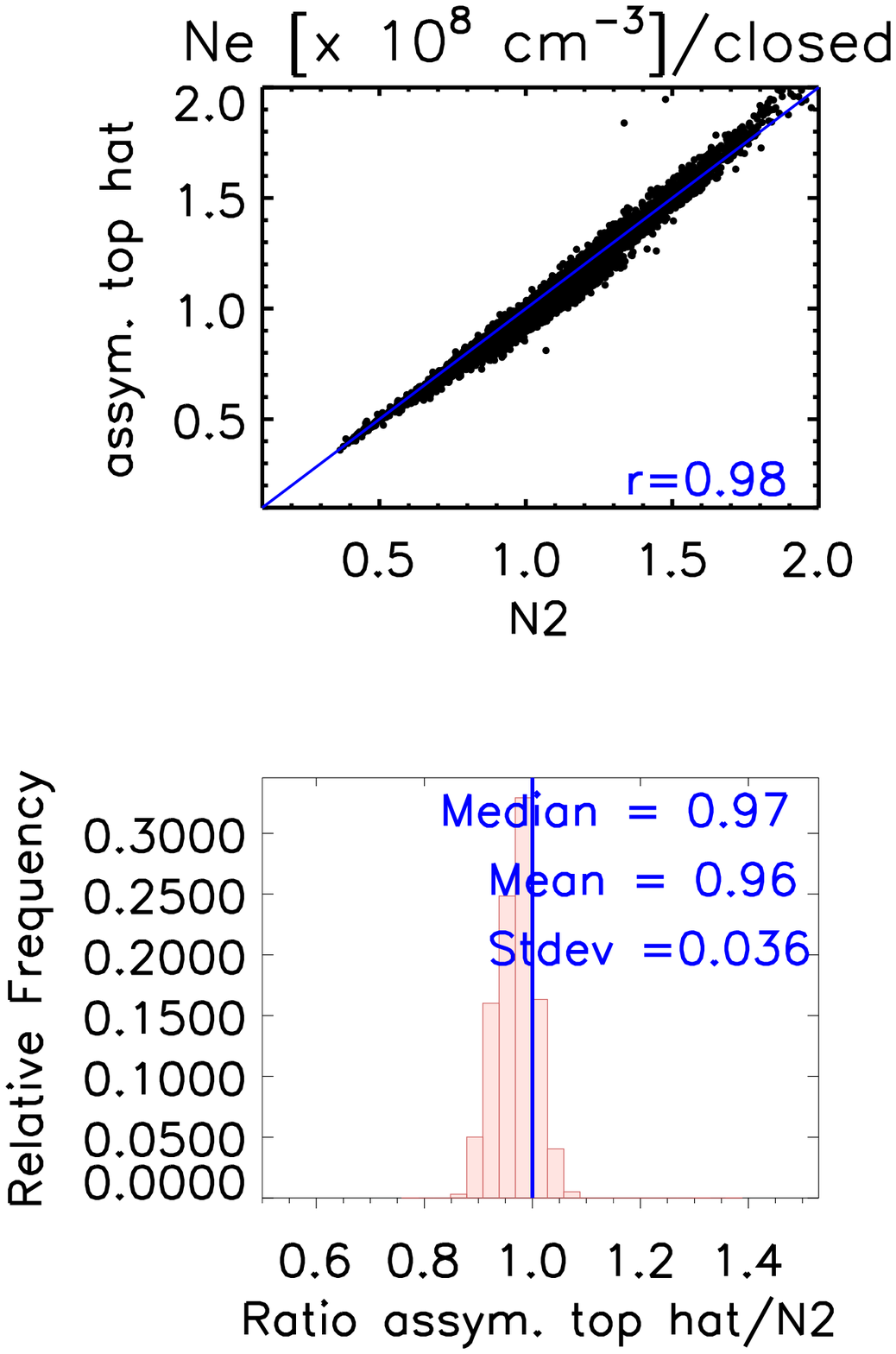}
\includegraphics[width=0.49\linewidth]{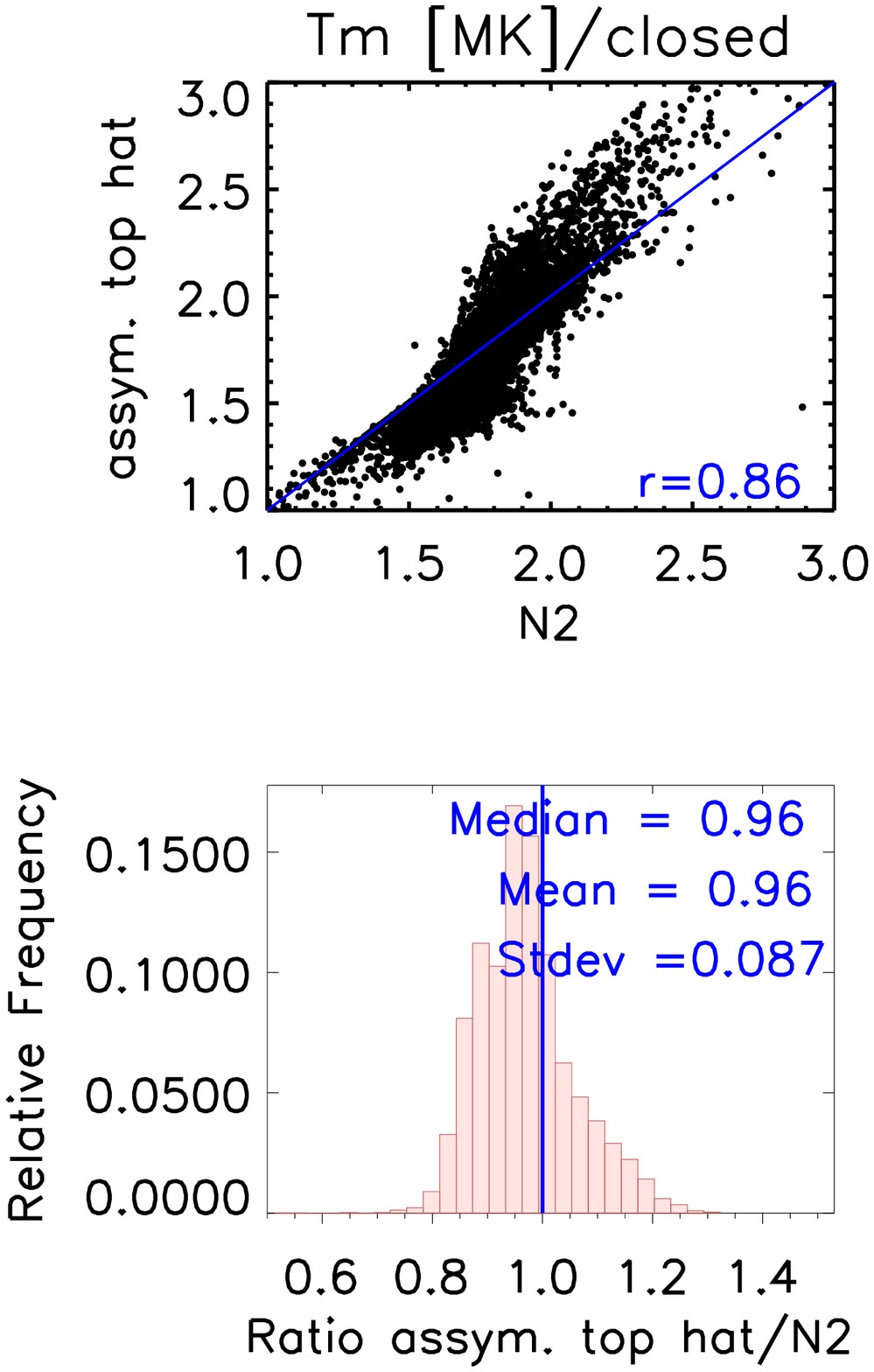}
\caption{{Comparison of DEMT results in the closed magnetic region between N2 (x-axis) and ATH models (y-axis) for the AIA-4 data set.}}
\label{scateN2vsATH}
\end{figure}

\subsection{Analysis of the Multimodal Corona}

In the previous section we established that the closed quiet corona is bimodal at the {spatiotemporal resolution of the tomographic grid and the temperature resolution of the AIA narrow band TRFs. In this section, we quantify how strong the bimodality of a given LDEM distribution is, by calculating the relative plasma fraction $\left(N_{e,j}/N_{e}\right)^2$ for each normal component} ($j=1,2$) of the N2 model. Figure \ref{AIA4_Nej_Ne} shows carrington maps of both $\left(N_{e,1}/N_{e}\right)^2$ and $\left(N_{e,2}/N_{e}\right)^2$, at 1.075 $\Rsun$. {It is readily seen that {the cooler component is dominant in most of the magnetically closed corona, except in localized regions, and in coronal holes}. The hotter component is significant in the  magnetically closed corona, and dominates the compact regions where the cooler component is more modest. To quantify the statistical relevance of the hotter component,} Figure \ref{histo_OC_Nej} shows the frequency histograms of $\left(N_{e,2}/N_{e}\right)^2$ at the same height in the open and closed regions, {separating the voxels within the QS (middle panel) from those in the ARs (bottom panel)}. {It is readily seen that the bimodality is much more important in the closed region than in the open region. Also, within the closed region, the bimodality is more important in ARs than in the QS}. {To quantify this, the fraction of voxels for which the hotter component is the dominant one, i.e. for which $\left(N_{e,2}/N_{e}\right)^2 \geqq 0.5$, is 48, 6 and 1 \% in the ARs, closed quiet corona, and open corona, respectively.}

\begin{figure}[ht]
\begin{center}
\includegraphics[width=0.75\textwidth]{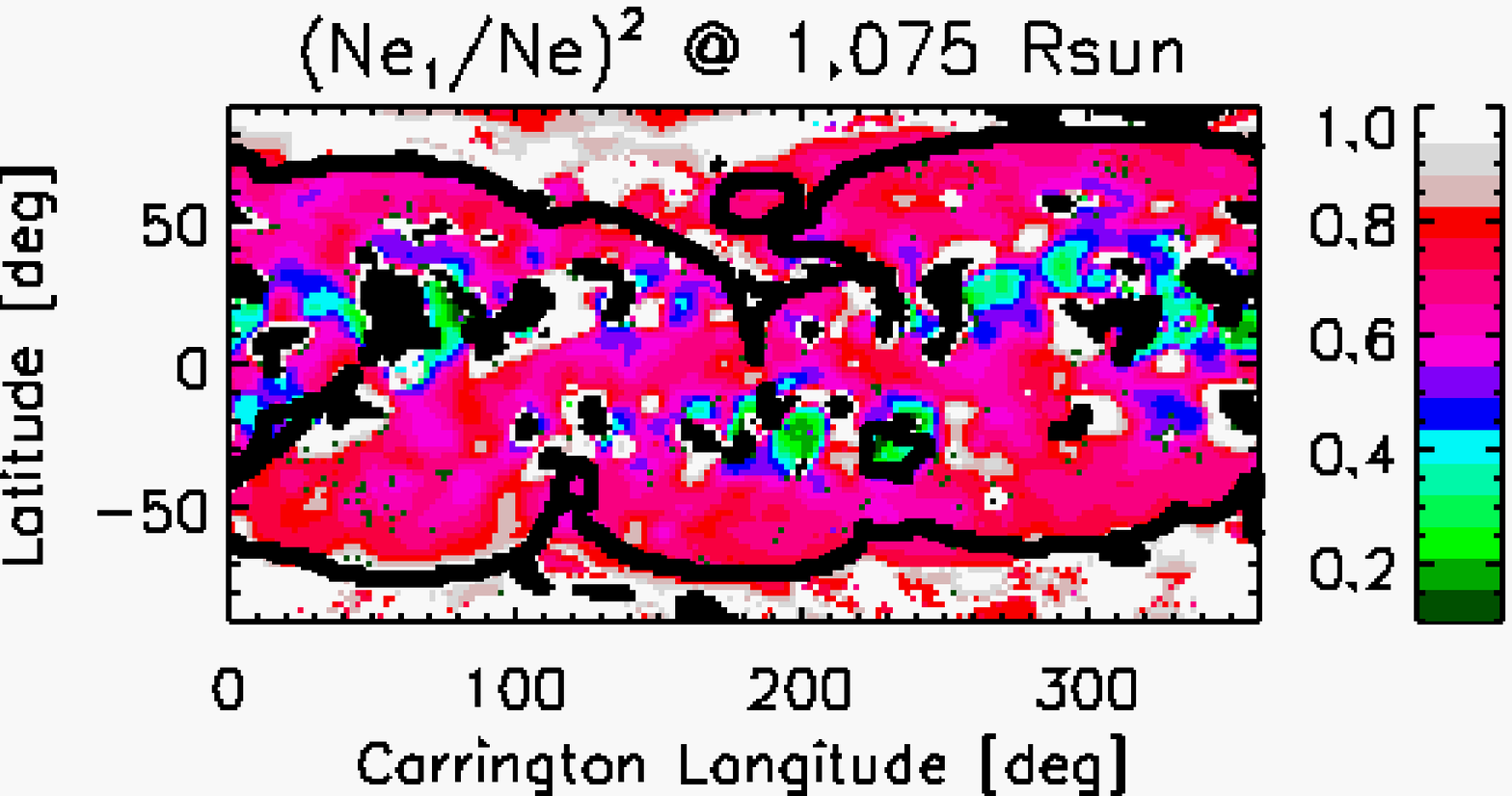}
\includegraphics[width=0.75\textwidth]{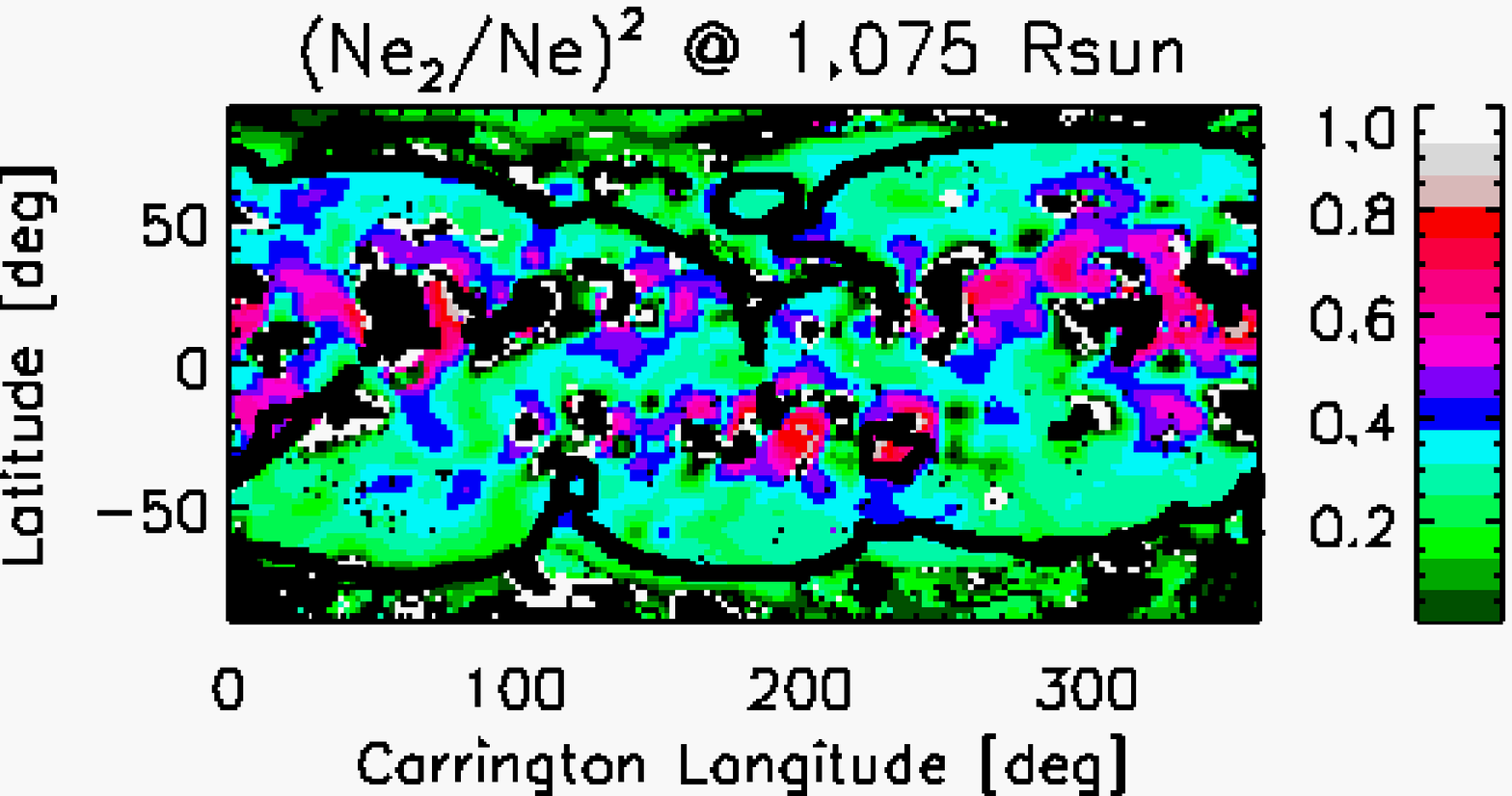}
\end{center}
\caption{For CR-2099 at 1.075 Rsun we show the relative fraction of plasma $\left(N_{e,j}/N_{e}\right)^2$ for each component ($j=1,2$) of the N2 LDEM when applied to the AIA-4 data.}
\label{AIA4_Nej_Ne}
\end{figure}

\begin{figure}
\begin{center}
\includegraphics[width=0.8\textwidth]{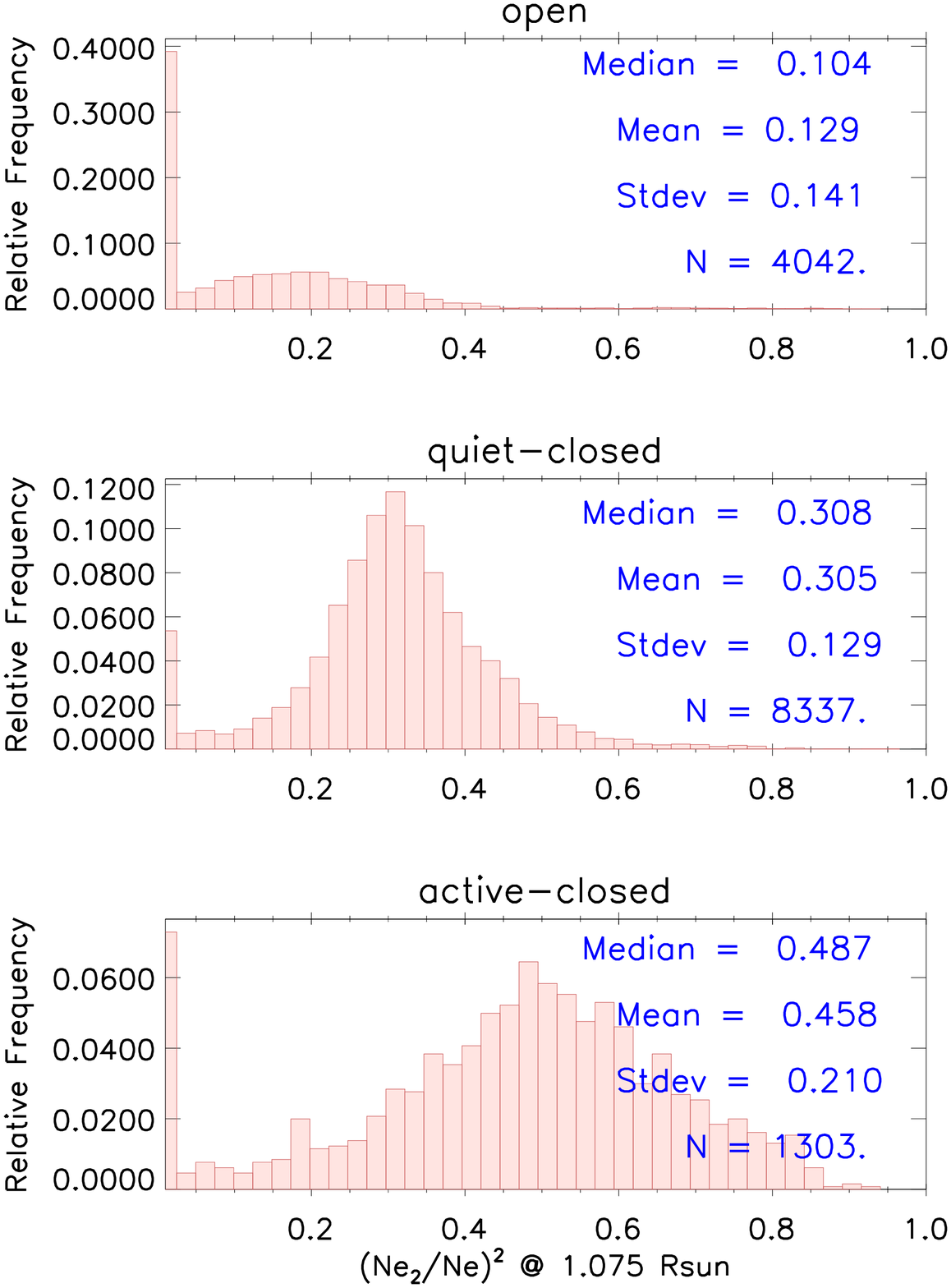}
\end{center}
\caption{{Histograms of $\left(N_{e,2}/N_{e}\right)^2$ at 1.075 $\Rsun$, which measures the fraction of plasma described by the hotter component of the parametric N2 model. The analysis is shown separately for the open region (top panel), the quiet corona in the closed region (middle panel), and the active corona in the closed region (bottom panel).}}
\label{histo_OC_Nej} 
\end{figure}

{Figure \ref{centroids} shows Carrington maps and histograms of the centroid $T_{0,1}$ of the cooler component and the centroid $T_{0,2}$ of the hotter one.} The values of $T_{0,1}$ are significantly {and sistematically} smaller in the open region, consistent with sub-MK temperatures in coronal holes reported by many studies {(V\'asquez et al. 2012; V\'asquez et al. 2011; Feng et al. 2009; Wilhelm, et al. 1998)}. On other hand, the values of the centroid $T_{0,2}$ distribute always around the maximum temperature sensitivity of the 335 \AA\ band (see table \ref{tabQklsensit}) throughout the closed corona. Within the open corona {the centroid $T_{0,2}$  tends to be systematically smaller, yet statistically close to the maximum temperature sensitivity of the 335 \AA\ band}. {As commented above in relation to Figure \ref{histo_OC_Nej}, the hotter component of the model N2 is much weaker in the open region. This is due to the colder temperatures that dominate the open corona and the correspondingly much weaker 335 \AA\ emissivity. Still, the model N2 is the most successful one also in the open corona, as it allows to make the hotter component as weak as needed, yet still present. Unimodal models can mimic this by increasing their width, at the cost of diminishing the synthetic-to-observed agreement for the colder bands, then increasing the value of the score $R$.}

\begin{figure}[ht]
\begin{center}
\includegraphics[width=0.49\textwidth]{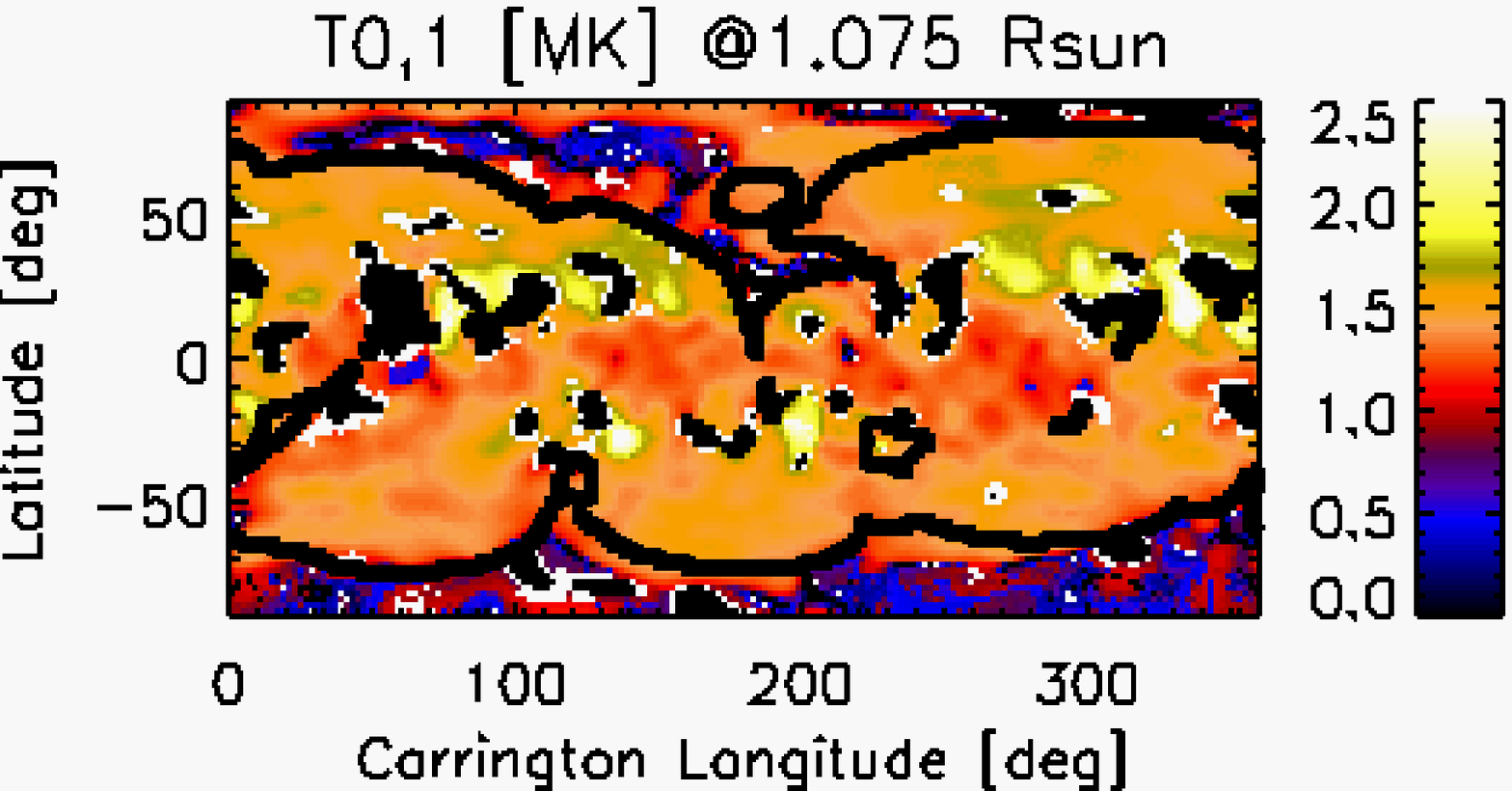}
\includegraphics[width=0.49\textwidth]{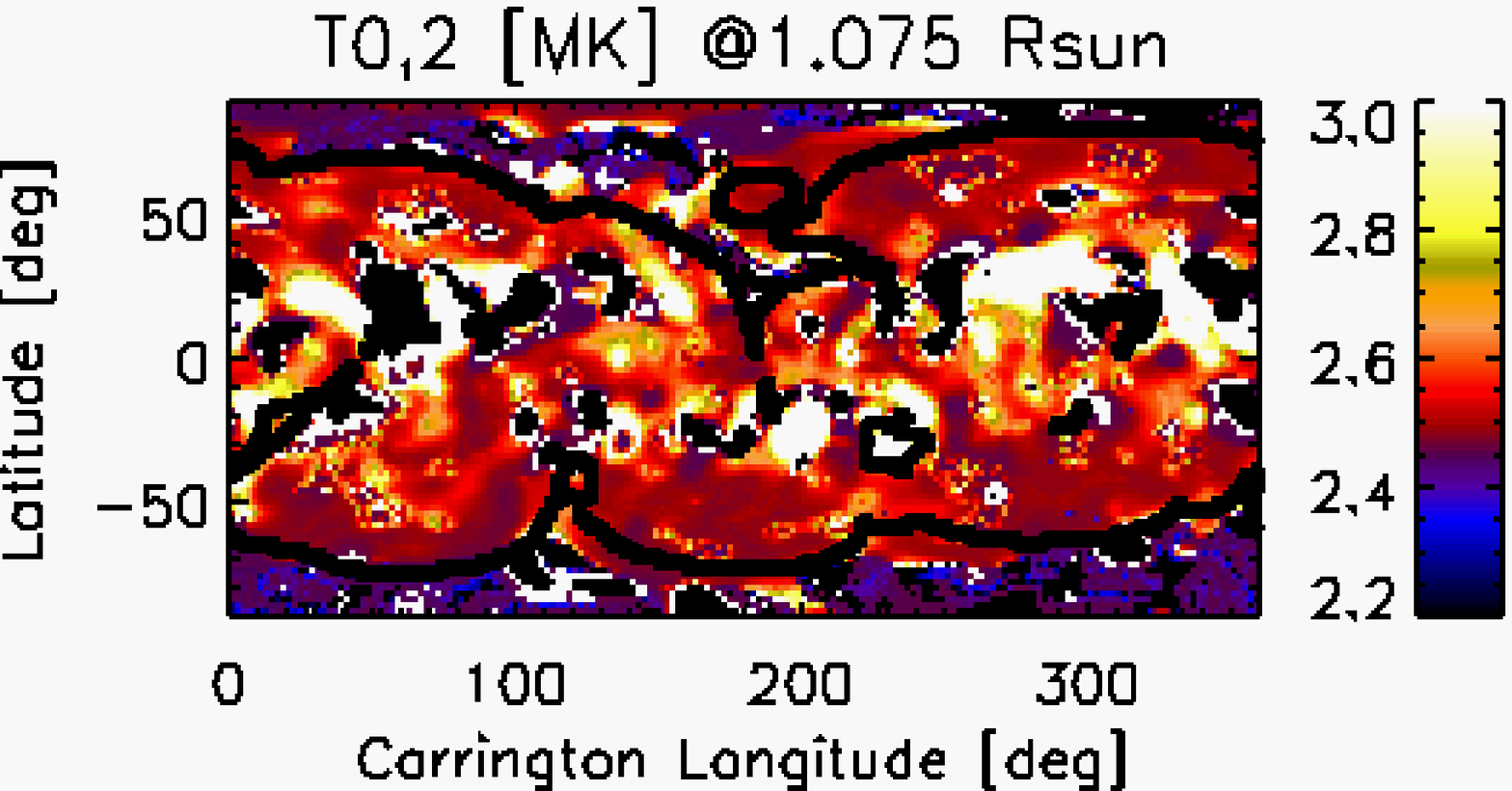}
\includegraphics[width=0.49\textwidth]{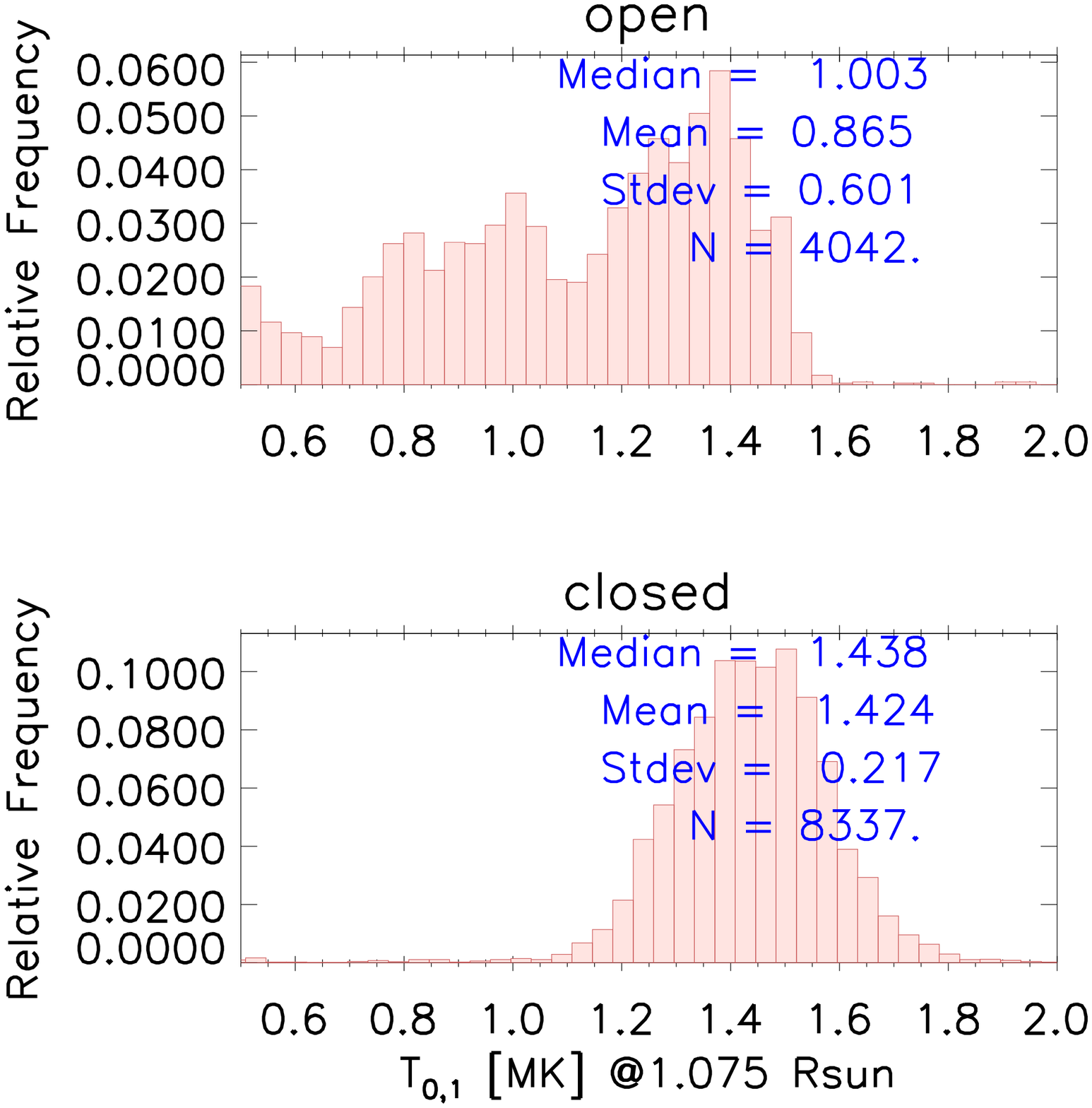}
\includegraphics[width=0.49\textwidth]{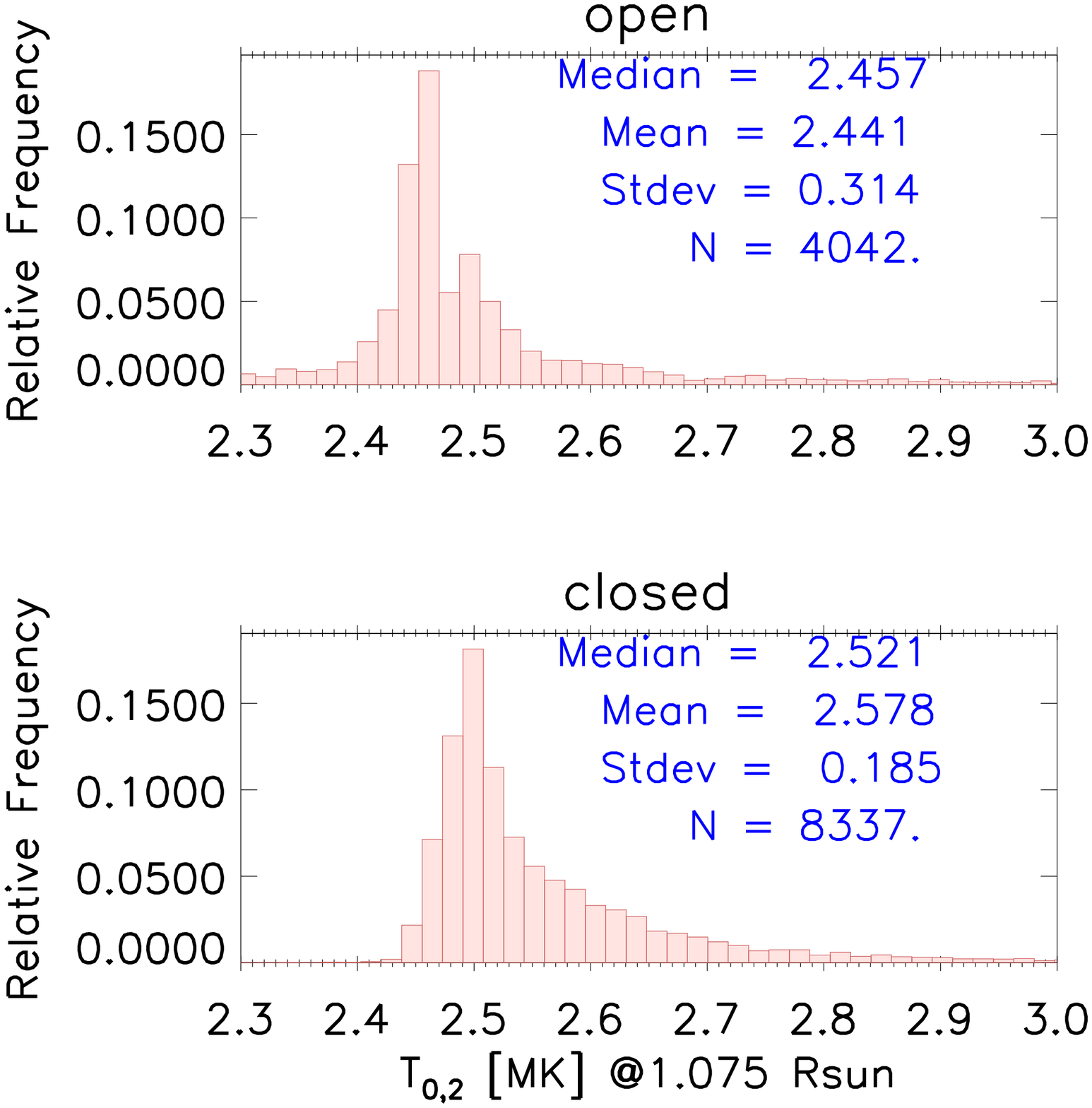}
\end{center}
\caption{Top panels: Carrington maps at 1.075 $\Rsun$ of the centroids of the first ($T_{0,1}$) and the second normal distribution ($T_{0,2}$).
Center and Bottom panels: Histograms of the centroids shown in the top panels in the open and closed magnetic region, respectively.} 
\label{centroids}
\end{figure}

{In order to gain understanding of which locations exhibit more or less bimodality, we correlated the results of the AIA-4 and AIA-3 studies in the quiet corona. The left panel in Figure \ref{scatter_Nej} shows a scatter plot of $\left(N_{e,2}/N_{e}\right)^2$ obtained with AIA-4/N2 versus the electron density value $N_e$ obtained with AIA-3/N1. The orange dots indicate voxels of the magnetically closed region, while blue dots indicate voxels of the magnetically open region. The right panel shows a scatter plot of $\left(N_{e,2}/N_{e}\right)^2$ obtained with AIA-4/N2 versus the electron mean temperature value $T_m$ obtained with AIA-3/N1. These plots show that, in the closed quiet corona, the bimodality is stronger for denser and hotter regions. it also shows that when the cooler component is denser the same happens with the hotter component. As we will see in the discussion section this fact places an important constraint on heating models of the quiet corona. Finally, in the open region the same trends are observed and, in addition, the bimodality is on the average much less important than in the closed region, as already shown in Figure \ref{histo_OC_Nej}.}

\begin{figure}[ht]
\begin{center}
\includegraphics[width=0.49\textwidth]{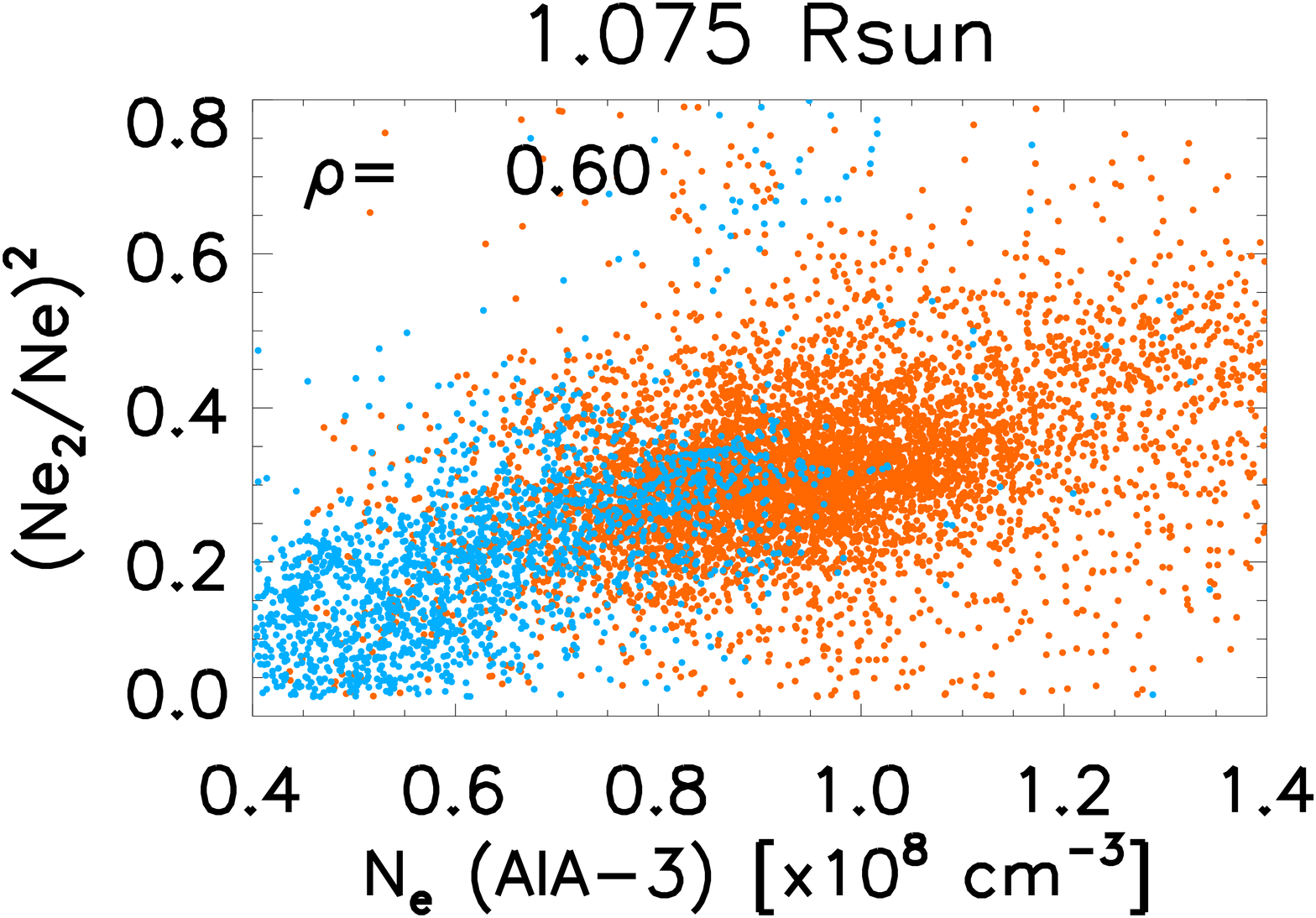}
\includegraphics[width=0.49\textwidth]{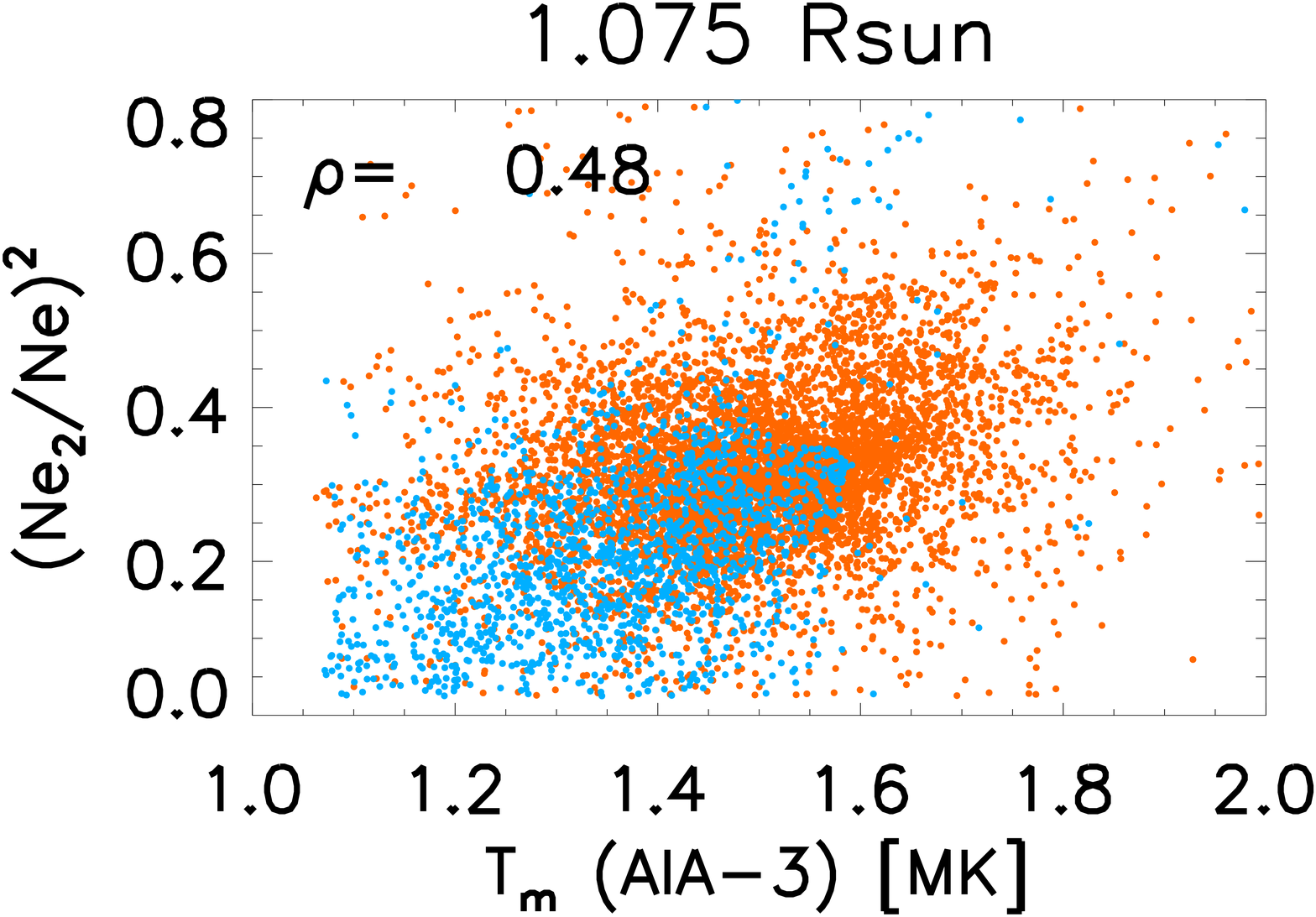}
\end{center}
\caption{Left panel: Scatter plot of $N_e$ of AIA-3 versus $\left(N_{e,2}/N_{e}\right)^2$ at 1.075 $\Rsun$. Right panel: Scatter plot of $T_m$ of AIA-3 versus $\left(N_{e,2}/N_{e}\right)^2$ at 1.075 $\Rsun$. In both panels, the blue dots corresponds to voxels in the open region and the orange dots correspond to voxels in the closed region. {In each panel the Pearson correlation coefficient $\rho$ between both plotted quantities is indicated.}}
\label{scatter_Nej}
\end{figure}

{For the analysis shown in this section we chose 1.075 $\Rsun$ as an intermediate sample height. Similar results were found at all tomographic heights. We investigated the dependence with height of both $N_{e,2}(r)$ and $N_{e,1}(r)$ within the quiet region of the streamer belt, similar to V\'asquez et al. (2011). The scale heights of both quantities differ by only about 12\%, so that the ratio $\left(N_{e,2}/N_{e,1}\right)^2$ is roughly constant with height, showing a variability of less than 10\% over the range 1.025 to 1.20 $\Rsun$.} 

{Spectral observations of the corona allow studying the temperature of the emitting plasma by means of the \emph{emission measure} (EM) technique. If the number of available spectral lines is large enough, the EM technique is a powerful tool to assess whether a plasma is isothermal or multithermal. Systematic studies of the temperature structure of the solar corona using this technique have been recently reviewed by Feldman \& Landi (2008). They conclude that the existing EM analysis of the corona support a scenario in which the coronal plasma is a combination of a few nearly-isothermal populations, with well defined temperatures. They find coronal holes to be characterized by temperatures around ${\rm log}(T)=5.95$, quiet sun regions by temperatures around ${\rm log}(T)=6.15$, and ARs appear characterized by two distinct populations with temperatures ${\rm log}(T)=6.15$ and ${\rm log}(T)=6.5$.  While these works did not show bi-modal distributions, the temperatures reported therein are consistent with our results. According to the histograms in the bottom panels of Figure \ref{centroids}, the closed quiet sun is characterized by two well defined temperatures given by the mean values ${\rm log}<T_{0,1}>=6.15$ and ${\rm log}<T_{0,2}>=6.41$. The mean temperature of the cooler component of the ubiquitous bimodal LDEM found by the DEMT study matches the characteristic temperature of the quiet sun reported by the EM studies. The mean temperature of the hotter component of the LDEM matches that of the hotter component of the EM studies in ARs.}

Those EM studies where there is only a warm component (close to $T_{0,1}$) may not have included enough hot lines, or may have been centered on regions where the hotter component (close to $T_{0,2}$) was smaller. A weaker signal of the hotter component may also be the result of the LOS-integration encompassing more regions with the cooler component being dominant over the hotter component, giving a cumulative larger importance to the former than to the latter, which then does not provide enough counts to be observed.  The case may also be that the reported EM studies of quiet sun regions correspond to epochs when the hotter component found in this DEMT study (that corresponds to the early rising phase of solar cycle 24) was particularly weak. In this regard, Landi \& Testa (2014) find that overall the thermal structure of streamers does not change much over the solar cycle, although with a few spectroscopic measurements. LOS-integrated spectral studies may also be consistent with more than one characteristic temperature (Ralchenko et al., 2007). The question rises in that case of whether such multithermality is an artifact of the LOS integration of different large-scale structures at very different places. In this context it should be highlighted that in the results of this paper the multithermality is a local property.

\subsection{2D DEM analysis}
\label{DEM}

The technique for the parametric inversion of the LDEM used in this work is the same as has been used for all the previously published DEMT work {(for example Frazin et al. 2009, V\'asquez et al. 2010, 2011, and Nuevo et al. 2013)}.  As explained in Section \ref{DEMT}, the same technique can be used to find the standard 2D DEM from simultaneous EUV image sets. With the aim of cross-validating our parametric inversion technique, we use it to perform a standard DEM analysis of an AR and the diffuse region around it. We used the AIA-4 images taken at {12:00} UT on {2010, July, 25}, and found the DEM at each pixel using the N2 model. By taking the zeroth moment of each DEM we find the emission measure (EM [${\rm cm}^{-5}$]) of each pixel.

{The left panel of Figure \ref{ARDEM} shows, in gray scale, an image of the EM of the analyzed region, which corresponds to AR {NOAA 11089}. The color boxes, of size $5\times 5$ pixels, identify selected regions in the image. Using the same color code, the right panel shows the average DEM model obtained within each color box. This figure can be directly compared with a similar analysis shown in Figure {6} of  Narukage et al. (2014), which performed a MCMC inversion. Their technique to determine the DEM, described in Winebarger et al. (2012), uses 7 spectral lines of the {{\emph{Extreme-ultraviolet Imaging Spectrograph}} (EIS) on board the Hinode} (1-6 MK) and 2 broadband images of the {{\emph{X-Ray Telescope}} (XRT) on board Hinode}. {In their Figure 6, the authors show the DEM obtained in each selected region, which are indicated in their Figure 7. Note that we have chosen the color code of this work in a similar fashion to that of Figures 6 and 7 of Narukage et al. (2014), with the red/yellow colors corresponding to the core of the AR core, dark/blue colors corresponding to the dark region around the AR, and green/light-blue colors corresponding to intermediate regions.}

There are several similarities between both works that we should highlight. In the comparison one should bear in mind that the specific ARs analyzed in both works is not the same. Firstly, the MCMC analysis of Narukage et al. (2014) shows in a bimodal DEM in virtually all regions, except for pixels in the deep core of the AR, where a third hotter component may arise. Secondly, in both analyses, as one moves from the dimmer regions surrounding the AR into the deep core of the AR, the area of both the cooler and the hotter component increases, the centroids of the two components gradually shift to larger temperatures, and the hotter component becomes dominant. In both works, the range of temperatures of the plasma described by the DEM exhibits similar values. For the MCMC analysis the centroid of the cooler component ranges from 1.2 to 1.5 MK, while the centroid of the hotter component ranges from 2.5 to 4.0 MK. In the parametric analysis of this work, the centroid of the cooler component ranges from 1.5 to 2.25 MK, while the centroid of the hotter component ranges from 2.5 to 4.0 MK. The absolute values of the DEM in units of [${\rm cm}^{-5}\,{\rm K}^{-1}$] ({that we will omit in the rest of this paragraph}) are also similar in both analyses, with comparable orders of magnitude. In the MCMC analysis, the amplitude of the cooler component is typically within the range $1-4\times10^{21}$, while the amplitude of the hotter component ranges from $0.5-10 \times 10^{21}$. The parametric analysis of this work, the amplitude of the cooler component ranges between $0.3-6\times 10^{21}$, while the amplitude of the hotter component ranges between $0.2-70 \times10^{21}$. {In the core of the AR, the DEMs by Narukage et al. (2014) predict the intensity of the key EIS spectral lines (Winebarger et al., 2012) within a 30\% precision. That level of agreement is comparable to the requirement $R<0.2$ we have set for the voxels that are included in our tomographic analysis.
}

\begin{figure}[ht]
\begin{center}
 \includegraphics[height=0.4\linewidth] {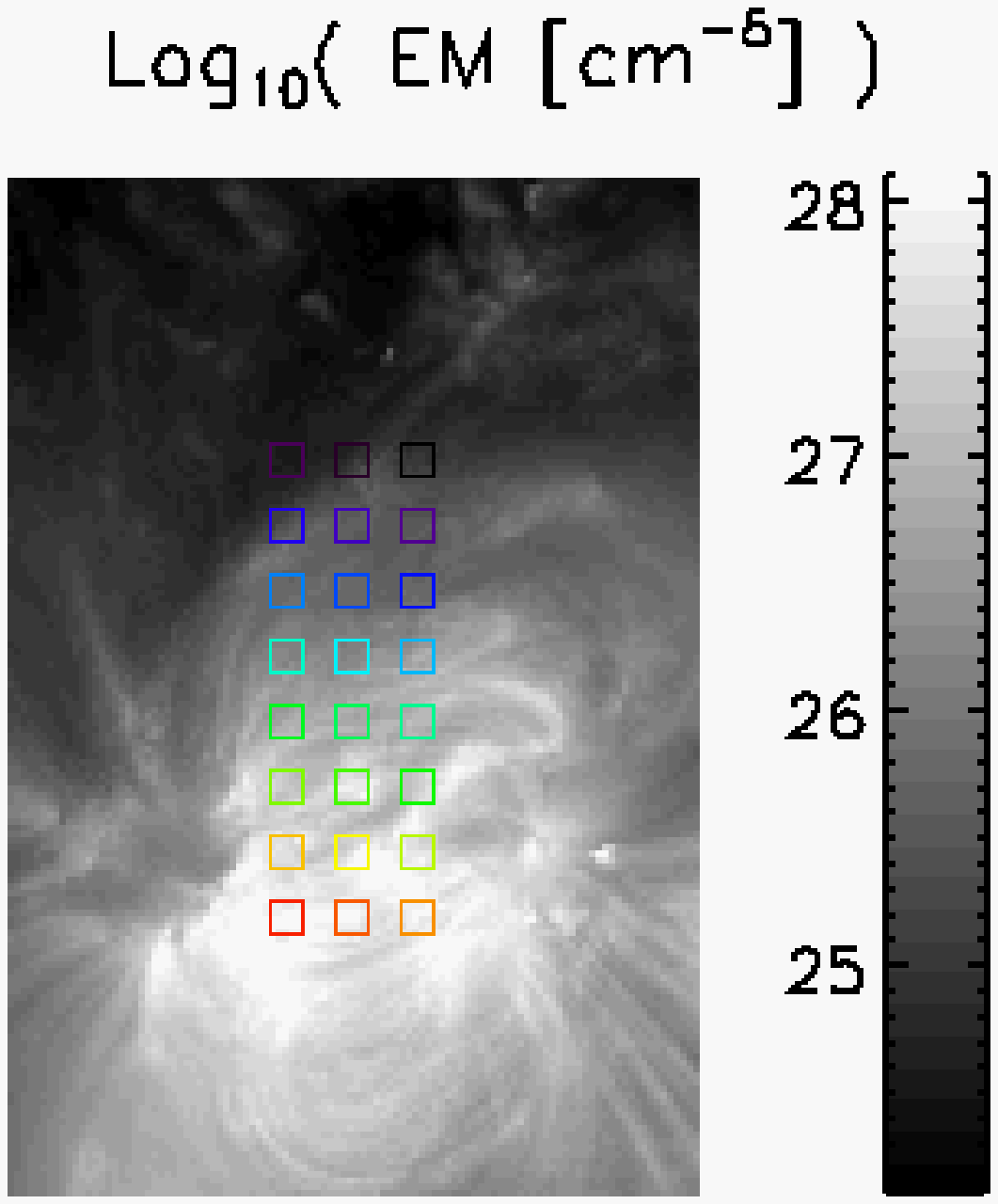}
\includegraphics[height=0.35\linewidth]{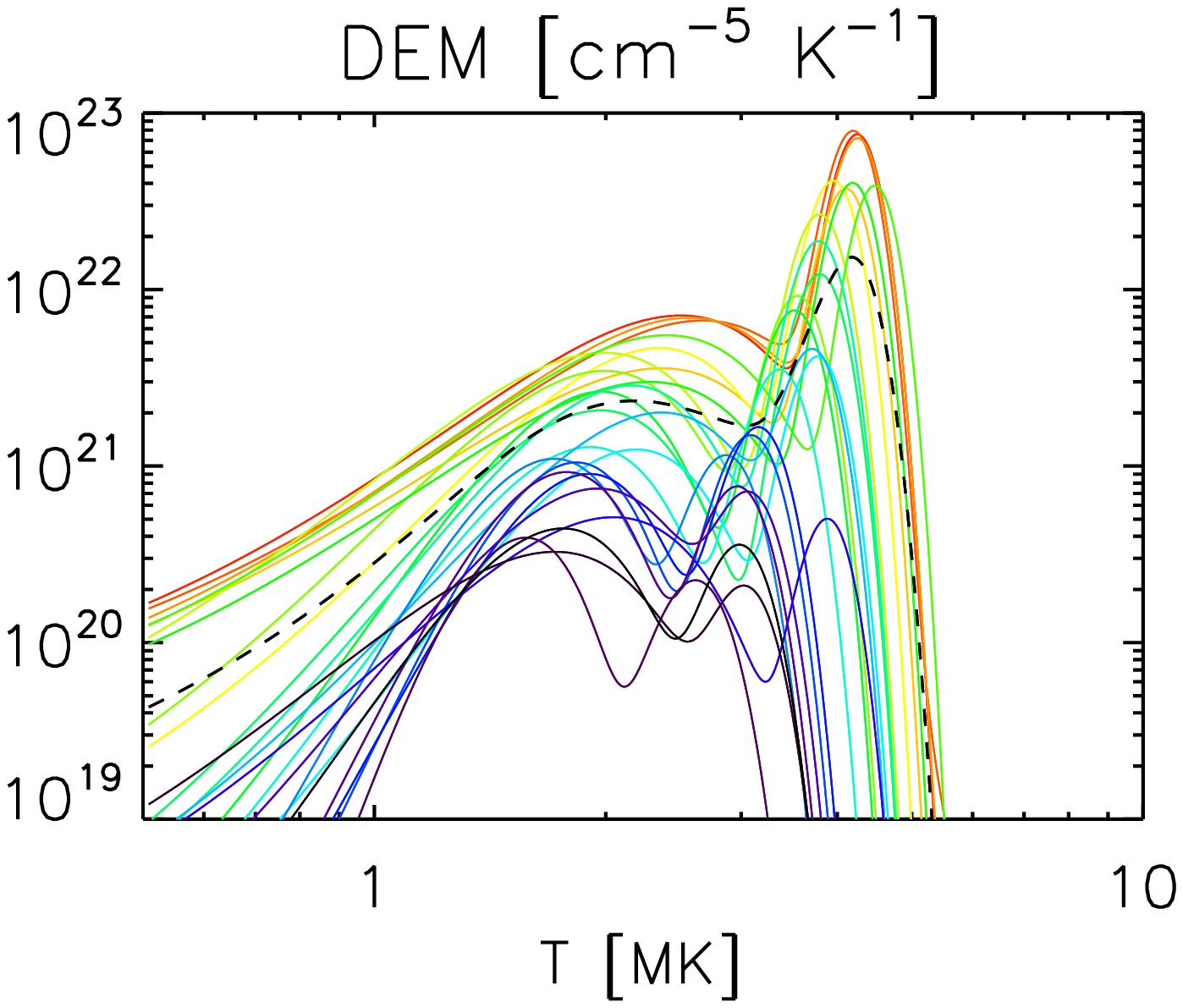}
\end{center}
\caption{{Left panel: EM image based on the parametric inversion of an AIA-4 image set of AR {NOAA 11089, taken on 2010, July, 25, 12:00 UT}. The color boxes, of size $5\times 5$ pixels, highlight selected subregions in the AR and the diffuse region around it. Right panel: Average DEM curve obtained for each color box of the top panel (same color code), using the parametric N2 model.}}
\label{ARDEM}
\end{figure}

\section{Discussion and Conclusions}

In this work the DEMT technique is applied for the first time to 4 coronal bands instead of the usual 3 bands used in all previous works. The 4 bands used are those of the SDO/AIA instrument that are more suitable for studying the quiet corona, namely the bands of 171, 193, 211, and 335 \AA. The analysis of simulated data (Section \ref{simul}) shows how the inverted LDEM is able to succesfully reproduce the electron density and mean temperature of an assumed LDEM (derived from its zeroth and first order moments), even when both LDEM parametric models are not the same. It also clearly illustrates how the EUV bands are incapable of detecting plasma at temperatures outside their sensitivity range, and thus how the LDEM aims at describing the plasma only in that sensitivity range.

This work is the first to apply DEMT to AIA data and, therefore, a comparison of results against those based on EUVI data is included. The analysis indicates that there is a high degree of  consistency between the DEMT results obtained with AIA-3 and those obtained with EUVI. DEMT quantitative results based on AIA-3 data can then be directly compared with previous published DEMT reconstructions based on EUVI data.

Using the AIA-3 data set, the single normal (N1) parametric model of the LDEM (3 free parameters) is able to succesfully reproduce the tomographic emissivities of all bands (171, 193, and 211 \AA), but it can not expain the AIA-4 data (adding the 335 \AA\ band). Other symmetric and asymmetric unimodal LDEM parameterizations were tested for AIA-4. In all cases the tomographic versus synthetic agreement is not nearly as good as when using 3 bands, despite having 4 or 5 free parameters. 

Multimodal LDEM were modeled using a variety of combinations of partially conditioned normal functions, an approach similar to {Aschwanden \& Boerner (2011)}. The model that consistently achieves the best goodness-of-fit score $R$ is the bi-normal (N2) model, which is a superposition of two normal distributions with distinct cool and hot components at $\left<T_{0,1}\right> \sim 1.4\,{\rm MK}$ and ​$\left<T_{0,2}\right> \sim 2.6\,{\rm MK}$. Our study shows that the quiet corona LDEM is multimodal at {the spatial resolution of the tomographic grid, which is $0.01 \ \Rsun \times 2^\circ \times 2^\circ$, or about $(7\times10^3 {\rm km}) \times [(2.44\times10^4 {\rm km})^2]$ for a representative voxel at a height of 0.1 $\Rsun$ above the phostosphere at the equator.} 

{The analysis in Section \ref{aia4} for AIA-4 shows that any choice of parametrization that reasonably reproduces the emissivity in all bands leads to very similar estimates of both density and mean temperature (see Figure \ref{ejemplos_LDEM}), which is often the reason to perform DEMT. Using the AIA-4 set it is found that the 335 \AA\ band reveals plasma previously not seen by instruments of 3 bands (EUVI), which provides for an increase of about 14\% of the mean electron density in the closed quiet corona when compared to results with AIA-3, as well as for rise of about 18\% of the estimated mean temperature.}

{The bimodality is found to be stronger for denser and hotter regions. The cooler component of the bimodal LDEM is dominant in coronal holes and most of the magnetically closed corona, except in compact active regions where the cooler component is more modest. The fraction of voxels for which the hotter component is the dominant one, i.e. for which $\left(N_{e,2}/N_{e}\right)^2 \geqq 0.5$, is 48, 6 and 1 \% in the ARs, closed quiet corona, and open corona, respectively. To quantify the bimodality of the plasma in the quiet diffuse corona, the density added by the hotter component of the N2 model when using the AIA-4 data was compared to: a) the electron density computed from the AIA-3 data set, and, b) the mean electron temperature computed from the AIA-3 data set. In the first case the Pearson correlation coefficient is $\rho=0.6$, while in the later one is $\rho=0.5$, revealing that the bimodality is stronger for denser and hotter regions (see Figure \ref{scatter_Nej}).} 

{Each tomographic cell is threaded by a large number of thermally isolated coronal loops and the LDEM obtained at any given voxel is due to the many small-isothermal pieces of loops, each one at its own temperature. The fact that the LDEM at any given voxel is bimodal indicates that the typical voxel contains two clearly distinct populations of loops: i) "warm" loops (described by the cooler component of the N2 model) and, ii) "hot" loops (described by the hotter component of the N2 model). Those two kinds of loops are ubiquitous in the quiet corona, with one or the other being dominant in any given region depending on the local average density and temperature.}

{Different kinds of physical phenomena have been proposed to explain coronal heating, such as dissipation of Alfv\'en waves, nanoflaring, in both ARs (Schmelz et al. 2010; Aschwanden \& Boerner 2011) and the diffuse quiet corona (Benz \& Krucker, 2002; Freij et al., 2014; Uritsky \& Davila, 2014; Hahn \& Savin, 2014). It has been suggested that nanoflare heating could be mostly operating not directly in the corona but rather at chromospheric levels, and then the heated plasma injected in the corona, forming the so called type II spicules \citep{depontieu20007, depontieu2011}. Assuming this scenario, \citet{klimchuk12} have explored its observational consequences for predicted EUV Fe spectral lines in an analytical fashion, while \citet{klimchuk14} have done the same numerically. Both studies have found very large discrepancies with the observations (of up to one to two orders of magnitude), in terms of line intensities, blue shifts, and blue-red asymmetries. They conclude that chromospheric nanoflare can only explain a very small fraction of the coronal heating, and that the mechanism should be acting locally in the corona. Moreover, based on geometric and energy release considerations, Klimchuk (2015) estimated that a typical frequency for nanoflaring in the quiet corona should be of order 200 sec. In such a scenario, the stable temperatures of the quiet corona are explained in terms of omnipresent high frequency nanoflaring occuring in a constant fashion, with coronal magnetic free energy being permanently "recharged" as field lines are driven by the constant motion of their footpoints.}

{Assuming its coronal nature then, nanoflare heating of the homogeneous plasma of the diffuse quiet sun should occur randomly throughout the whole volume of any given tomographic voxel, mixing neighboring magnetic threads inside each coronal loop within the voxel. In such a scenario, a broad distribution of temperatures is to be expected within each magnetic loop ({Klimchuk 2006, Aschwanden \& Boerner, 2011}), which should be detected by the LDEM of the tomographic voxel. In fact, since the LDEM provides a time-averaged description of the thermodynamics of loops threading the quiet corona, and the result in each tomographic voxel is in itself an average of the plasma contained in such a voxel, it seems reasonable to expect that the LDEM in each voxel should be very well described by broad unimodal distributions, such as the TH or ATH models. The observed bimodality of the LDEM seems then difficult to explain in a nanoflare heating scenario.}

{The cooler component of the model N2 of the LDEM is an average description of the thermodynamics of the warm loops populating the voxel, while the hotter component is an average description of the hot loops within the same voxel. In a wave-dissipation dominated heating scenario, the reason for having neighboring warm and hot loops within any given voxel is to be found in the inhomogeneous chromospheric boundary conditions, where the loops are rooted and the injection of energy in waves takes place. Inhomogeneities will lead to injection of more energy in waves into some loops (hot ones) and less energy into others (warm ones). In this context, we speculate that the observed bimodality of the LDEM may be the result of some sort of effective thermodynamical selection effect, allowing only loops with very specific characteristics (plus some range of variability) to survive long enough.}

The multimodality is even stronger in ARs, but the tomography is not well suited to study such inhomogenoeus and quickly evolving regions. In Section \ref{DEMT} we showed how 2D DEM analysis can be performed from instantaneous image sets using the same parametric inversion technique for the DEM (Equation \ref{I_DEM}). We performed such study on a sample AR using images in all bands of the AIA-4 set, assuming the N2 parametrization for the DEM. We studied the core of the AR and its surrouding diffuse, selecting regions in a similar fashion to a recent study by {Narukage et al. (2014)}, who used a MCMC approach to derive DEM models for an AR. Our models in the AR and diffuse region exhibit several similarities to Narukage's results. Firstly, Narukage's models are typically bimodal, with the values of the centroids and amplitudes being similar to our results. Also, both models show a stronger bimodality in the core of the AR than in its surroundings, with the centroid temperature and area  of the second normal component increasing as one gets deeper in the core of the AR (see Figure \ref{ARDEM}). {This comparison serves as a validation of the parametric inversion of the LDEM that is used in DEMT studies. It also gives support to the bimodal LDEM ubiquitously present in the quiet diffuse corona as a real physical property that heating models should be able to reproduce, in particular when integrated over the spatiotemporal resolution of the tomographic grid.}

\acknowledgments
{F.A.N. acknowledges the CONICET Type I pre-doctoral fellowship that supports his participation in this research. The work of E.L. was supported by several NSF and NASA grants.}

\end{document}